\documentclass[a4paper,11pt]{article}
\pdfoutput=1 

\usepackage{jheppub} 

\usepackage[T1]{fontenc} 

\usepackage{amssymb}
\usepackage{graphicx}
\usepackage{color}
\usepackage{tabls}
\usepackage{hyperref}
\usepackage{bm}
\DeclareMathOperator{\SU}{\mathrm{SU}}

\newcommand{\eq}[1]{\begin{equation}\label{#1}}
\newcommand{\en}{\end{equation}}
\newcommand{\eqar}[1]{\begin{eqnarray}\label{#1}}
\newcommand{\enar}{\end{eqnarray}}

\title{\boldmath Lattice QCD study of inclusive semileptonic decays of heavy mesons}

\author[a]{Paolo~Gambino,}
\author[b,c]{Shoji~Hashimoto,}
\author[a,d]{Sandro~M\"achler,}
\author[a]{Marco~Panero,}
\author[e]{Francesco~Sanfilippo,}
\author[e]{Silvano~Simula,}
\author[a]{Antonio~Smecca}
\author[f]{and Nazario~Tantalo}

\affiliation[a]{Dipartimento di Fisica, Universit\`a di Torino \& INFN, Sezione di Torino,\\Via Pietro Giuria 1, I-10125 Turin, Italy}
\affiliation[b]{Theory Center, Institute of Particle and Nuclear Studies, High Energy Accelerator Research Organization (KEK), Tsukuba 305-0801, Japan}
\affiliation[c]{School of High Energy Accelerator Science, The Graduate University for Advanced Studies (SOKENDAI), Tsukuba 305-0801, Japan}
\affiliation[d]{Physikinstitut, Universit\"at Z\"urich, Winterthurerstrasse 190, CH-8057 Z\"urich, Switzerland}
\affiliation[e]{INFN, Sezione di Roma Tre, Via della Vasca Navale 84, I-00146 Rome, Italy}
\affiliation[f]{Dipartimento di Fisica, Universit\`a di Roma ``Tor Vergata'' \& INFN, Sezione di Roma ``Tor Vergata'', Via della Ricerca Scientifica 1, I-00133 Rome, Italy}

\emailAdd{paolo.gambino@unito.it}
\emailAdd{shoji.hashimoto@kek.jp}
\emailAdd{sandro.machler@unito.it}
\emailAdd{marco.panero@unito.it}
\emailAdd{francesco.sanfilippo@infn.it}
\emailAdd{silvano.simula@roma3.infn.it}
\emailAdd{antonio.smecca@unito.it}
\emailAdd{nazario.tantalo@roma2.infn.it}

\abstract{We present an \emph{ab initio} study of inclusive semileptonic decays of heavy mesons from lattice QCD. Our approach is based on a recently proposed method, that allows one to address the study of these decays from the analysis of smeared spectral functions extracted from four-point correlators on the lattice, where the smearing is defined in terms of the phase-space integration relevant to the inclusive decays. We present results obtained from gauge-field ensembles from the JLQCD and ETM collaborations, and discuss their relation with theoretical predictions from the operator-product expansion.}

\begin{document} 

\begin{flushright}
  KEK-CP-0390
\end{flushright}

\maketitle
\flushbottom

\section{Introduction}
\label{sec:introduction}
The theoretical study of semileptonic decays of $B$ mesons continues to be an important and very active area of research in high-energy physics: this interest is mainly driven by the fact that these decays encode direct information on the modulus of two of the elements of the Cabibbo-Kobayashi-Maskawa (CKM) quark mixing matrix~\cite{Cabibbo:1963yz, Kobayashi:1973fv}, namely $|V_{ub}|$ and $|V_{cb}|$, and may be a sensitive probe to new physics beyond the Standard Model (SM). As a matter of fact, many different types of extensions of the SM are expected to affect flavour physics, inducing new flavour-changing interactions, complex phases in the CKM matrix, possible violations of lepton-flavour universality, etc. Even if the mass scales of new particles beyond the SM turned out to be very high, quantum effects of the associated fields could leave detectable imprints onto the physics of bottom and charm quarks.

On the experimental side, recent results from $B$ factories reveal some tension with SM predictions, but also exhibit puzzling discrepancies between exclusive and inclusive channels~\cite{ParticleDataGroup:2020ssz, HFLAV:2019otj, Gambino:2019sif,Gambino:2020jvv}. For theorists, this provides further motivation to improve the understanding of these decays and to refine their predictions. Currently, the most powerful tool to obtain theoretical predictions from the first principles of QCD is the one based on numerical simulations in the lattice regularisation of the theory~\cite{Wilson:1974sk}. It is an intrinsically non-perturbative approach, that allows one to obtain accurate and systematically improvable predictions for a variety of quantities, including those relevant for decays of heavy mesons: for an up-to-date world review of lattice results relevant to flavour physics, see ref.~\cite{FlavourLatticeAveragingGroup:2019iem}. It should be emphasized, however, that most lattice calculations focus on \emph{exclusive} decays: in a nutshell, this is due to the fact that inclusive processes consist of a potentially very large number of physical states---including states featuring multiple hadrons, which pose their own challenges---and their systematic analysis in numerical calculations is very impractical, if possible at all. 

Recently, however, novel approaches have been put forward, that allow one to address inclusive decays in lattice QCD. As an example, in ref.~\cite{Hashimoto:2017wqo} it was pointed out that the differential rate for inclusive decays of the type $B\to X\ell\nu$ (where $X$ denotes all hadronic states that are compatible with the semileptonic decay of the bottom quark) could be evaluated by relating the hadronic tensor
\begin{equation}
W_{\mu\nu}(p,q) = \frac{4\pi^3}{E_B} \sum_X \delta^4(p-q-p_X) \langle B(p)|J_\mu^\dagger|X(p_X)\rangle \langle X(p_X)|J_\nu|B(p)\rangle
\end{equation} 
(where $J_\mu$ is the weak current associated with the $b$ quark decay, $p$ and $p_X$ respectively denote the four-momenta of the $B$ meson and of the $X$ state, while $q$ is the transferred four-momentum) to the forward scattering matrix element $T_{\mu\nu}(p,q)$~\cite{Manohar:1993qn, Blok:1993va}, and by extracting the latter through an analytical continuation of lattice results obtained for this quantity in an unphysical region, where the decay is forbidden by kinematics.

In ref.~\cite{Hansen:2017mnd}, on the other hand, it was proposed to study decay and transition rates into final states with an arbitrary number of hadrons by reconstructing the spectral function associated with a Euclidean four-point function in a finite volume from lattice correlators, with an appropriate smoothing protocol. A closely related approach was discussed in ref.~\cite{Hansen:2019idp} (see also refs.~\cite{Bulava:2019kbi,Bulava:2021fre}).

Finally, in ref.~\cite{Gambino:2020crt} it was suggested to study inclusive decays on the lattice by computing a suitable ``smeared'' spectral density $\rho(w)$ of hadron correlators, where the smearing is defined by the integration over the allowed phase-space region. Also in this case, the strategy involves the lattice determination of a class of four-point correlation functions. This technique allows one to bypass the need for analytical continuation, and, at least in principle, paves the way for the determination of the total semileptonic width as well as of the moments 
of any kinematic distribution associated with general $B\to X \ell \nu$ decays.

In the present work, we focus on the method proposed in ref.~\cite{Gambino:2020crt}, presenting the results of explicit lattice calculations based upon this framework. We discuss results from two different types of ensembles of lattice QCD configurations, and we also compare them with an analytical calculation based on the operator-product expansion (OPE)~\cite{Wilson:1969zs, Kadanoff:1969zz} within the framework of an expansion in inverse powers of the heavy-quark mass~\cite{Bigi:1992su, Bigi:1993fe, Blok:1993va}.

The structure of this article is the following. In section~\ref{sec:formulation_of_the_method_and_application_to_observables}, we recapitulate the formulation of the method, extending the presentation in ref.~\cite{Gambino:2020crt} with additional remarks, and commenting on its application to observables of particular interest (including differential distributions and moments). In section~\ref{sec:numerical_implementation_in_lattice_QCD}, we present an explicit implementation of the method in lattice QCD calculations, using two different ensembles of configurations, generated by the JLQCD~collaboration and by the ETM~collaboration; the final part of the section is devoted to a technical discussion about the extrapolation to the limit in which the smearing parameter $\sigma$ tends to zero. The following section~\ref{sec:OPE_and_comparison_with_lattice} presents the analytical calculation based on the OPE, and  compares its predictions with the results from lattice QCD. Finally, in section~\ref{sec:discussion_and_future_prospects} we summarize our results and discuss future prospects.

\section{Formulation of the method and application to observables}
\label{sec:formulation_of_the_method_and_application_to_observables}

\subsection{Spectral representation of the inclusive decay rate}

Here we review the formalism to calculate the inclusive semileptonic decay rate in lattice QCD~\cite{Gambino:2020crt}. To be specific, we consider the semileptonic decay of a $B$ meson to charmed final states $X_c$ with a pair of  massless leptons ($\ell\bar{\nu}$)
through the flavour-changing current $J_\mu=V_\mu-A_\mu=\bar{c}\gamma_\mu(1-\gamma_5)b$.

We start from the differential decay rate
\begin{flalign}
  \frac{d\Gamma}{dq^2dq^0dE_\ell}=
  \frac{G_F^2|V_{cb}|^2}{8\pi^3} L_{\mu\nu}W^{\mu\nu}\;,
  \label{eq:differential}
\end{flalign}
where $G_F$ is the Fermi constant and $|V_{cb}|$ is the relevant CKM matrix element. Here we work in the rest frame of the initial $B$ meson,
so that 
\begin{flalign}
p=(m_B,\bm{0})\;,
\qquad
q=p_\ell+p_{\bar{\nu}}=(q_0,\bm{q})\;,
\qquad
r=p-q=(\omega,-\bm{q})\;,
\end{flalign}
where the differential decay rate is a function of the three kinematical variables $q^2$, $q_0$ and the lepton energy $E_\ell$, and is given by the product of the leptonic tensor,
\begin{flalign}
  L^{\mu\nu}=p_\ell^\mu p_{\bar{\nu}}^\nu
  -p_\ell \cdot p_{\bar{\nu}}g^{\mu\nu}
  +p_\ell^\nu p_{\bar{\nu}}^\mu
  -i\epsilon^{\mu\alpha\nu\beta}p_{\ell,\alpha}p_{\bar{\nu},\beta}\;,
\end{flalign}
and the hadronic tensor,
\begin{flalign}
  W^{\mu\nu}(p,q) = \sum_{X_c}(2\pi)^3\delta^{(4)}(p-q-r)
  \frac{1}{2E_{B}(\bm{p})}
  \langle \bar B(\bm{p})|J^{\mu\dagger}(0)|X_c(\bm{r})\rangle
  \langle X_c(\bm{r})|J^\nu(0)|\bar B(\bm{p})\rangle.
  \label{eq:Wmunu}
\end{flalign}
The sum over the charmed states $X_c(\bm{r})$ actually includes an integral over $\bm{r}$.\footnote{
More precisely, the sum over the charmed states $X_c(\bm{r})$ should be written as
  \begin{displaymath}
    \sum_{X_c} |X_c(\bm{r})\rangle\langle X_c(\bm{r})|
    \to
    \sum_{X_c} \int\frac{d^3\bm{r}}{(2\pi)^3}
    \frac{1}{2E_{X_c(\bm{r})}}|X_c(\bm{r})\rangle\langle X_c(\bm{r})|
  \end{displaymath}
when the standard relativistic normalization for a single-particle state is employed for $X_c$.
}

Performing the integral over the lepton energy $E_\ell$ in its kinematical range, i.e.\ from $(q_0-\sqrt{\bm{q}^2})/2$ to $(q_0+\sqrt{\bm{q}^2})/2$, and changing the remaining kinematical variables from $(q_0,q^2)$ to $(\omega,\bm{q}^2)$, the total rate can be written as 
\begin{flalign}
  \Gamma=\frac{G_F^2|V_{cb}|^2}{24\pi^3}
  \int_0^{\bm{q}^2_{\mathrm{max}}} d\bm{q}^2 \sqrt{\bm{q}^2}
  \bar X(\bm{q}^2)\;,
  \qquad
  \bar X(\bm{q}^2) \equiv
  \int_{\omega_{\mathrm{min}}}^{\omega_{\mathrm{max}}}
  d\omega\, X(\omega,\bm{q}^2)\;,  
  \label{eq:q2integ}
\end{flalign}
where
\begin{flalign}
\bm{q}^2_{\mathrm{max}} = \frac{(m_B^2-m_D^2)^2}{4 m_B^2}\;,
\qquad
\omega_{\mathrm{min}}=\sqrt{m_{D}^2+\bm{q}^2}\;,
\qquad
\omega_{\mathrm{max}}=m_{B}-\sqrt{\bm{q}^2}\;.
\label{eq:omega_integ}
\end{flalign}
The quantity $X(\omega, \bm{q}^2)$ appearing above is a linear combination, with coefficients depending on the kinematical variables, of the different components of the hadronic tensor. Indeed, Lorentz invariance and time-reversal symmetry allow one to decompose $W^{\mu\nu}$ into invariant structure functions according to
\begin{flalign}
W^{\mu\nu}(p,q) 
= &-g^{\mu\nu} W_1(\omega,\bm{q}^2) 
+ \frac{p^\mu p^\nu}{m_B^2} W_2(\omega,\bm{q}^2) 
-i \epsilon^{\mu\nu\alpha\beta}\frac{p_\alpha q_\beta}{m_B^2} W_3(\omega,\bm{q}^2) 
\nonumber \\
&+\frac{q^\mu q^\nu}{m_B^2} W_4(\omega,\bm{q}^2) 
+ \frac{p^\mu q^\nu + p^\nu q^\mu}{m_B^2} W_5(\omega,\bm{q}^2)\;.
\label{eq:Wdecomp}
\end{flalign}
By introducing the following basis for three-dimensional space, 
\begin{flalign}
\bm{\hat n}= \frac{\bm{q}}{\sqrt{\bm{q}^2}}\;,
\qquad
\bm{\epsilon}^{(a)}\cdot \bm{\hat n}=0\;,
\qquad
\bm{\epsilon}^{(a)}\cdot \bm{\epsilon}^{(b)} = \delta^{ab}\;,
\qquad
\{a,b\}=\{1,2\}\;,
\end{flalign}
and the hadronic quantities
\begin{flalign}
&Y^{(1)}=-\sum_{a=1}^2\sum_{i,j=1}^3 \epsilon^{(a)}_i \epsilon^{(a)}_j W^{ij}\;,
\qquad
Y^{(2)}=W^{00}\;,
\qquad
Y^{(3)}=\sum_{i,j=1}^3 \hat{n}^i \hat{n}^j W^{ij}\;,
\nonumber \\
&Y^{(4)}=\sum_{i=1}^3 \hat{n}^i (W^{0i}+W^{i0})\;,
\qquad
Y^{(5)}=\frac{i}{2}\sum_{i,j,k=1}^3 \varepsilon^{ijk}\hat{n}^k W^{ij}\;,
\label{eq:Ydefs}
\end{flalign}
it is easy to see that, in the rest frame of the $B$ meson, the information contained in $W^{\mu\nu}$ can be equivalently parametrized in terms of $Y^{(i)}\equiv Y^{(i)}(\omega,\bm{q}^2)$. A convenient representation of $X(\omega,\bm{q}^2)$ is then given by
%
%
%
\begin{flalign}
&X(\omega,\bm{q}^2)=\sum_{l=0}^2 (\sqrt{\bm{q}^2})^{2-l} (m_{B}-\omega)^l X^{(l)}(\omega,\bm{q}^2)\;, \nonumber\\
&
X^{(0)}=Y^{(1)}+Y^{(2)}\;,
\qquad
X^{(1)}=-Y^{(4)}\;,
\qquad
X^{(2)}=Y^{(3)}-Y^{(1)}\;. \label{eq:XY}
\end{flalign}
At this point, some observations are in order. First, we notice that the parity-violating structure function $W_3$ (or equivalently $Y^{(5)}$) does not contribute to the differential decay rate after the integral over $E_\ell$ has been performed (this will not be the case for the moments considered below). Then, by rewriting eq.~(\ref{eq:Wmunu}) as
\begin{flalign}
  W_{\mu\nu}(\omega,\bm{q}) = \frac{(2\pi)^3}{2m_{B}}\langle\bar{B}(\bm{0})|
  J_\mu^\dagger(0) \delta(\hat{H}-\omega)\delta^3(\hat{\bm{P}}+\bm{q})
  J_\nu(0)|\bar{B}(\bm{0})\rangle\;,
  \label{eq:Wasspectre}
\end{flalign}
where $\hat{H}$ and $\hat{\bm{P}}$ are the QCD Hamiltonian and momentum operators, we explicitly see that in the rest frame of the $B$ meson the different components of the hadronic tensor are functions of $\omega$ and $\bm{q}$ (we already used this information in changing the integration variables from $(q_0,q^2)$ to $(\omega,\bm{q}^2)$). In the following we refer to eq.~(\ref{eq:Wasspectre}) as the \emph{spectral} representation of the hadronic tensor. Flavour and momentum conservation imply that the hadronic tensor vanishes identically for energies $\omega<\omega_{\mathrm{min}}$. From this observation one obtains $\omega_{\mathrm{min}}=\sqrt{m_{D}^2+\bm{q}^2}$. This means that by introducing the kernels
\begin{flalign}
  K^{(l)}(\omega,\bm{q}^2) = (m_B-\omega)^l \theta(\omega_{\mathrm{max}}-\omega)\;,
  \label{eq:kernel_exact}
\end{flalign}
the $\omega$ integral in eq.~(\ref{eq:q2integ}) can be rewritten as
\begin{flalign}
\bar X(\bm{q}^2)
&=\sum_{l=0}^2 (\sqrt{\bm{q}^2})^{2-l}\,  \int_0^\infty d\omega\, K^{(l)}(\omega,\bm{q}^2)\, X^{(l)}(\omega,\bm{q}^2)\;.
\label{eq:Xint1}
\end{flalign}
%
We close this subsection by providing an equivalent representation of $X(\omega,\bm{q}^2)$ which will be useful later. 
It is given by
\begin{flalign}
&X(\omega,\bm{q}^2)=\sum_{l=0}^2 (\sqrt{\bm{q}^2})^{2-l} (\omega_{\mathrm{max}}-\omega)^l Z^{(l)}(\omega,\bm{q}^2)\;,
\end{flalign}
where the $Z^{(l)}(\omega,\bm{q}^2)$ are again linear combinations of the $Y^{(l)}$,
\begin{flalign}
Z^{(0)}=Y^{(2)}+Y^{(3)}-Y^{(4)}\;,
\qquad
Z^{(1)}=2Y^{(3)}-2Y^{(1)}-Y^{(4)}\;,
\qquad
Z^{(2)}=Y^{(3)}-Y^{(1)}\;.
\label{eq:ZfromY}
\end{flalign}
By introducing the kernels 
\begin{flalign}
  \Theta^{(l)}(x) = x^l \theta(x)\;,
  \label{eq:kernel_theta}
\end{flalign}
that are functions of the single variable $x=\omega_{\mathrm{max}}-\omega$, we thus have
\begin{flalign}
\bar X(\bm{q}^2)
&=\sum_{l=0}^2 (\sqrt{\bm{q}^2})^{2-l}\, Z^{(l)}(\bm{q}^2)\;,
\qquad
Z^{(l)}(\bm{q}^2)
=
\int_0^\infty d\omega\, \Theta^{(l)}(\omega_{\mathrm{max}}-\omega)\, Z^{(l)}(\omega,\bm{q}^2)\;.
\label{eq:Xint2}
\end{flalign}

\subsection{Decay rate from Euclidean correlators}
In order to calculate $\bar X(\bm{q}^2)$, as given in eq.~(\ref{eq:Xint1}) or eq.~(\ref{eq:Xint2}), we need to evaluate the integral over $\omega$ of the different components of the hadronic spectral density (\ref{eq:Wasspectre}) with the kernels $K^{(l)}(\omega,\bm{q}^2)$ or $\Theta^{(l)}(\omega_{\mathrm{max}}-\omega)$. To this end, following ref.~\cite{Hashimoto:2017wqo}, we first establish the connection between suitably chosen correlation functions that can be calculated on the lattice, and $W_{\mu\nu}$.

We start by considering the Euclidean correlator
\begin{flalign}
  C_{\mu\nu}(t_{\mathrm{snk}},t_2,t_1,t_{\mathrm{src}};\bm{q}) =	
  \int d^3x\, e^{i\bm{q}\cdot\bm{x}}\, 
  T\langle 0\vert\, \tilde{\phi}_B(\bm{0};t_{\mathrm{snk}})
  J_\mu^\dagger(\bm{x};t_2) J_\nu(\bm{0};t_1)
  \tilde{\phi}_B^\dagger(\bm{0};t_{\mathrm{src}})\, \vert 0\rangle\;,
  \label{eq:4pt}
\end{flalign}
where $\tilde{\phi}_B(\bm{0};t)$ is a $B$-meson creation/annihilation operator projected onto zero spatial momentum by integrating over space at
a time $t$. A zero-momentum $B$ meson is thus created at time $t_{\mathrm{src}}$ and annihilated at $t_{\mathrm{snk}}$. The two currents are inserted in between, at times $t_2$ and $t_1$. The charmed hadrons are created at time $t_1$ with a momentum insertion $-\bm{q}$ and propagate until they are transformed back to the $B$-meson state at time $t_2$.

The four-point function $C_{\mu\nu}$ is saturated by the $B$-meson non-local matrix element 
\begin{flalign}
  M_{\mu\nu}(t;\bm{q}) = e^{-m_B t}\, \int d^3x\, 
  \frac{e^{i\bm{q}\cdot\bm{x}}}{2m_B}
  \langle \bar B(\bm{0})|J_\mu^\dagger(\bm{x},\!t)J_\nu(\bm{0},\!0)
  |\bar B(\bm{0})\rangle\;,
  \label{eq:forward-scatteringME}
\end{flalign}
when the double limit $t_{\mathrm{src}}\to-\infty$, $t_{\mathrm{snk}}\to\infty$ is taken. To include a proper normalization, one can analyse
\begin{flalign}
  M_{\mu\nu}(t_2-t_1;\bm{q}) =
  Z_B\,
  \lim_{\substack{t_{\mathrm{snk}}\to+\infty\\t_{\mathrm{src}}\to-\infty}}
  \frac{C_{\mu\nu}(t_{\mathrm{snk}},t_2,t_1,t_{\mathrm{src}};\bm{q})}{
    C(t_{\mathrm{snk}}-t_2)C(t_1-t_{\mathrm{src}})}\;,
  \label{eq:Mmunu}
\end{flalign}
where $C(t)$ is the $B$-meson two-point function
\begin{flalign}
  C(t) =
  T\langle0\vert\, \tilde{\phi}_B(\bm{0};t) \tilde{\phi}_B^\dagger(\bm{0};0) \vert 0\rangle
  \label{eq:C2t}
\end{flalign}
and $Z_B$ is its residue when a large time separation is taken, $C(t)\to Z_B e^{-m_Bt}$. 

Starting from eq.~(\ref{eq:forward-scatteringME}) we can establish the connection between $M_{\mu\nu}(t;\bm{q})$ and the hadronic tensor given in eq.~(\ref{eq:Wasspectre}). We have
\begin{flalign}
  M_{\mu\nu}(t;\bm{q}) 
  &= 
  \int d^3x\, 
  \frac{e^{i\bm{q}\cdot\bm{x}}}{2m_B}
  \langle \bar B(\bm{0})|J_\mu^\dagger(\bm{0},\!0)
  e^{-t \hat H +i \bm{\hat P}\cdot \bm{x}}
  J_\nu(\bm{0},\!0)
  |\bar B(\bm{0})\rangle
  \nonumber \\
  &= 
  \frac{(2\pi)^3}{2m_B}
  \langle \bar B(\bm{0})|J_\mu^\dagger(\bm{0},\!0)
  e^{-t \hat H}\delta^3(\bm{\hat P}+\bm{q})
  J_\nu(\bm{0},\!0)
  |\bar B(\bm{0})\rangle  
  \nonumber \\
  &= 
  \int_0^\infty d\omega\, W_{\mu\nu}(\omega,\bm{q})\, e^{-\omega t}\;.
\end{flalign}
The problem of the calculation of $\bar X(\bm{q}^2)$ is now reduced to that of trading the integral of $W_{\mu\nu}(\omega,\bm{q})$ with the kernels $e^{-t\omega}$ for the integral with the kernels $\Theta^{(l)}(\omega_{\mathrm{max}}-\omega)$ (or $K^{(l)}(\omega,\bm{q}^2)$).

The general inverse problem represented by the extraction of hadronic spectral densities from Euclidean correlators is notoriously ill-posed. Recently, methods to cope with these problems have been proposed, and they treat the above mentioned integrals with some kernels. In this paper we use two approaches proposed in refs.~\cite{Hansen:2019idp,Bailas:2020qmv}. The differences between the two methods will be discussed in detail in the following sections. Here we concentrate on the common starting point of the two approaches, which are actually closely related to each other.

We start by introducing an arbitrary length scale $a$. On the lattice this will be identified with the lattice spacing. The correlators $M_{\mu\nu}(t;\bm{q})$ will be computed at times $t=a \tau$ where $\tau\ge 0$ is an integer. By introducing the variable $x=e^{-a\omega}$ (and its inverse mapping $\omega=-\log(x)/a$), standard theorems of numerical analysis guarantee that any $C_\infty$ function $f(\omega)\equiv g(x)$ in the interval $\omega \in [0,\infty]$ (corresponding to $x\in [0,1]$), vanishing at $\omega=\infty$ ($x=0$), can be approximated with arbitrary precision in terms of polynomials in $x$ according to
\begin{flalign}
f(\omega)=\sum_{\tau=1}^\infty\, g_\tau\, x^\tau \equiv \sum_{\tau=1}^\infty\, g_\tau\, e^{-a\omega \tau}\;.
\label{eq:fwapprox1}
\end{flalign}
This implies that the integral of the product of $W_{\mu\nu}(\omega,\bm{q})$ with $f(\omega)$ can be computed, once the coefficients $g_\tau$ are known, by using the linear relation
\begin{flalign}
\int_0^\infty d\omega\, W_{\mu\nu}(\omega,\bm{q})\, f(\omega)
=
\sum_{\tau=1}^\infty\, g_\tau\, M_{\mu\nu}(a\tau;\bm{q})\;.
\end{flalign}
This procedure cannot be applied straightforwardly to the calculation of $X(\bm{q}^2)$ because the kernels  $\Theta^{(l)}(\omega_{\mathrm{max}}-\omega)$ (or $K^{(l)}(\omega,\bm{q}^2)$) are not smooth, i.e. they contain a discontinuity due to the $\theta$-function. In this case, a sequence of 
polynomials can still converge to the kernels {\it in mean}, which would be sufficient for our purposes,  but a reasonable approximation would imply a very large number of terms.
However, the problem can be solved by introducing smeared $C_\infty$ versions of the $\theta$-function, $\theta_\sigma$, such that the sharp step-function is recovered in the limit in which the smearing parameter  $\sigma$ is sent to zero,
$\lim_{\sigma\to 0} \theta_\sigma(x) = \theta(x)$.

By considering, as suggested in ref.~\cite{Gambino:2020crt}, the corresponding smeared versions of the kernels entering the definition of $\bar X(\bm{q}^2)$, that we call $\Theta^{(l)}_\sigma(\omega_{\mathrm{max}}-\omega)$ and  $K^{(l)}_\sigma(\omega,\bm{q})$, we then have
\begin{flalign}
\Theta^{(l)}_\sigma(\omega_{\mathrm{max}}-\omega) = m_B^l \sum_{\tau=1}^\infty\, g_\tau^{(l)}(\omega_{\mathrm{max}},\sigma)\, e^{-a\omega \tau}\;,
\label{eq:coeff1}
\end{flalign}
and
\begin{flalign}
\int_0^\infty d\omega\, W_{\mu\nu}(\omega,\bm{q})\, \Theta^{(l)}(\omega_{\mathrm{max}}-\omega)
=
\lim_{\sigma\to 0}
m_B^{l}\sum_{\tau=1}^\infty\, g_\tau^{(l)}(\omega_{\mathrm{max}},\sigma)\, M_{\mu\nu}(a\tau;\bm{q})\;,
\label{eq:coeff2}
\end{flalign}
as well as similar relations in the case of the kernels $K^{(l)}(\omega,\bm{q}^2)$.

A few observations are now in order. The first concerns a subtle theoretical issue. The smearing procedure, which is algorithmically required to implement the procedure just outlined, is also necessary for theoretical reasons. Hadronic spectral densities, and therefore also $W_{\mu\nu}(\omega,\bm{q})$, are elements in the space of distributions and their product with another distribution, such as the $\theta$-function, can only be defined through a regularization procedure (when it exists). The issue is particularly important in the case of lattice simulations because they are necessarily performed on a finite volume. Finite-volume spectral functions, due to the quantization of the energy spectrum, are sums of isolated $\delta$-function singularities and their connection with the corresponding physical quantities requires an ordered double-limit procedure: first the infinite volume limit has to be taken and only \emph{after} that, if the quantity is non-singular, can one take the $\sigma\to 0$ limit.

The second observation is related to the fact that the problem we are addressing is particularly hard from the computational point of view. In the limit of very small $\sigma$ the coefficients $g_\tau^{(l)}(\omega_{\mathrm{max}},\sigma)$ of eq.~(\ref{eq:coeff1}) tend to become arbitrarily large in modulus and oscillate in sign. Since lattice correlators are unavoidably affected by statistical and systematic errors, in these cases the resulting uncertainties on the sums on the left-hand side of eq.~(\ref{eq:coeff2}) tend to explode. The two approaches of refs.~\cite{Hansen:2019idp,Bailas:2020qmv} differ for the procedures used to determine the coefficients $g_\tau^{(l)}(\omega_{\mathrm{max}},\sigma)$, once the series is truncated at $\tau=\tau_{\mathrm{max}}$, in such a way to keep both statistical and systematic errors under control.

\subsection{Kernel approximation}
In this subsection we review the methods of refs.~\cite{Hansen:2019idp,Bailas:2020qmv} by highlighting the differences in the procedures used to approximate the smearing kernels. To simplify the formulae, we shall consider a generic kernel $f(\omega)$, that will then be identified with the kernels $\Theta^{(l)}(\omega_{\mathrm{max}}-\omega)/m_B^l$ or $K^{(l)}(\omega,\bm{q})/m_B^l$, and a generic correlator
\begin{flalign}
C(t)=\int_0^\infty d\omega\, \rho(\omega)\, e^{-\omega t}\;,
\end{flalign}
to be identified with $M_{\mu\nu}(t;\bm{q})$, so that $\rho(\omega)$ will correspond to $W_{\mu\nu}(\omega,\bm{q})$. In this work we shall not address the systematics associated with the finiteness of the extent of the lattice in the temporal direction, see refs.~\cite{Hansen:2019idp,Bulava:2021fre} for an extended discussion of this issue and, in general, for more details concerning the algorithm and its applications. 

In the method of ref.~\cite{Hansen:2019idp} the coefficients $g_\tau$ corresponding to the approximation of $f(\omega)$ are determined by minimizing the functional
\begin{align}\label{e:Wfunc}
W_{\lambda}[g] = (1-\lambda)\frac{A[g]}{A[0]} + \lambda B[g] \,,
\end{align}
where $\lambda\in[0,1]$ is the so-called ``trade-off parameter'' (see below) and the functionals $A[g]$ and $B[g]$ are given by
\begin{align}
A[g] =
a\int_{E_0}^\infty d\omega \left\{ f(\omega) - \sum_{\tau=1}^{\tau_{\rm max}} g_\tau \, e^{-a\omega\tau}\right\}^2\,,
\qquad
B[g] = \sum_{\tau,\tau^\prime=1}^{\tau_{\rm max}} g_{\tau} g_{\tau^\prime} \, 
\frac{{\rm Cov} \left[C(a\tau), C(a\tau^\prime) \right]}{\left[C(0)\right]^2} \,.
\label{eq:AandB}
\end{align}
Here ${\rm Cov} \left[C(t), C(t^\prime) \right]$ is the statistical covariance of the correlator $C(t)$ and, consequently, the functional $B[g]$ is positive definite. The functional $A[g]$ is also a positive definite quadratic form in the coefficients $g_\tau$. Therefore, the minimum conditions
\begin{align}
\left.\frac{\partial W_{\lambda}[g]}{\partial g_\tau}\right\vert_{g_\tau=g_\tau^\lambda}=0 \,
\label{eq:Wminimumg}
\end{align}
are a linear system of equations to be solved for the coefficients $g_\tau^\lambda$. These coefficients define the approximation of $f(\omega)$ and the associated estimator for the integral of $\rho(\omega)$ with $f(\omega)$ according to
\begin{align}
f^\lambda(\omega) = \sum_{\tau= 1}^{\tau_{\rm max}} g^\lambda_\tau \, e^{-a\omega \tau}\;,
\qquad
\rho^\lambda[f] = \sum_{\tau=1}^{\tau_{\rm max}} g^\lambda_{\tau} C(a\tau)
= \int_0^\infty d\omega \, f^\lambda(\omega) \, \rho(\omega)\;.
\label{eq:mainHLT}
\end{align}
The functional $B[g^\lambda]$ is the statistical variance of $\rho^\lambda[f]$ normalized with the square of the correlator in zero and, therefore, vanishes in the ideal case of infinitely precise input data. On the other hand, $A[g^\lambda]$ measures the distance between the target kernel $f(\omega)$ and its approximation $f^\lambda(\omega)$ in the range\footnote{The parameter $E_0$ can be adjusted by exploiting the fact that $\rho(\omega)$ has support only for $\omega> \omega_{\rm min}$, so that $\rho[f]$ is insensitive to $f(\omega)$ for $\omega<\omega_{\rm min}$. The same holds for $\rho^\lambda[f]$ so that the functional form of $f^\lambda( \omega)$ can be left unconstrained for $\omega<\omega_{\rm min}$. Any $E_0 < \omega_{\rm min}$ is therefore a viable choice in determining the coefficients $g_t^\lambda$ so $E_0$ can be chosen to improve the numerical stability of the minimization procedure.} $\omega\in [E_0,\infty]$. In fact $A[g^\lambda]$ is the squared $L_2$-norm in function space of the difference $f^\lambda(\omega)-f(\omega)$ and can only vanish in the limit $t_{\rm max}\rightarrow \infty$. 

In the absence of errors, the coefficients $g_\tau^\lambda$ that minimize $A[g]$ provide the best polynomial approximation of $f(\omega)$ with respect to the $L_2$-norm. This has to be compared with the method of ref.~\cite{Bailas:2020qmv} that provides the best polynomial approximation of $f(\omega)$ with respect to the $L_\infty$-norm (see below). In the presence of errors, the coefficients $g^\lambda_\tau$ that minimize $W_{\lambda}[g]$ represent a particular balance between statistical and systematic errors, as dictated by the $\lambda$ parameter. For small $\lambda$ the estimator $\rho^\lambda[f]$ is close to $\rho[f]$ but with a large statistical uncertainty. Conversely, for large $\lambda$ the estimator $\rho^\lambda[f]$ has a small statistical error but differs significantly from $\rho[f]$. When evaluated at the minimum, the functional $W_\lambda[g]$ is a function of $\lambda$ only, thus defining $W(\lambda)\equiv W_\lambda[g^\lambda]$. The prescription suggested in ref.~\cite{Hansen:2019idp} to choose the optimal value of the trade-off parameter defines $\lambda_\star$ such that
\begin{align}
\left. \frac{\partial W(\lambda)}{\partial \lambda}\right\vert_{\lambda=\lambda_\star} =0\;.
\label{eq:Wmax}
\end{align}
From eq.~\eqref{eq:Wminimumg} it follows that at $\lambda_\star$ (the maximum of $W(\lambda)$ where $g_\tau^{\lambda}=g_\tau^\star$) one has $A[g^\star]=A[0]B[g^\star]$. This can be understood as the condition of ``optimal balance'' between statistical and systematic errors. The numerical results discussed in subsection~\ref{subsec:ETMC_lattice_calculation} have been obtained using this method, also monitoring the stability of the results with respect to $\lambda\le \lambda_\star$.

In ref.~\cite{Bailas:2020qmv}, on the other hand, the function $f(\omega)$
is approximated using the Chebyshev approximation as
\begin{equation}
  f(\omega)\simeq \frac{c_0^*}{2} + \sum_{j=1}^N c_j^* T_j^*(e^{-a\omega})\;,
\end{equation}
where $T_j^*(x)$ is a (shifted) Chebyshev polynomial of the $j$-th
order.
The coefficients $c_j^*$ are determined only by the function $f(\omega)$:
\begin{equation}
  c_j^* = \frac{2}{\pi}\int_0^\pi d\theta\,
  f\left(-\ln\frac{1+\cos\theta}{2}\right) \cos(j\theta)\;.
\end{equation}
This yields the best approximation in the sense of the $L_\infty$-norm\footnote{More precisely, the Chebyshev approximation of a generic function $f(y)$ for $y\in [-1,1]$ is in practice equivalent, although not identical, to the optimal polynomial approximation of $f(\vec g;y)=\sum_{\tau=0}^N g_\tau y^\tau$ obtained by minimizing the $L_\infty$-norm $\|f(y)-f(\vec g;y)\|_\infty=\mbox{max}_{y\in [-1,1]}\vert f(y)-f(\vec g;y)\vert$ with respect to the coefficients $\vec g$. In fact, the Chebyshev approximation is obtained by minimizing the \emph{weighted} squared $L_2$-norm given by $\int_{-1}^1 dy\, w(y) \vert f(y)-f(\vec g;y)\vert^2$ with $w(y)=1/\sqrt{1-y^2}$. By setting instead $w(y)=1$, as done in the case of the method of ref.~\cite{Hansen:2019idp}, one gets the Legendre polynomial approximation.
}.
The approximation of the $\omega$-integral is then constructed as
\begin{equation}
  \int d\omega f(\omega)\rho(\omega) =
  \frac{\langle\psi_\mu|f(\hat{H})|\psi_\nu\rangle}{
    \langle\psi_\mu|\psi_\nu\rangle} \simeq
  \frac{c_0^*}{2} + \sum_{j=1}^N c_j^*
  \frac{\langle\psi_\mu|T_j^*(e^{-a\hat{H}})|\psi_\nu\rangle}{
    \langle\psi_\mu|\psi_\nu\rangle}\;,
  \label{eq:Chebyshev_expansion}
\end{equation}
where $|\psi_\mu\rangle \equiv e^{-\hat{H}t_0} J_\mu|B\rangle$
is defined such that the state is evolved for some small time $t_0$
after applying the current insertion: this allows one to avoid any ultraviolet divergence due to contact terms of two currents.
To reflect this change, the kernel $f(\omega)$ is multiplied by
$e^{2\omega t_0}$ to cancel the time evolution.
The right-hand side of eq.~(\ref{eq:Chebyshev_expansion}) can be
reconstructed from the matrix elements
(\ref{eq:forward-scatteringME}) using
$M_{\mu\nu}(t+2t_0)/M_{\mu\nu}(2t_0)$.

An advantage of this construction is that the matrix element appearing
on the right-hand side of eq.~(\ref{eq:Chebyshev_expansion}), 
$\langle\psi_\mu|T_j^*(e^{-a\hat{H}})|\psi_\nu\rangle/
\langle\psi_\mu|\psi_\nu\rangle$,
is strictly bound between $-1$ and $+1$, by construction of the
Chebyshev polynomial.
This corresponds to the condition that the eigenvalues of $e^{-\hat{H}}$ lie between $0$ and $1$, or equivalently that the eigenvalues of $\hat{H}$ are
positive semi-definite. 
Then, the convergence of the series appearing in eq.~(\ref{eq:Chebyshev_expansion}) is
dictated by that of the coefficients $c_j^*$.
Since $c_j^*$ can be easily calculated for arbitrarily large $j$'s,
the error due to the truncation in (\ref{eq:Chebyshev_expansion}) can
be rigorously estimated.

The constraint 
$|\langle\psi_\mu|T_j^*(e^{-a\hat{H}})|\psi_\nu\rangle/
\langle\psi_\mu|\psi_\nu\rangle|\le 1$
is not automatically satisfied in the presence of statistical errors.
Since the Chebyshev polynomial $T_j^*(x)$ is a sign-alternating series
of growing powers of $x$ with (exponentially) large coefficients, this 
constraint is satisfied after huge cancellations for large $j$.
Therefore, even a small statistical error of the lattice correlator can
easily violate the constraint. 
In the numerical analysis, one should add the constraint when 
the Chebyshev matrix elements are determined by a fit, see ref.~\cite{Bailas:2020qmv} for details.
The higher-order terms are then masked by the statistical uncertainties and 
become basically undetermined within $\pm 1$, so that they only
contribute to the truncation error.

In both methods, a good approximation is obtained only when the
kernel function is sufficiently smooth.
If this is not the case, the truncation error becomes significant, e.g. due to
unsuppressed higher-order coefficients $c_j^*$ in the case of the
Chebyshev approximation.
Unfortunately, the kernel functions $K^{(l)}(\omega,\bm{q})$ or $\Theta^{(l)}(\omega_{\mathrm{max}}-\omega)$ are
not smooth, because they contain the Heaviside function
$\theta(\omega_{\mathrm{max}}-\omega)$.
We therefore introduce smeared versions of the $\theta$-function and then we take the limit of $\sigma\to 0$ to recover the unsmeared kernel. This has been done by considering three different smeared $\theta$-functions,
\begin{flalign}
\theta_\sigma^{\mathtt{s}}(x) = \frac{1}{1+e^{-\frac{x}{\sigma}}}\;,
\qquad
\theta_\sigma^{\mathtt{s1}}(x) = \frac{1}{1+e^{-\sinh\left(\frac{x}{r^{\mathtt{s1}}\sigma}\right)}}\;,
\qquad
\theta_\sigma^{\mathtt{e}}(x) = \frac{1+\mbox{erf}\left(\frac{x}{r^{\mathtt{e}}\sigma}\right)}{2}\;,
\label{eq:thetasss}
\end{flalign}
and by extrapolating the numerical data to the $\sigma \to 0$ limit.
In the following we shall refer to $\theta_\sigma^{\mathtt{s}}(x)$ as the ``sigmoid function'', to $\theta_\sigma^{\mathtt{s1}}(x)$ as the ``modified sigmoid function'' and to $\theta_\sigma^{\mathtt{e}}(x)$ as the ``error function''. 
 Any choice of the parameters $r^{\mathtt{s1}}$ and $r^{\mathtt{e}}$ appearing in the previous formulae corresponds to a legitimate definition of the smearing kernels that approach the same $\sigma\to 0$ limit, i.e. the $\theta$-function. By adjusting the values of these parameters one can change the rate of convergence to the $\theta$-function and balance between statistical and systematic errors. In the following we set $r^{\mathtt{s1}}=2.2$ and $r^{\mathtt{e}}=2.0$. 
This (empirical) choice gives statistical errors of the same order of magnitude for the three kernels at fixed $\sigma$ and similar (although not identical) shapes for $\theta_\sigma^{\mathtt{s}}(x)$ and $\theta_\sigma^{\mathtt{e}}(x)$ while $\theta_\sigma^{\mathtt{s1}}(x)$ results into a smoother approximation of the $\theta$-function. A combined analysis of smearing kernels that have rather different shapes at fixed $\sigma$ is in fact helpful in order to quantify the systematics associated with the $\sigma\to 0$ extrapolations (see also ref.\cite{Bulava:2021fre}).

\subsection{Decomposition of the total rate}
\label{sec:decomposition}
The expression of the total rate in eq.~(\ref{eq:q2integ}) can also be used to compute 
the differential decay rate in  $\bm{q}^2$, i.e. 
$d\Gamma/d\bm{q}^2=G_F^2|V_{cb}|^2/(24\pi^3)
|\bm{q}| \bar X 
(\bm{q}^2)$.
This can be further decomposed into its contributions from parallel ($\parallel$)
and perpendicular ($\perp$) components, where 
the $\perp$ components are defined as those involving the polarization vector
$\epsilon^{*(\alpha)}$, while the $\parallel$ ones are the rest.
In addition, we also separate the contributions from vector ($V$) and
axial-vector ($A$) current insertions.
Since two currents are inserted, we have $VV$, $AA$, as well
as $VA$ and $AV$ contributions.
Among them, $VA$ and $AV$ do not contribute to the differential rate
after integrating over $E_\ell$, and thus to the total decay rate.
We therefore analyze four components:
$VV_\parallel$, $VV_\perp$, $AA_\parallel$, $AA_\perp$.
For the lepton energy moments, the $VA$ and $AV$ insertions can also appear (see below). 

\subsection{Moments}
It is also interesting to consider the moments of various kinematical quantities.
In particular, two types of moments have been studied experimentally:
the hadronic mass moments $\langle (M_X^2)^n\rangle$
and the lepton energy moments $\langle E_\ell^{n_\ell}\rangle$.
They are defined as
\begin{eqnarray}
  \langle (M_X^2)^n\rangle
  & = & \frac{
        \displaystyle
        \int d\bm{q}^2 dq_0 dE_\ell \, (\omega^2-\bm{q}^2)^n
        \left[\frac{d\Gamma}{d\bm{q}^2dq_0dE_\ell}\right]
        }{
        \displaystyle
        \int d\bm{q}^2 dq_0 dE_\ell
        \left[\frac{d\Gamma}{d\bm{q}^2dq_0dE_\ell}\right]
        }\;,
        \label{eq:MX2}
  \\
  \langle E_\ell^{n_\ell}\rangle
  & = & \frac{
        \displaystyle
        \int d\bm{q}^2 dq_0 dE_\ell \, E_\ell^{n_\ell}
        \left[\frac{d\Gamma}{d\bm{q}^2dq_0dE_\ell}\right]
        }{
        \displaystyle
        \int d\bm{q}^2 dq_0 dE_\ell
        \left[\frac{d\Gamma}{d\bm{q}^2dq_0dE_\ell}\right]
        }\;.
        \label{eq:El}
\end{eqnarray}
The strategy to compute these moments on the lattice is the same as in 
the method described above.
For the hadronic mass moments defined in eq.~(\ref{eq:MX2}), the numerator contains
extra powers of $\omega^2-\bm{q}^2$, with which
the $\omega$-dependence of $X^{(0)}$, $X^{(1)}$, $X^{(2)}$ is modified.
Otherwise, the basic procedure remains the same.
Beside these quantities which require an integration over the whole $\bm q^2$ range,
we will also consider  moments at fixed values of  $\bm q^2$, i.e. {\it differential moments}:
\begin{eqnarray}
H_n(\bm{q}^2)\equiv  \langle (M_X^2)^n\rangle_{\bm{q}^2}
  & = & \frac{
        \displaystyle
        \int  dq_0 dE_\ell \, (\omega^2-\bm{q}^2)^n
        \left[\frac{d\Gamma}{d\bm{q}^2dq_0dE_\ell}\right]
        }{
        \displaystyle
        \int  dq_0 dE_\ell
        \left[\frac{d\Gamma}{d\bm{q}^2dq_0dE_\ell}\right]
        }\;,
        \label{eq:MX2diff}
  \\
L_{n_\ell}(\bm{q}^2)\equiv \langle E_\ell^{n_\ell}\rangle_{\bm{q}^2}  & = & \frac{
        \displaystyle
        \int  dq_0 dE_\ell \, E_\ell^{n_\ell}
        \left[\frac{d\Gamma}{d\bm{q}^2dq_0dE_\ell}\right]
        }{
        \displaystyle
        \int  dq_0 dE_\ell
        \left[\frac{d\Gamma}{d\bm{q}^2dq_0dE_\ell}\right]
        }\;,
        \label{eq:Eldiff}
\end{eqnarray}
and the second central moment or variance of the lepton energy distribution
\[
 L_{2c}(\bm{q}^2)= L_2(\bm{q}^2)-\Big(L_1(\bm{q}^2)\Big)^2\;.
\]

In the case of leptonic moments, the $E_\ell$ integral is modified with respect to (\ref{eq:q2integ}). The  integrand in the denominator is the same as in (\ref{eq:XY});
if we set the $\bm{q}$ momentum direction $\bm{n}$ along the $k$-th axis, the two vectors
$\bm{\epsilon}^{a}$ can be chosen in the perpendicular directions of the $i$-th and $j$-th axes,
and we can re-express $X(\omega, \bm{q}^2)$ as 
\begin{equation}
  X_{n_\ell=0} =
  \bm{q}^2(W^{00}-W^{ii}-W^{jj})-q_0q_k(W^{0k}+W^{k0})+q_0^2(W^{kk}+W^{ii}+W^{jj})\;,
\end{equation}
where repeated indices are not summed. The integrand in the numerators of eq.~(\ref{eq:El}) and eq,~(\ref{eq:Eldiff}) depends on the exponent $n_\ell$. For  $n_\ell=1$, it reads
\begin{eqnarray}
  X_{n_\ell=1}
  & = & \frac{q_0}{2} \left[
        \bm{q}^2(W^{00}-W^{ii}-W^{jj})-q_0q_k(W^{0k}+W^{k0})+q_0^2(W^{kk}+W^{ii}+W^{jj})
        \right]
        \nonumber\\
  & & + \frac{i}{2} q_k (q_0^2-\bm{q}^2)W^{ij}\;,
  \label{eq:X1ugly}
\end{eqnarray}
where the last term corresponds to the insertion of
$VA$ or $AV$.
The other terms are the same as ${X}_{n_\ell=0}$, up to a factor
$q_0/2$.
The next orders are more involved:
\begin{eqnarray}
  X_{n_\ell=2}
  & = & \frac{1}{4} \left\{
        \left(q_0^2\bm{q}^2+\frac{1}{5}|\bm{q}|^4\right) W^{00} +
        \left(2q_0^4-\frac{6}{5}q_0^2\bm{q}^2-\frac{4}{5}|\bm{q}|^4\right)
        W^{ii} +
        \left( q_0^4+\frac{1}{5}q_0^2\bm{q}^2 \right) W^{kk}
        \right.\nonumber
  \\
  & & \left.
      -\left(q_0^3|\bm{q}|+\frac{1}{5}q_0|\bm{q}|^3\right)
      (W^{0k}+W^{k0}) +
      \frac{i}{2} q_0|\bm{q}|(q_0^2-\bm{q}^2) W^{ij}
      \right\}\;,
  \\
  X_{n_\ell=3}
  & = & \frac{1}{8} \left\{
        \left( q_0^3\bm{q}^2+\frac{3}{5}q_0|\bm{q}|^4
        \right) W^{00} +
        \left( q_0^5+\frac{3}{5}q_0^3\bm{q}^2 \right) W^{kk} +
        \left(
        2q_0^5+\frac{2}{5}q_0^3\bm{q}^2-\frac{12}{5}q_0|\bm{q}|^4
        \right) W^{ii}
        \right.\nonumber
  \\
  & & \left.
      -\left( q_0^4|\bm{q}|+\frac{3}{5}q_0^2|\bm{q}|^3 \right)
      (W^{0k}+W^{k0}) +
      i \left(3q_0^2|\bm{q}|+\frac{3}{5}|\bm{q}|^3\right)
      (q_0^2-\bm{q}^2) W^{ij}
      \right\}\;.
\end{eqnarray}
Again, the term with $W^{ij}$ survives for $VA$ and $AV$ insertions,
while the others are from $VV$ or $AA$.

The contributions in eq.~(\ref{eq:X1ugly}) can be rearranged in such a way that the $\omega$-integral contributing to the numerator of eq.~(\ref{eq:Eldiff}) takes the form
\begin{flalign}
\bar X_{n_\ell=1}(\bm{q}^2)
&=\sum_{l=0}^3 (\sqrt{\bm{q}^2})^{3-l}\, Z^{(l)}_{n_\ell=1}(\bm{q}^2)\;, \nonumber \\
\qquad
 Z^{(l)}_{n_\ell=1}(\bm{q}^2)&
=
\int_0^\infty d\omega\, \Theta^{(l)}(\omega_{\mathrm{max}}-\omega)\, Z^{(l)}_{n_\ell=1}(\omega,\bm{q}^2)\;,
\label{eq:X1int2}
\end{flalign}
where  the $Z^{(l)}_{n_\ell=1}(\omega,\bm{q}^2)$ are given by
\begin{flalign}
&
Z^{(0)}_{n_\ell=1}=\frac{Y^{(2)}+Y^{(3)}-Y^{(4)}}{2}\;,
\qquad
Z^{(1)}_{n_\ell=1}=\frac{-2Y^{(1)}+Y^{(2)}+3Y^{(3)}-2Y^{(4)}+2Y^{(5)}}{2}\;,
\nonumber \\
&
Z^{(2)}_{n_\ell=1}=\frac{-3Y^{(1)}+3Y^{(3)}-Y^{(4)}+Y^{(5)}}{2}\;,
\qquad
Z^{(3)}_{n_\ell=1}=\frac{-Y^{(1)}+Y^{(3)}}{2}\;.
\end{flalign}
The previous expressions are analogous to the corresponding expressions for the differential decay rate, eq.~(\ref{eq:Xint2}) and eq.~(\ref{eq:ZfromY}), but include the sum of four terms with the one corresponding to $l=3$ that involves the kernel $\Theta^{(3)}(\omega_{\mathrm{max}}-\omega)$. In this basis the second leptonic moment is given by  
\begin{flalign}
\bar X_{n_\ell=2}(\bm{q}^2)
&=\sum_{l=0}^4 (\sqrt{\bm{q}^2})^{4-l}\, Z^{(l)}_{n_\ell=2}(\bm{q}^2)\;,
\nonumber \\
Z^{(l)}_{n_\ell=2}(\bm{q}^2)&
=
\int_0^\infty d\omega\, \Theta^{(l)}(\omega_{\mathrm{max}}-\omega)\, Z^{(l)}_{n_\ell=2}(\omega,\bm{q}^2)\;,
\label{eq:X2int2}
\end{flalign}
where
\begin{flalign}
&
Z^{(0)}_{n_\ell=2}=3\frac{Y^{(2)}+Y^{(3)}-Y^{(4)}}{10}\;,
\nonumber \\
&
Z^{(1)}_{n_\ell=2}=\frac{7 Y^{(1)} - 5 Y^{(2)} - 11 Y^{(3)} + 8 Y^{(4)} - 10 Y^{(10)}}{10}\;,
\nonumber \\
&
Z^{(2)}_{n_\ell=2}=\frac{-27 Y^{(1)} + 5 Y^{(2)} + 31 Y^{(3)} - 15 Y^{(4)} + 30 Y^{(5)}}{20}\;,
\nonumber \\
&
Z^{(3)}_{n_\ell=2}=\frac{4 Y^{(1)} - 4 Y^{(3)} + Y^{(4)} - 2 Y^{(5)}}{4}\;,
\nonumber \\
&
Z^{(4)}_{n_\ell=2}=\frac{-Y^{(1)} + Y^{(3)}}{4}\;,
\end{flalign}
and the first hadronic moment is
\begin{flalign}
\bar X_{n=1}(\bm{q}^2)
&=\sum_{l=0}^4 Z^{(l)}_{n=1}(\bm{q}^2)\;,
\qquad
Z^{(l)}_{n=1}(\bm{q}^2)
=
\int_0^\infty d\omega\, \Theta^{(l)}(\omega_{\mathrm{max}}-\omega)\, Z^{(l)}_{n=1}(\omega,\bm{q}^2)\;,
\label{eq:XH1int2}
\end{flalign}
where the $Z^{(l)}_{n=1}(\omega,\bm{q}^2)$ are given by
\begin{flalign}
Z^{(0)}_{n=1}&=m_B\vert\bm{q}\vert^3(m_B-2\vert\bm{q}\vert) \left(Y^{(2)} + Y^{(3)} - Y^{(4)}\right)\;,
\nonumber \\
\nonumber \\
Z^{(1)}_{n=1}
&=
2 \vert\bm{q}\vert^4 \left(Y^{(2)} + Y^{(3)} - Y^{(4)}\right) + m_B^2 \vert\bm{q}\vert^2 \left(-2 Y^{(1)} + 2 Y^{(3)} - Y^{(4)}\right) 
\nonumber \\
&
+ m_B \vert\bm{q}\vert^3 \left[-2 \left(Y^{(2)} + Y^{(3)} - Y^{(4)}\right) + 2 \left(2 Y^{(1)} - 2 Y^{(3)} + Y^{(4)}\right)\right]\;,
\nonumber \\
\nonumber \\
Z^{(2)}_{n=1}&=
m_B^2 \vert\bm{q}\vert \left(-Y^{(1)} + Y^{(3)}\right) 
+ \vert\bm{q}\vert^3 \left[Y^{(2)} + Y^{(3)} - Y^{(4)} - 2 \left(2 Y^{(1)} - 2 Y^{(3)} + Y^{(4)}\right)\right] 
\nonumber \\
&
+ m_b \vert\bm{q}\vert^2 \left[-2 \left(-Y^{(1)} + Y^{(3)}\right) + 2 \left(2 Y^{(1)} - 2 Y^{(3)} + Y^{(4)}\right)\right]\;,
\nonumber \\
\nonumber \\
Z^{(3)}_{n=1}&=
-2 m_B\vert\bm{q}\vert \left(-Y^{(1)} + Y^{(3)}\right) 
+ \vert\bm{q}\vert^2 \left[-2 Y^{(1)} + 2 Y^{(3)} + 2 \left(-Y^{(1)} + Y^{(3)}\right) - Y^{(4)}\right]\;,
\nonumber \\
\nonumber \\
Z^{(4)}_{n=1}&=\vert\bm{q}\vert \left(-Y^{(1)} + Y^{(3)}\right)\;.
\end{flalign}

\section{Numerical implementation in lattice QCD}
\label{sec:numerical_implementation_in_lattice_QCD}

In this section, we discuss in detail two different implementations of the method in lattice QCD calculations. First, in subsection~\ref{subsec:JLQCD_lattice_calculation} we present an implementation based on configurations generated within the JLQCD~collaboration. Then, in subsection~\ref{subsec:ETMC_lattice_calculation} we discuss an analogous calculation based on an ensemble generated by the ETM~collaboration (ETMC). In both cases, we specify the technical details of the lattice calculations, and discuss the different types of uncertainties affecting the results. Finally, in subsection~\ref{subsec:sigma_to_zero_extrapolation}, we discuss a few technical aspects related to the extrapolation to the $\sigma \to 0$ limit.

\subsection{Lattice implementation with JLQCD configurations}
\label{subsec:JLQCD_lattice_calculation}
One dataset used to demonstrate the lattice computation of the
inclusive semileptonic decay rate is based on the ensemble generated
by the JLQCD collaboration.
See the supplementary materials of \cite{Colquhoun:2022atw} for details of the gauge configurations.
It employs M\"obius domain-wall fermions for both valence and sea
quarks. 
In the sea, 2+1 flavors of light and strange quarks are included.
The light quark mass corresponds to a pion of mass around 300~MeV;
the strange quark mass is slightly heavier than its physical value.
The gauge action is the tree-level $O(a^2)$-improved Symanzik at
$\beta$ = 4.35.
The corresponding lattice spacing is $a\simeq$ 0.055~fm, corresponding to 
inverse lattice spacing $1/a$ = 3.610(9)~GeV.
The lattice volume is $48^3\times 96$, so that the spatial volume is
about $L^3=(2.6\mbox{~fm})^3$.
This ensemble corresponds to ``M-$ud$3-$s$a'' of \cite{Colquhoun:2022atw}.
(See also \cite{Nakayama:2017lav}.)

The valence quarks are also described by M\"obius domain-wall
fermions. 
The charm quark mass is tuned to its physical value (see
ref.~\cite{Nakayama:2017lav} for details), while the bottom quark mass
is set at $1.25^4\simeq 2.44$ times the charm quark mass.
The spectator quark is a strange quark, so the process corresponds to the
inclusive semileptonic decay of a $B_s$ meson, albeit with a light $B_s$
meson mass of $\simeq$ 3.45~GeV.

The measurement is carried out on 100 gauge configurations and is
replicated four times on each configuration with shifted position of
the initial source.
The $B_s$ meson is created by a interpolating pseudoscalar operator,
which is spatially smeared by a gauge-invariant operator
$(1-(\alpha/N)\Delta)^N$ with a discretized Laplacian $\Delta$ and
parameters $\alpha$ = 20 and $N$ = 200.
The source points are spread over the source time slice
$t_{\mathrm{src}}=0$ with $\mathbb{Z}_2$ noises to improve statistics.
The initial $B_s$ meson is thus projected to zero spatial momentum.
The $B_s$ meson on the other end is created by another pseudoscalar
operator of the same type placed at the time slice $t_{\mathrm{snk}}$ = 42
using a sequential source from the spectator strange quark propagator.
The bottom quark propagates from there to a time slice $t_2$, where
the first $b\to c$ current is inserted with momentum $\bm{q}$ and
is fixed at $t_2=26$.
The charm quark propagator then connects the time slice $t_2$ to
$t_1$ where the other $b\to c$ current is contracted with momentum
insertion $-\bm{q}$.
The charm quark propagator is computed repeatedly for each choice of
the current operator and momentum insertion at $t_1$.
We fix the time separation between $t_{\mathrm{src}}$ and $t_{\mathrm{snk}}$ under an assumption that the ground-state $B_s$ meson state dominates the signal between $t_{\mathrm{src}}$ and $t_1$ or between $t_2$ and $t_{\mathrm{snk}}$.
This separation is at least 16 in the lattice unit, which corresponds to 0.9~fm. The saturation is confirmed in \cite{Hashimoto:2017wqo}.

The matrix element (\ref{eq:forward-scatteringME}) is then constructed
as in eq.~(\ref{eq:Mmunu}). 
For the analysis of this ensemble we applied the Chebyshev polynomial approximation, following \cite{Gambino:2020crt}. The polynomial order is set to $N=15$, but the results are unchanged within the statistical error with other choices beyond $N=12$.
The $\sigma\to 0$ limit is taken for each point assuming a polynomial in $\sigma$ with data points at $\sigma a$ = 0.02, 0.05, 0.10 and 0.20. For all the cases, the extrapolation is small compared to the statistical error on the finite values of $\sigma$.

\begin{figure}[tbp]
  \centering
  \includegraphics[width=7.7cm]{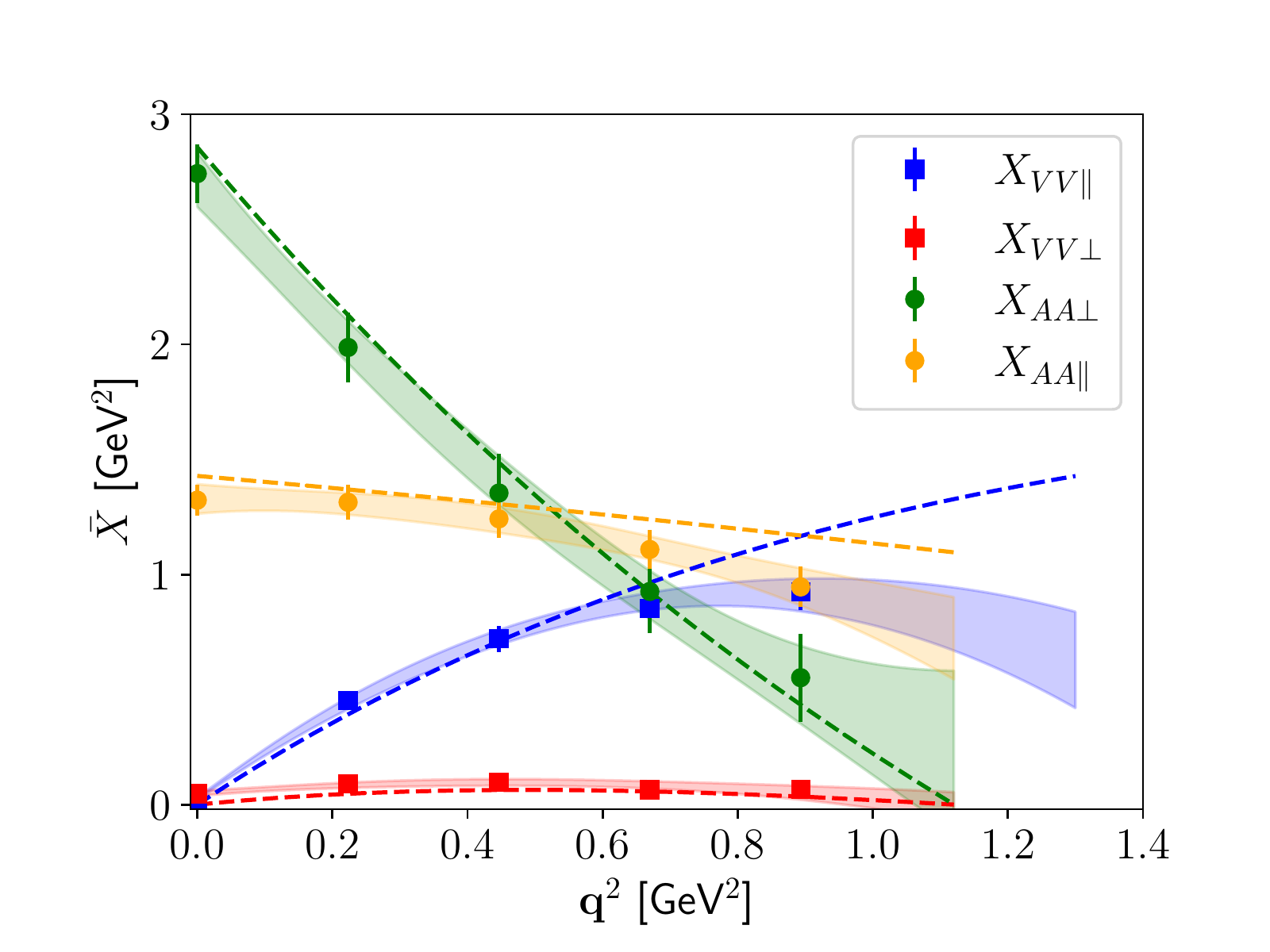}\!\!\!\!\!\!\!\!
  \includegraphics[width=7.7cm]{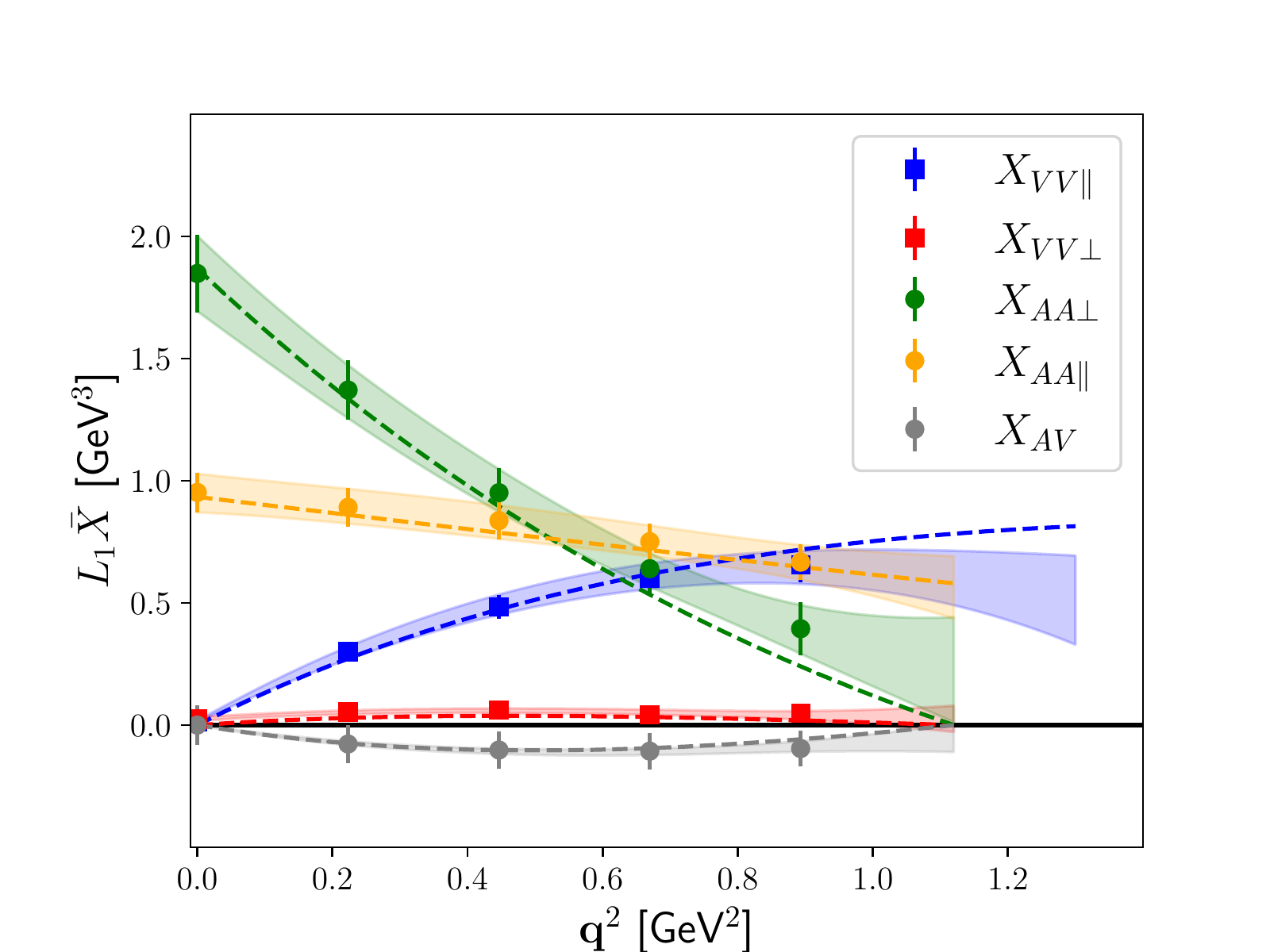}
  \caption{
    $\bar{X}$ (left panel) and $L_1\bar{X}$ (right panel, corresponding to the numerator of eq.~(\ref{eq:El})) as functions of $\bm{q}^2$.
    The results are shown for each channel.
    $X^{AV}$ is non-vanishing only for $L_1\bar{X}$.
    The dashed curves are the estimated contributions from the ground
    states of $D_s$ and $D_s^*$.
  }
  \label{fig:Xbar}
\end{figure}

The results are shown in Figure~\ref{fig:Xbar}.
The left panel is $\bar{X}$ as a function of $\bm{q}^2$, while the
integrand to produce the numerator of $\langle E_\ell\rangle$ is shown
in the right panel.
The lattice data are obtained at momentum transfer $\bm{q}$ at
(0,0,0), (0,0,1), (0,1,1), (1,1,1), (0,0,2) in units of $2\pi/L$.
Data points represent different channels as discussed in
section~\ref{sec:decomposition}.  

Also shown in figure~\ref{fig:Xbar} are dashed curves which represent
the contributions from the ground-states, i.e. $D_s$ and $D_s^*$
mesons. 
They are computed using the form factors obtained by JLQCD for the same quark mass parameters.
The necessary formulae and the lattice data are presented in the appendix~\ref{app:ground_state}.

The lattice data with different momentum insertion $\bm{q}$ are
analyzed together to account for the statistical correlations
among them.
We then fit $\bar{X}$ in a polynomial of $\bm{q}^2$ including terms up
to $(\bm{q}^2)^2$.

We observe that the inclusive results for each channel are consistent
with the expected ground-state contributions.
This means that the excited-state contributions are small, which is
consistent with our expectation from the $B\to D^{**}\ell\nu$ form
factors based on heavy-quark effective theory (HQET)~\cite{Leibovich:1997em}.
Also phenomenologically, it is plausible because the mass of the initial bottom
quark is smaller than its physical value.
The heavy quark symmetry predicts that the wave-function overlap is 1
at zero recoil when the initial and final masses are degenerate.

\begin{figure}[tbp]
  \centering
  \includegraphics[width=7.7cm]{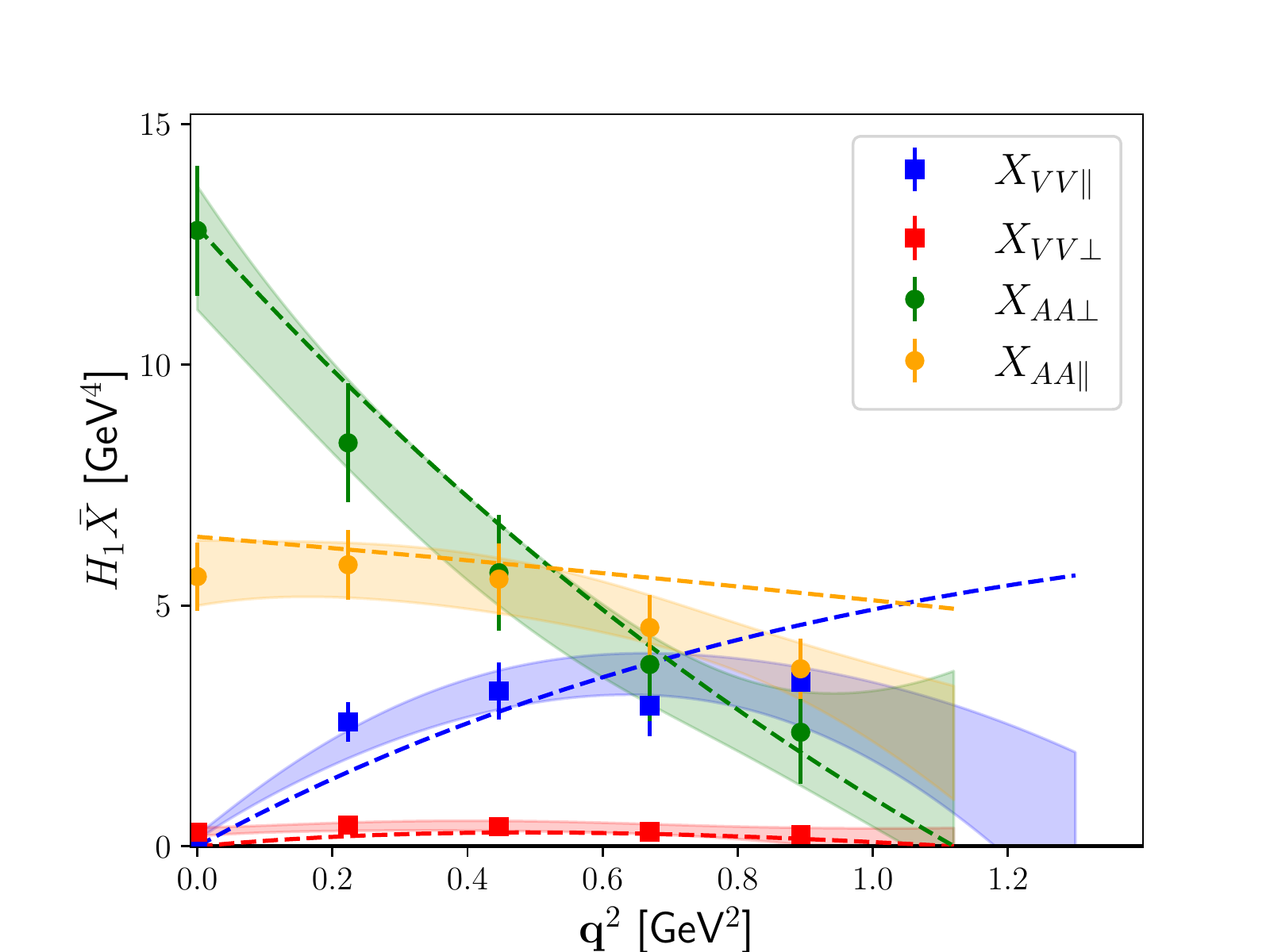}\!\!\!\!\!\!\!\!
  \includegraphics[width=7.7cm]{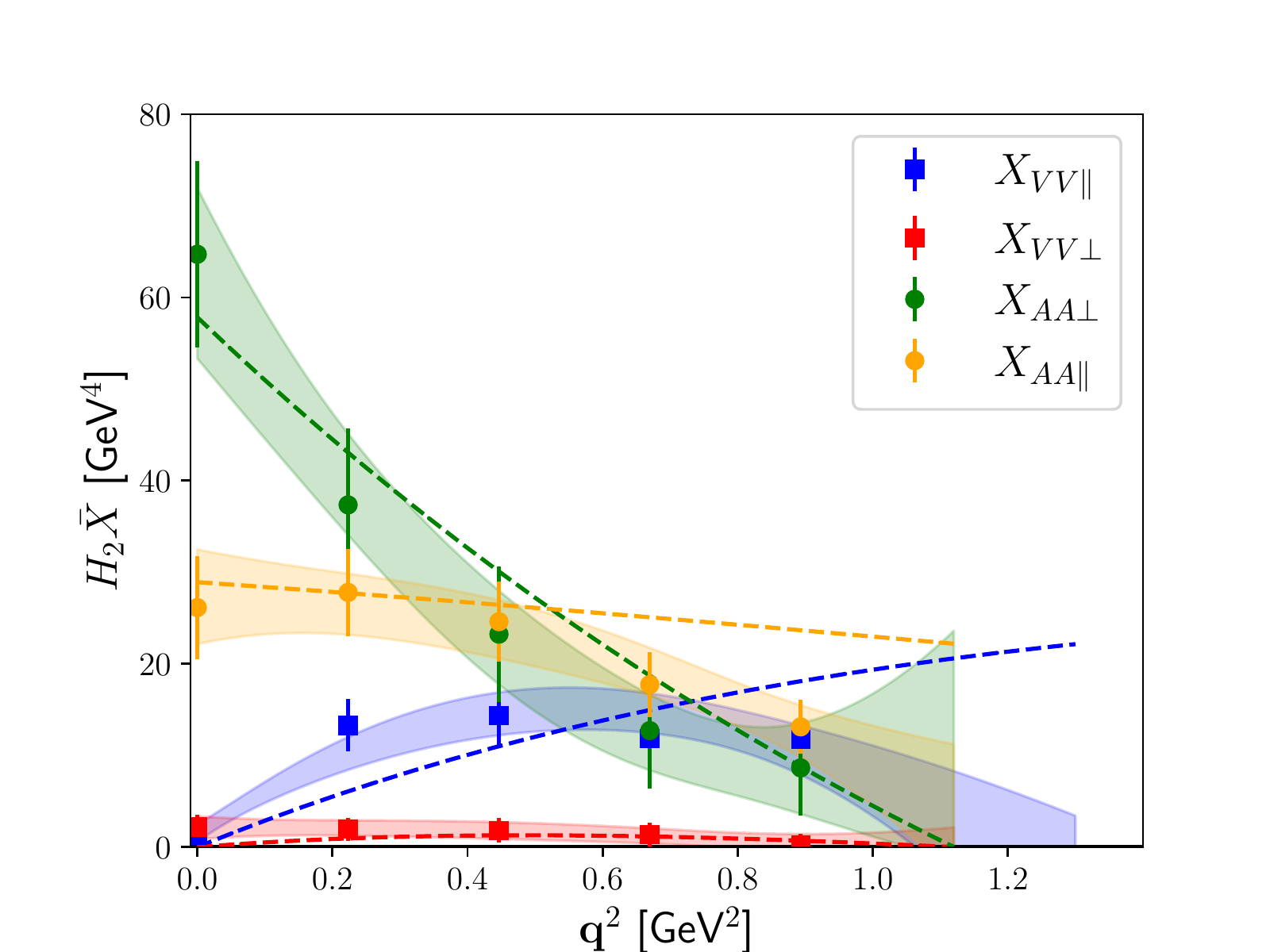}
  \caption{
    $H_1\bar{X}$ (left panel) and $H_2\bar{X}$ (right panel),
    which are numerators of eq.~(\ref{eq:MX2}),
    as a function of $\bm{q}^2$.
    The results are shown for each channel.
    The dashed curves are estimated contributions from the ground
    state of $D$ and $D^*$.
  }
  \label{fig:XbarMX}
\end{figure}

\begin{figure}[tbp]
  \centering
  \includegraphics[width=7.7cm]{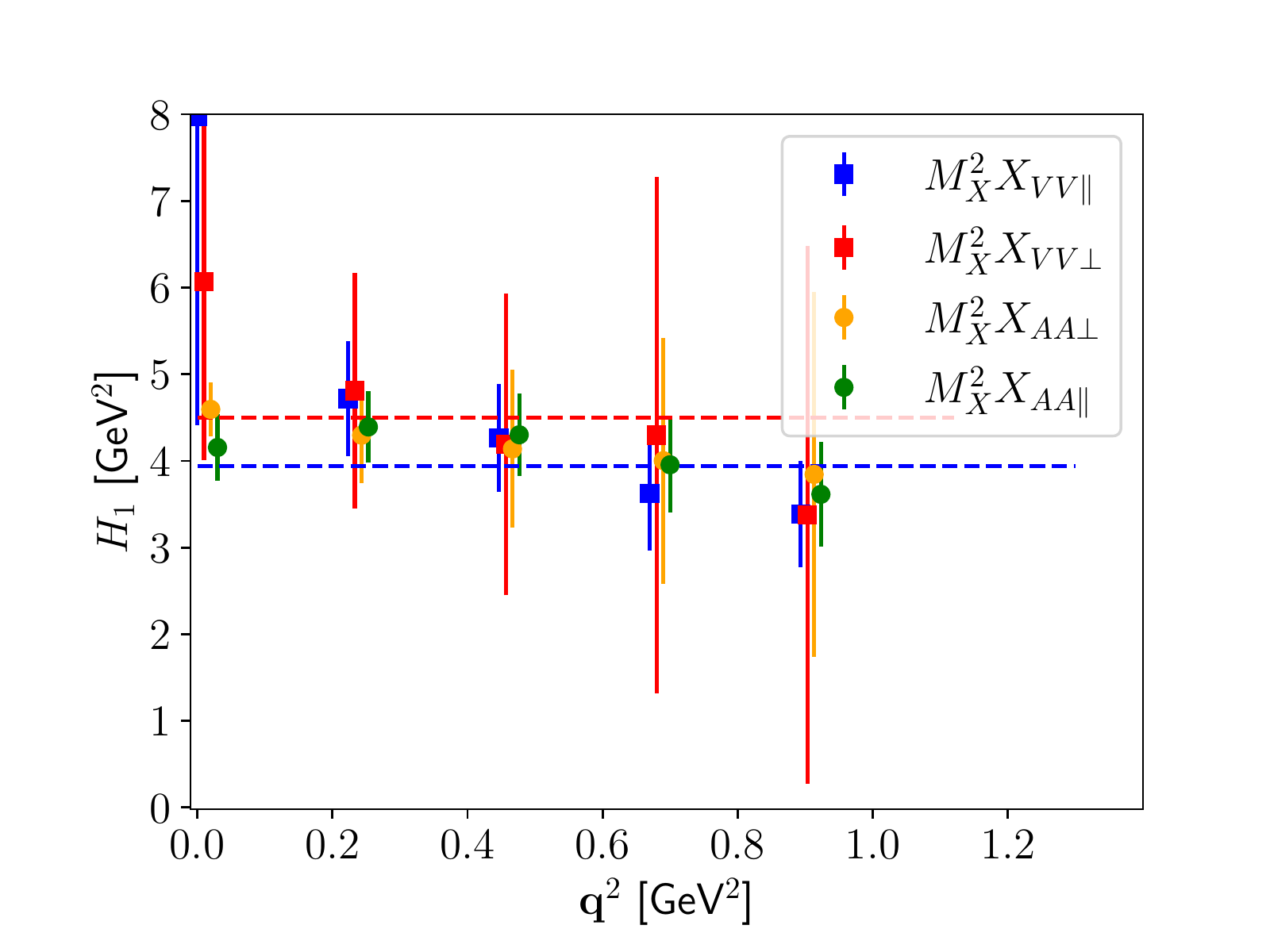}\!\!\!\!\!\!\!\!
  \includegraphics[width=7.7cm]{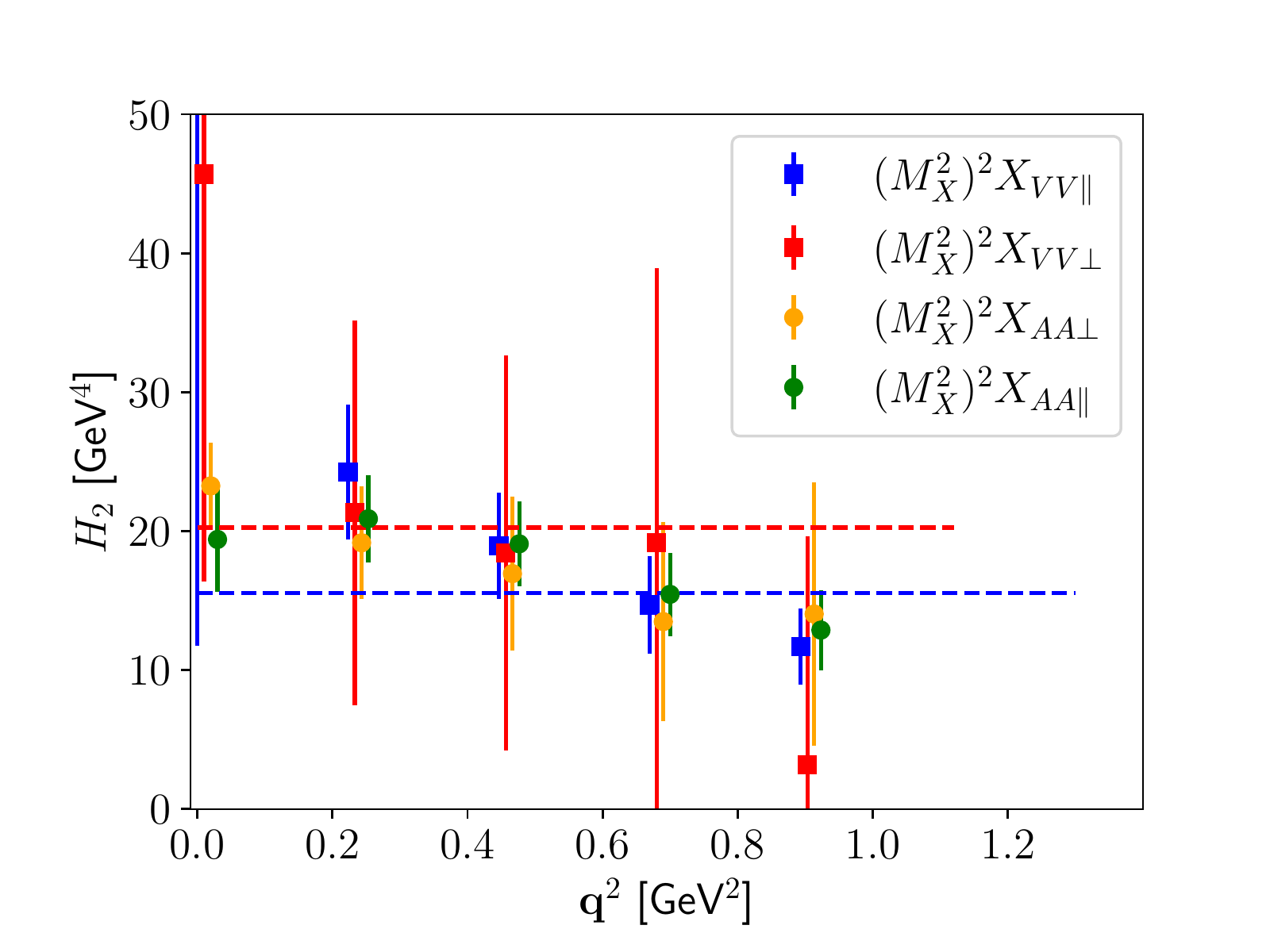}
  \caption{
    $\langle H_1\rangle_{\bm{q}^2}$ (left panel) and $\langle H_2\rangle_{\bm{q}^2}$
    (right panel) for each channel.
    The dashed lines are those of the expected contribution from the
    ground state.
  }
  \label{fig:MX}
\end{figure}

\begin{figure}[tbp]
  \centering
  \includegraphics[width=8cm]{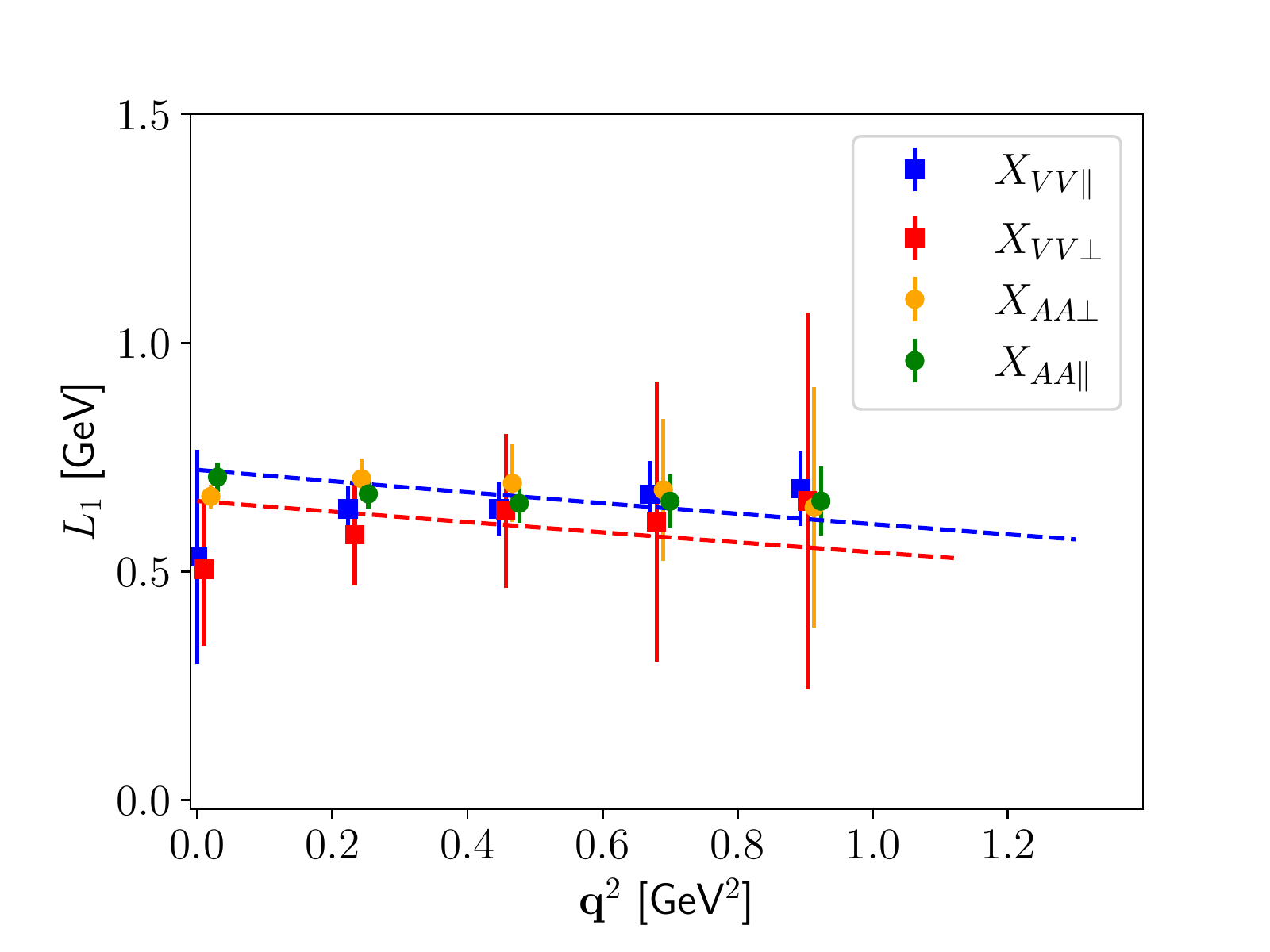}
  \caption{
    $\langle L_1\rangle_{\bm{q}^2}$ for each channel.
    The dashed lines are those of the expected contribution from the
    ground state.
  }
  \label{fig:El}
\end{figure}

We also calculate the differential moments.
The numerators for the hadronic mass moments $\langle M_X^2\rangle$
and $\langle(M_X^2)^2\rangle$ are shown in figure~\ref{fig:XbarMX},
while that for $\langle E_\ell\rangle$ is in figure~\ref{fig:Xbar}
(right panel).
The corresponding differential moments, evaluated for each channel at individual
momentum $\bm{q}^2$, are shown in figure~\ref{fig:MX} and in figure~\ref{fig:El}. 

\subsection{Lattice implementation with ETMC configurations}
\label{subsec:ETMC_lattice_calculation}

The ETMC gauge ensemble used in this work is the one named B55.32, generated by ETMC together with other 14 ensembles with $N_f = 2+1+1$ dynamical quarks in refs.~\cite{Baron:2010bv,ETM:2010cqp} for determining the average up/down, strange and charm quark masses.
The Iwasaki action~\cite{Iwasaki:1985we} and the Wilson twisted-mass action~\cite{Frezzotti:2000nk,Frezzotti:2003xj,Frezzotti:2003ni} were used for gluons and sea quarks, respectively. 
Using the mass renormalization constants determined in ref.~\cite{EuropeanTwistedMass:2014osg} the physical light, strange, and charm quark masses were found to be $m_{ud}^{\textrm{phys}}(\overline{\rm MS}, 2\,\mbox{{\rm GeV}}) = 3.70 (17)$~MeV, $m_s^{\textrm{phys}}(\overline{\rm MS}, 2\,\mbox{{\rm GeV}}) = 99.6 (4.3)$~MeV, and $m_c^{\textrm{phys}}(\overline{\rm MS}, 2\,\mbox{{\rm GeV}}) = 1176 (39)$~MeV, respectively.

In order to avoid the mixing of $K$- and $D$-meson states in the correlation functions a non-unitary setup~\cite{Frezzotti:2004wz} is used in the valence sectors: the strange and the charm valence quarks are regularised as Osterwalder-Seiler fermions~\cite{Osterwalder:1977pc}, while the up and down valence quarks have the same action as the sea.
 Working at maximal twist, such a setup guarantees an automatic ${\cal{O}}(a)$-improvement~\cite{Frezzotti:2003ni,Frezzotti:2004wz}.

The ensemble B55.32 has a lattice volume $L^3 \times T = (32^3 \times 64) ~ a^4$ with a lattice spacing equal to $a$ = 0.0815(30)~fm and a bare light-quark mass equal to $a \mu_\ell = 0.0055$, corresponding to a simulated pion mass $m_\pi$ = 375(13)~MeV~\cite{EuropeanTwistedMass:2014osg} with $m_\pi L \simeq 5.0$.
The number of analyzed gauge configurations, separated by $20$ trajectories, is $150$.
We have carried out our simulations using 
the values $a \mu_s = 0.021$ and $a \mu_c = 0.25$
for the bare valence strange and charm quark masses, which correspond to renormalised strange and charm quark masses very close to their physical values.

We have calculated the two-point function $C(t)$, defined in eq.~(\ref{eq:C2t}), using the interpolating operator $\overline{b}(x) \gamma_5 s(x)$ with a simulated $b$-quark mass equal to twice the physical charm mass, i.e. $m_b(\overline{\rm MS}, 2\,\mbox{{\rm GeV}}) \simeq 2.4$~GeV, and a physical strange quark. 
We set opposite Wilson parameters for the two valence quarks in order to guarantee that cutoff effects on the pseudoscalar mass are $O(a^2 \mu_f)$~\cite{Frezzotti:2003ni,Frezzotti:2005gi,Dimopoulos:2009qv}.
To improve the statistical precision we have made use of the ``one-end trick'' stochastic method~\cite{Foster:1998vw,McNeile:2006bz} and employed 10 spatial stochastic sources at a randomly chosen time-slice per gauge configuration.
Moreover, in order to suppress contributions of the excited states in the $B_s$-meson correlation function, we have used Gaussian smeared interpolating quark fields~\cite{Gusken:1989qx} both at the source and at the sink. 
For the values of the smearing parameters we set $k_{G}=4$ and $N_{G}=30$. In addition, we apply APE smearing 
to the gauge links~\cite{APE:1987ehd} in the interpolating fields with parameters $\alpha_{\mathrm{APE}}=0.5$ and $N_{\mathrm{APE}}=20$. 

\begin{figure}[tbp]
  \centering
  \includegraphics[width=7.7cm]{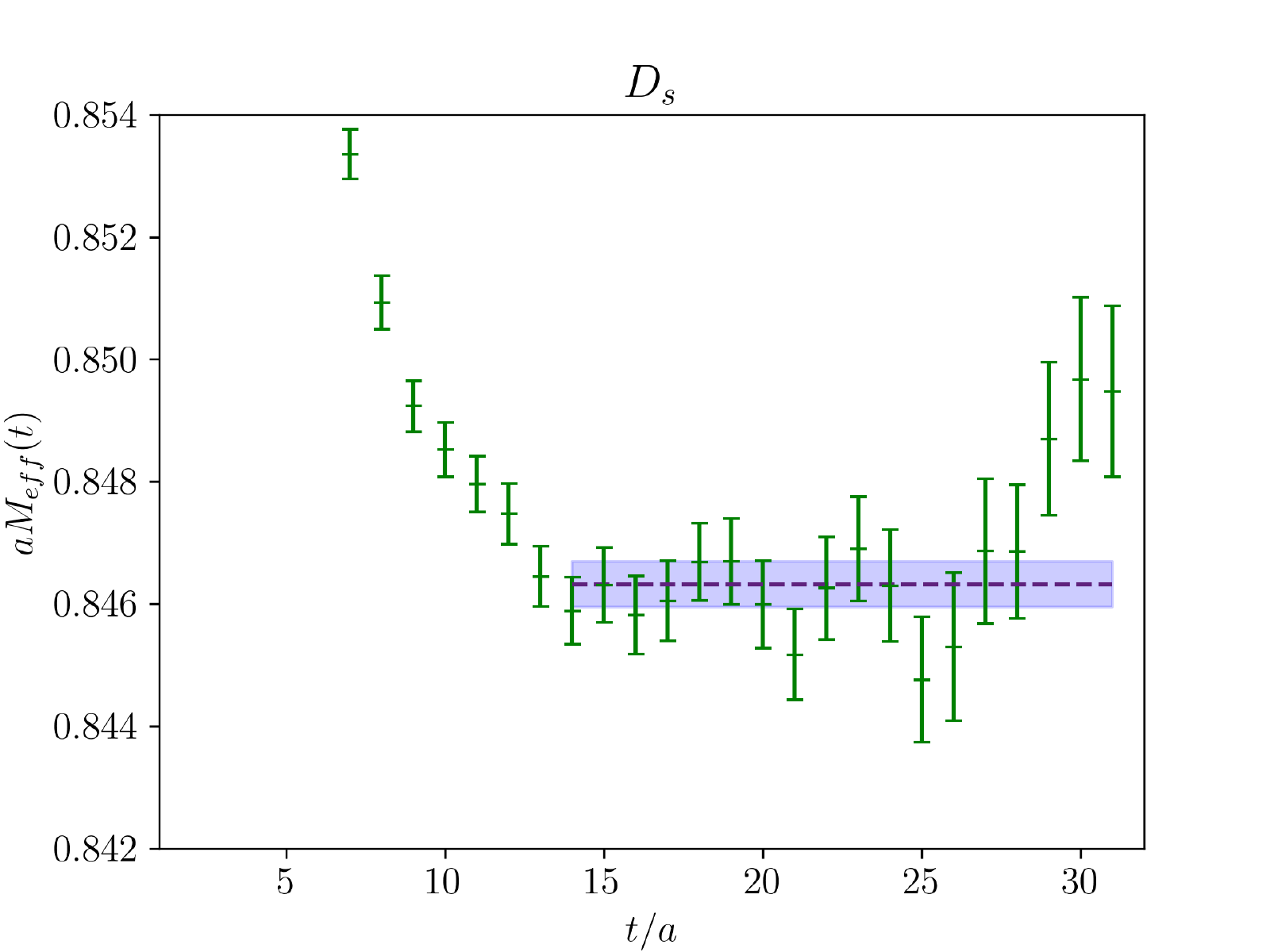}\!\!\!\!\!\!\!\!
  \includegraphics[width=7.7cm]{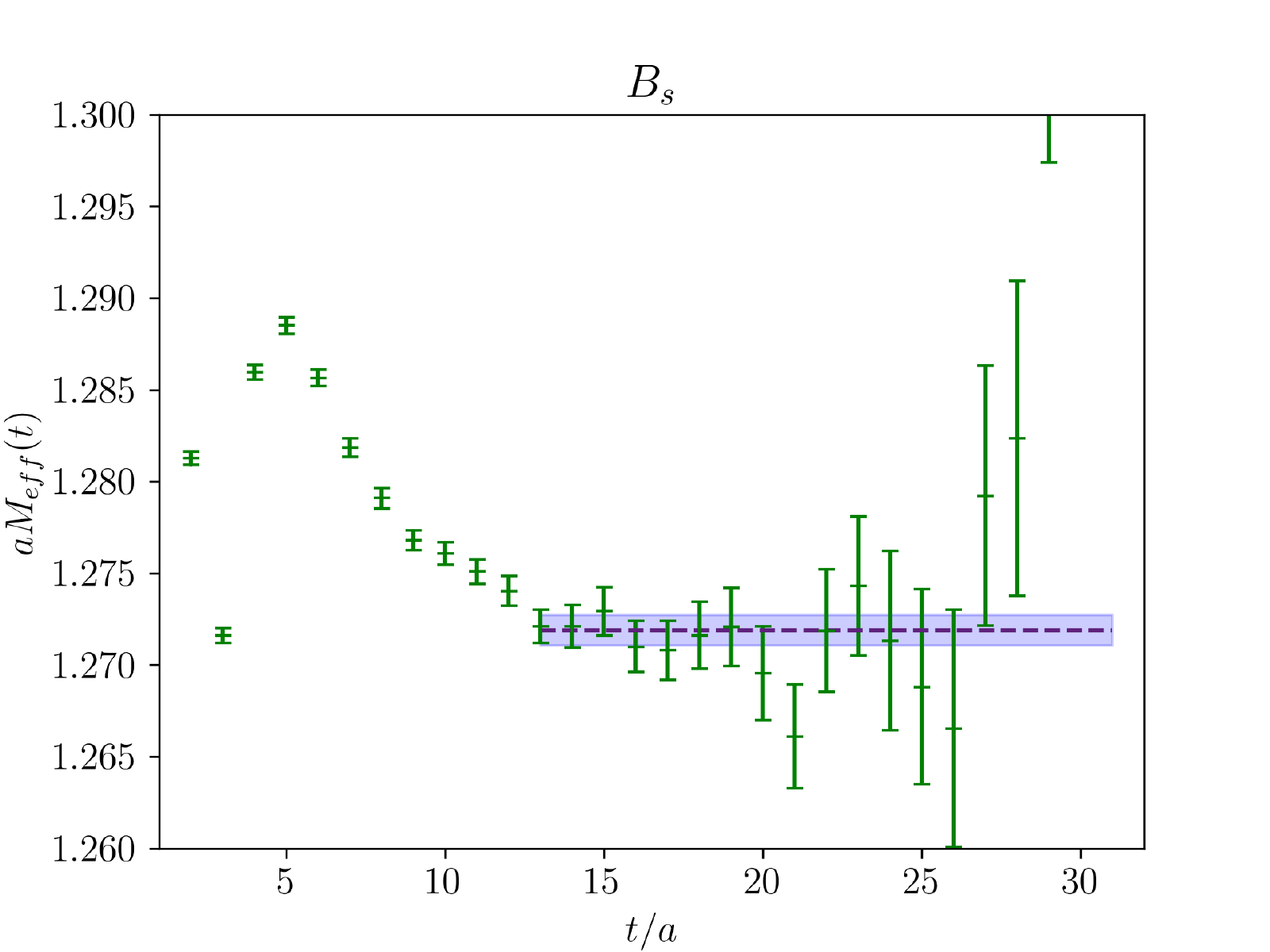}
\caption{Effective mass $aM_{\mathrm{eff}}(t) \equiv \log\left( C(t) / C(t+a) \right)$ in lattice units for the $D_s$-meson (left panel) and the $B_s$-meson (right panel) correlation function~(\ref{eq:C2t}), evaluated using the ETMC gauge ensemble B55.32 for bare quark masses equal to $a \mu_b = 0.50$, $a \mu_c = 0.25$ and $a \mu_s = 0.021$, corresponding to renormalised quark masses $m_b(\overline{\rm MS}, 2\,\mbox{{\rm GeV}}) \simeq 2.4$~GeV, $m_c(\overline{\rm MS}, 2\,\mbox{{\rm GeV}}) \simeq 1.2$~GeV and $m_s(\overline{\rm MS}, 2\,\mbox{{\rm GeV}}) \simeq 100$~MeV. The values of the Wilson $r$-parameter of the two valence quarks are opposite, i.e. $r_c = - r_s$ in the $D_s$ meson and $r_b = - r_s$ in the $B_s$ meson.}
\label{fig:Meff}
\end{figure} 
Smearing leads to improved projection onto the lowest-energy eigenstate at smaller Euclidean time separations. 
As shown by the effective mass $aM_{\mathrm{eff}}(t) \equiv \log\left( C(t) / C(t+a) \right)$ in fig.~\ref{fig:Meff}, the dominance of the ground-state signal starts around $t /a \simeq 13$ for both the $D_s$ and $B_s$ mesons. By averaging over the plateau regions shown in fig.~\ref{fig:Meff} the ground-state masses are respectively found to be $m_{D_s}$ = 2.05(8)~GeV and $m_{B_s}$ = 3.08(11)~GeV.

We have calculated the four-point function $C_{\mu\nu}(t_{\mathrm{snk}}, t_2, t_1, t_{\mathrm{src}}; \bm{q}) $, given by eq.~(\ref{eq:4pt}), as a function of $t_1$, the time at which the first weak current is inserted with momentum $\bm{q}$, for fixed values of $t_2$, where the second weak current is contracted with momentum insertion $-\bm{q}$, fixing $t_{\mathrm{src}} =0$ and $t_{\mathrm{snk}} = T/2 = 32 a$. The momentum $\bm{q}$ is inserted along one spatial direction, namely $\bm{q} = (0, 0, q)$ and we have considered eleven values for $q$ ranging from $q = 0$ up to $q = q_{\mathrm{max}} \simeq 0.9$~GeV. 
On the lattice these values are injected through the use of twisted boundary conditions (BC's)~\cite{Bedaque:2004kc,deDivitiis:2004kq,Guadagnoli:2005be} in the spatial directions and anti-periodic BC's in time. 
The sea dynamical quarks, on the contrary, are simulated with periodic BC's in the spatial directions and anti-periodic ones in time. 
The twisted BC's for the valence quark fields lift the severe limitations, arising from the use of periodic BC's, on the accessible kinematical regions of momentum-dependent quantities.
Furthermore we remark that, as shown in refs.~\cite{Sachrajda:2004mi,Bedaque:2004ax}, for physical quantities which do not involve final-state interactions (like, e.g., meson masses, decay constants and form factors), the use of different BC's for valence and sea quarks produces only finite-size effects that are exponentially small.

For the $ b \to c$ weak current we use the local vector and axial-vector quark currents, $\overline{b}(x) \gamma_\mu c(x)$ and $\overline{b}(x) \gamma_\mu \gamma_5 c(x)$. The value of the Wilson $r$-parameter for the charm quark is chosen to be opposite to that of the $b$ quark, i.e. $r_c = - r_b$, and therefore in our maximally twisted setup the vector and axial-vector currents renormalise respectively with the axial and vector renormalization constants, $Z_A$ and $Z_V$, determined in ref.~\cite{EuropeanTwistedMass:2014osg}.
\begin{figure}[tbp]
\centerline{\includegraphics[width=10cm]{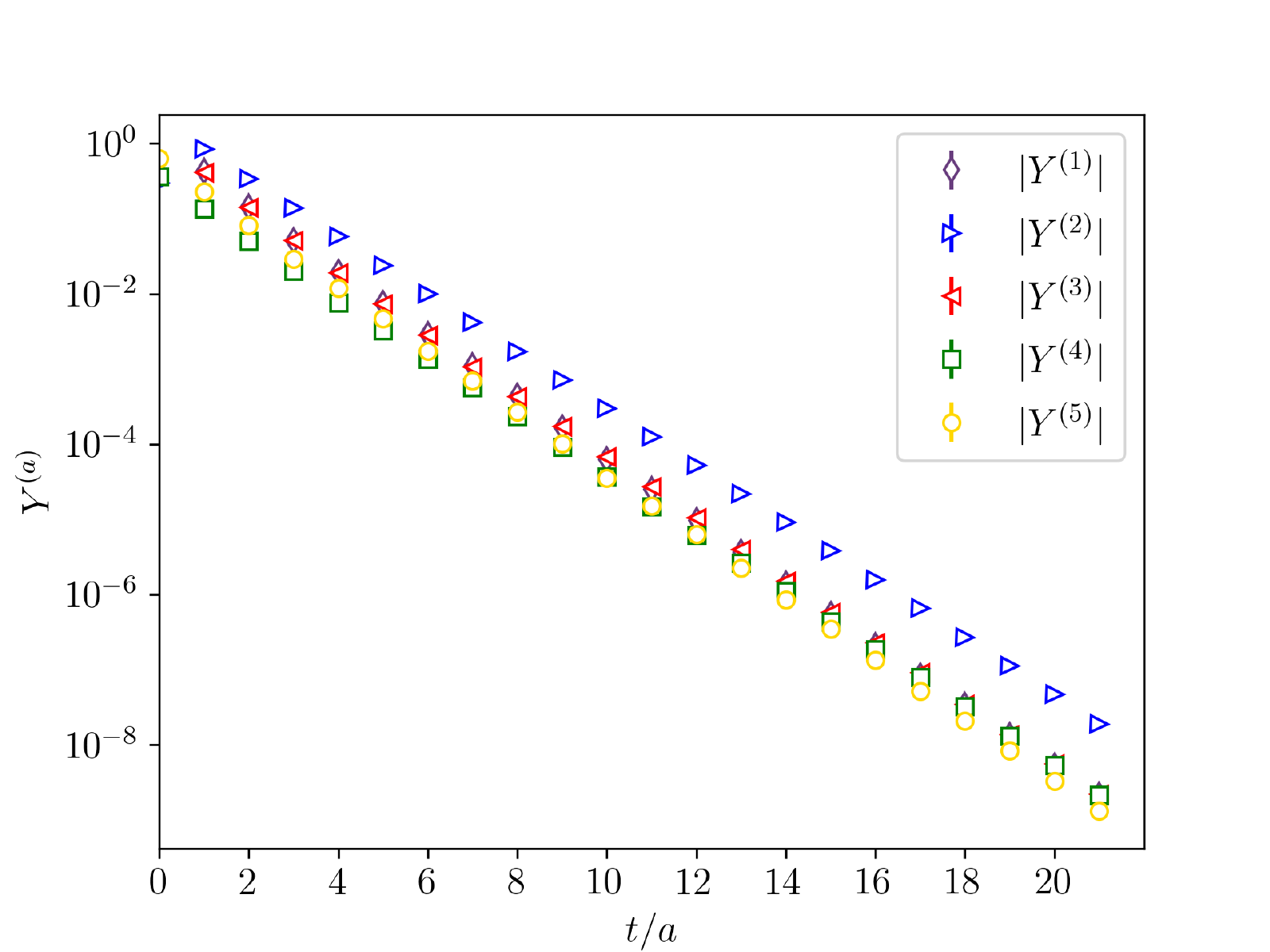}}
\caption{Time dependence of the correlators $Y^{(a)}(t;\bm{q})$ for $\vert\bm{q}\vert \simeq 0.5$~GeV calculated on the ETMC ensemble B55.32. The error bars are smaller than the point markers on this scale and a similar quality of the numerical signal is observed for the other momentum values considered in this work.}
\label{fig:Ycorrs}
\end{figure}

We extract the matrix elements $M_{\mu\nu}(t_2 - t_1; \bm{q})$ using eq.~(\ref{eq:Mmunu}). 
In order to calculate $\bar X(\bm{q}^2)$, as defined in eq.~(\ref{eq:Xint2}), we apply the smearing kernel $\Theta^{(l)}(\omega_{\mathrm{max}}-\omega)$ to the quantities $Z^{(l)}(\omega,\bm{q}^2)$. These in turn are defined in terms of the quantities $Y^{(a)}(\omega,\bm{q})$ in eq.~(\ref{eq:ZfromY}). To this end we start from the linear combinations of the correlators $M_{\mu\nu}(t;\bm{q})$ with the kinematical coefficients of eqs.~(\ref{eq:Ydefs}). We call these objects
\begin{flalign}
&
Y^{(a)}(t;\bm{q}^2) =
\int_0^\infty d\omega\, Y^{(a)}(\omega,\bm{q}^2)\, e^{-\omega t}\;,
\qquad
a=1,\cdots,5\;,
\nonumber \\
&
Z^{(l)}(t;\bm{q}^2) =
\int_0^\infty d\omega\, Z^{(l)}(\omega,\bm{q}^2)\, e^{-\omega t}\;,
\qquad
l=0,1,2\;.
\end{flalign}
To show the quality of the numerical data, in fig.~\ref{fig:Ycorrs} we plot the correlators $Y^{(a)}(t;\bm{q}^2)$ corresponding to $\vert\bm{q}\vert \simeq 0.5$~GeV. Notice that the correlators $Z^{(l)}(t;\bm{q}^2)$ are linear combinations of the $Y^{(a)}(t;\bm{q}^2)$'s, see eq.~(\ref{eq:ZfromY}). Similar results are obtained for the other momenta considered in this work.

\begin{figure}[tbp]
\centerline{\includegraphics[width=\textwidth]{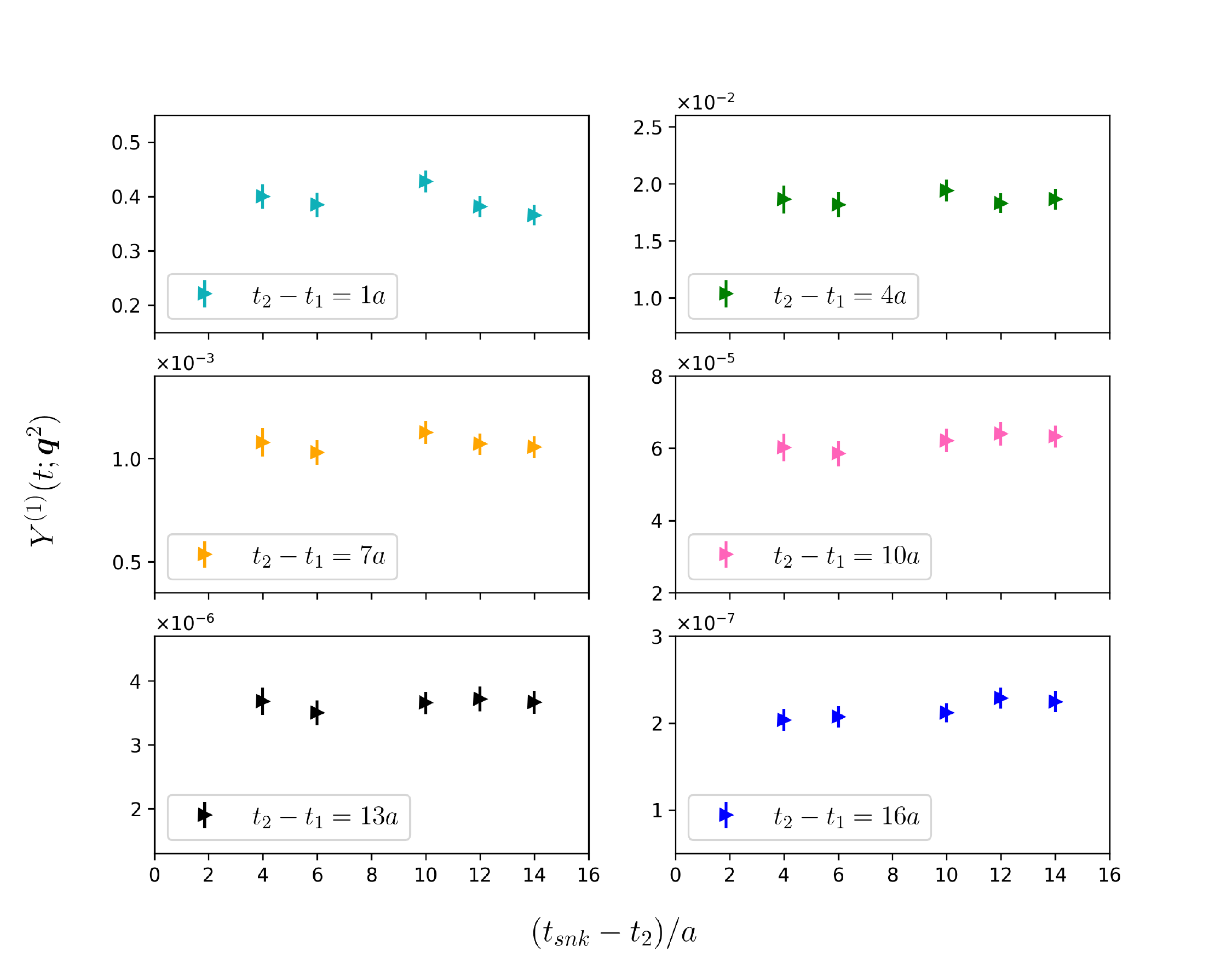}}
\caption{Correlator $Y^{(1)}(t,\bm{q}^2)$ at various time separations $t_2-t_1$ for $\vert \boldsymbol{q} \vert \simeq 0.5$~GeV. The points in each subplot are obtained for different values of $t_2$, with the $x$-axis showing the distance between $t_{\mathrm{snk}}$ and the time $t_2$ at which the current is inserted.}
\label{fig:Y1corrs}
\end{figure}

The central values for all the physical quantities extracted from the $Y^{(a)}(t,\bm{q})$ correlators have been extracted by setting $t_2=22a$ in eq.~(\ref{eq:Mmunu}) and by using the data up to $t=18a$, which corresponds to $t_1-t_{\mathrm{src}}=4a$. To check the approach to the $t_{\mathrm{src}}\to-\infty$ and $t_{\mathrm{snk}}\to\infty$ limits we have repeated the analysis by setting $t_2=\{18a, 20a, 22a, 26a, 28a\}$ and by varying the maximum value of $t$ used to reconstruct the smearing kernels. Figure~\ref{fig:Y1corrs} shows the comparison of the correlator $Y^{(1)}(t,\bm{q})$ at $\vert \boldsymbol{q} \vert \simeq 0.5$~GeV for different values of $t=t_2-t_1$ and $t_2$.
In the following analysis, we chose the value $(t_{\mathrm{snk}}-t_2)=10a$, corresponding to $t_2=22a$. Similar results are obtained for the other correlators ($Y^{(2)}$, $Y^{(3)}$, $Y^{(4)}$ and $Y^{(5)}$), and, in all cases, we observe that the onset of the $t_{\mathrm{snk}}\to\infty$ limit is reached within the uncertainties already for $t_{\mathrm{snk}}-t_2=4a$.   

\begin{figure}[tbp]
  \centering
  \includegraphics[width=8cm]{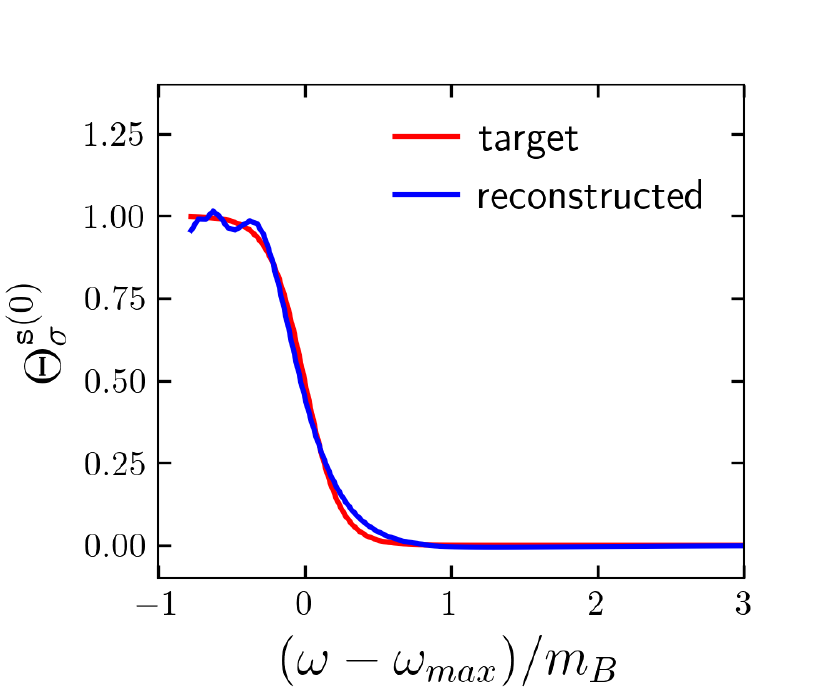}
  \includegraphics[width=8cm]{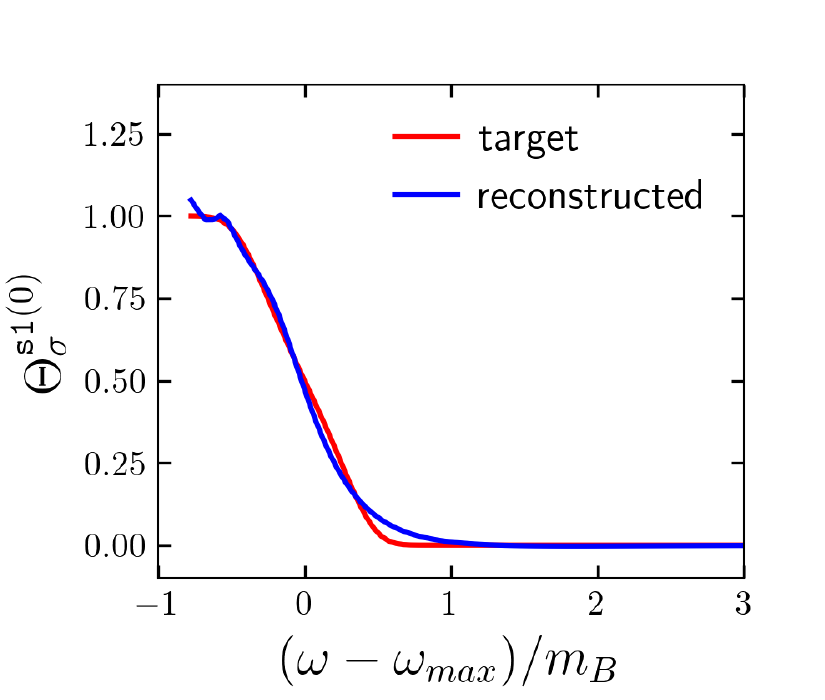}
  \includegraphics[width=8cm]{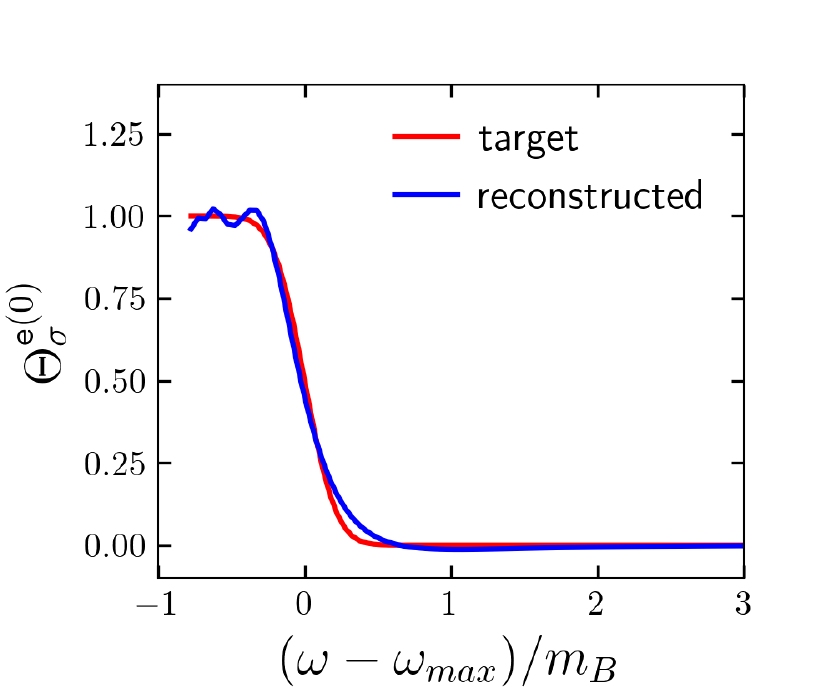}
  \caption{Reconstruction of the kernels $\Theta_\sigma^{(0)}(\omega_{\mathrm{max}}-\omega)$ defined with the three smearing types $\mathtt{s}$, $\mathtt{s1}$ and $\mathtt{e}$, see eq.~(\ref{eq:thetasss}), at $\lambda=\lambda_\star$.
  The data correspond to $\vert \bm{q} \vert\simeq 0.7$~GeV and $\sigma=0.12 m_{B_s}$, the smallest value of the smearing parameter that we used.}
\label{fig:dG0reco_approx}
\end{figure} 

\begin{figure}[tbp]
  \centering
  \includegraphics[width=\textwidth]{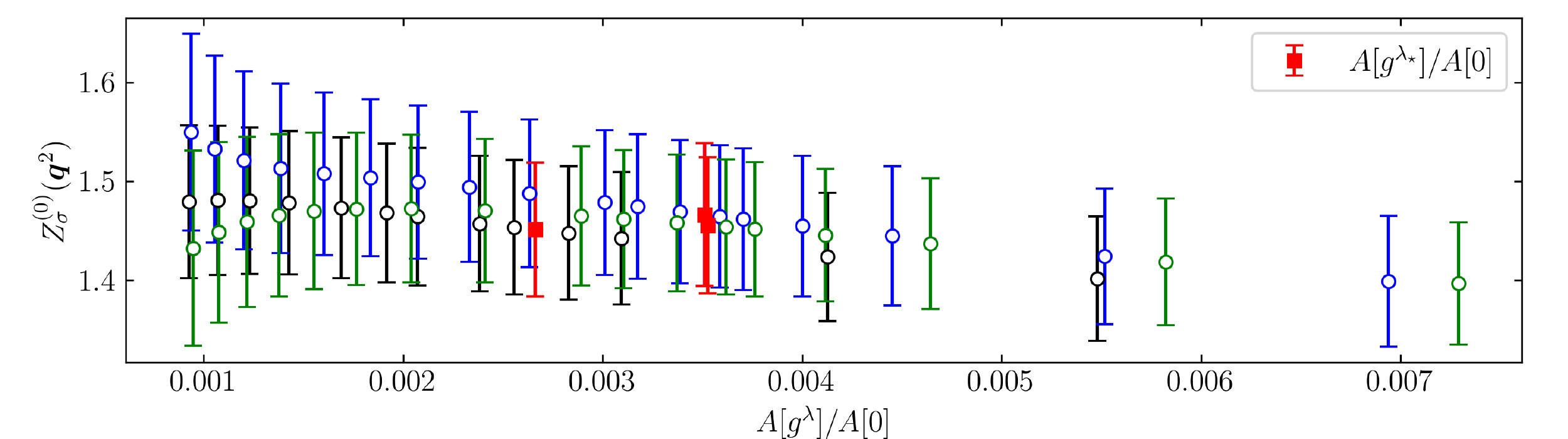}
\caption{
  Integral $\bar Z^{(0)}_\sigma(\bm{q})$ of the hadronic correlator with three kernels, plotted as a function of $A[g^\lambda]/A[0]$.
  No significant difference is observed within the statistical errors for values $A[g^\lambda]/A[0]$ smaller than $A[g^{\lambda_\star}]/A[0]$.
}
\label{fig:dG0reco}
\end{figure} 

%

We now turn to the discussion of the systematics associated with the approximation of the kernels of eq.~(\ref{eq:thetasss}) by using the method of ref.~\cite{Hansen:2019idp}. This is an important issue because, on the one hand, the reconstruction of a given kernel can never be exact with a finite number of time-slices and in the presence of errors. On the other hand, one can (and must) quantify the systematic error associated with an approximate reconstruction.

In order to illustrate this point we consider the quantity $Z^{(0)}_\sigma(\bm{q}^2)$ (see eq.~(\ref{eq:Xint2})) for three smooth approximations of the $\theta$-function given in eq.~(\ref{eq:thetasss}). The kernels are approximated as described in section~\ref{sec:formulation_of_the_method_and_application_to_observables}, see in particular eq.~(\ref{eq:mainHLT}), with $\tau_\mathrm{max}=18$. The quantity $Z^{(0)}_\sigma(\bm{q}^2)$ is then obtained by applying the coefficients $g_\tau^\lambda$ that represent the approximated kernel at a fixed value of $\lambda$ to the correlator $Z^{(0)}(t;\bm{q}^2)$. Figure~\ref{fig:dG0reco_approx} shows the comparison of the reconstructed kernels with the target ones for $\vert \bm{q} \vert\simeq 0.7$~GeV and $\sigma=0.12 m_{B_s}$ at the values $\lambda=\lambda_\star$ determined independently for each kernel. The values of $\lambda_\star$ are marked with red points in fig.~\ref{fig:dG0reco}, where we show the dependence of $Z^{(0)}_\sigma(\bm{q}^2)$ on the normalised $L_2$-norm $A[g^\lambda]/A[0]$. As explained in section~\ref{sec:formulation_of_the_method_and_application_to_observables}, for smaller values of $\lambda$ one obtains a more accurate reconstruction of the kernels and thus smaller $A[g^\lambda]/A[0]$ values.
There is no significant difference on the final results for $Z^{(0)}_\sigma(\bm{q})$ by decreasing $\lambda$ with respect to $\lambda_\star$.


By implementing this strategy, proposed in ref.~\cite{Bulava:2021fre}, we have checked that the estimated errors on the different quantities that enter our determinations of the physical observables discussed below properly take into account the systematics associated with the kernel approximation.  

\begin{figure}[tbp]
\centerline{\includegraphics[width=10cm]{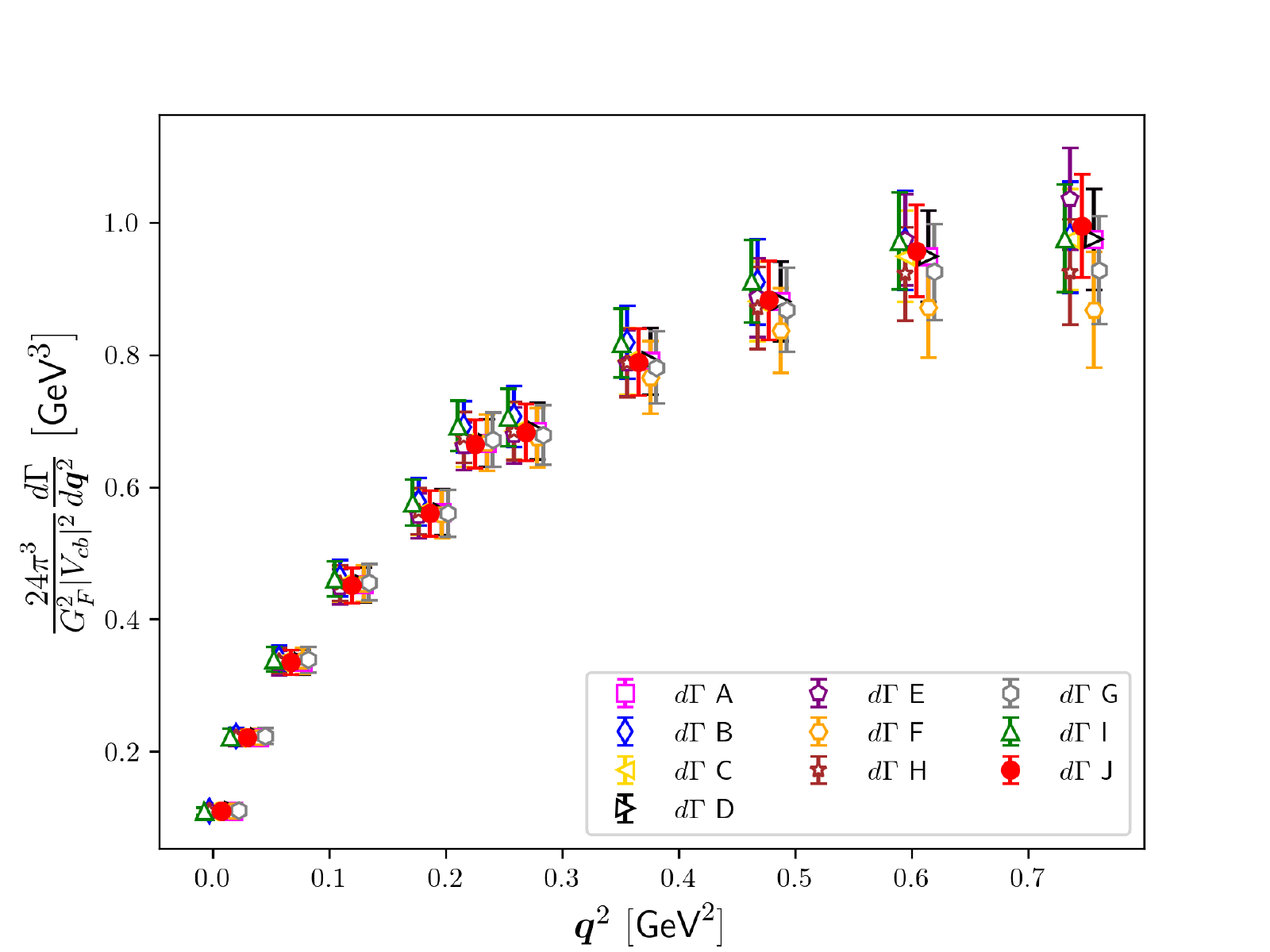}}
\caption{ Results for $\frac{24\pi^3}{G_F^2|V_{cb}|^2}\frac{d\Gamma}{d\boldsymbol{q}^2}$, obtained changing the parameters given as input to our analysis. The default values are: $A_{tr}=1\times10^{-3}$, $\tau_{max}=18$, extrapolations to $\sigma=0$ using $5$ values of $\sigma$. The letters in the legend stand for: \textbf{A)} All parameter equal to default, the final result is given by extrapolating to $\sigma=0$ the single components $X^{(i)}$ and then summing the extrapolations together. \textbf{B)} The same as case \textbf{(A)} but with extrapolations done employing all $10$ values of $\sigma$, as quoted in the caption of fig.~\ref{fig:sigmato0}. \textbf{C)} A threshold changed to $A_{tr}=1\times10^{-2}$. \textbf{D)} A threshold changed to $A_{tr}=5\times10^{-3}$. \textbf{E)} All parameters equal to default, final result given by summing all the single contributions $X^{(i)}$ together and then extrapolation the sum to $\sigma=0$. \textbf{F)} $\tau_{max}$ changed to $\tau_{max}=15$. \textbf{G)} $\tau_{max}$ changed to $\tau_{max}=16$. \textbf{H)} $\tau_{max}$ changed to $\tau_{max}=17$. \textbf{I)} Same as default, analysis performed using the bootstrap method. \textbf{J)} Final results obtained considering all previous results listed here. Central value and  standard deviation are calculated using the average procedure given by eq.~(28) of ref.~\cite{EuropeanTwistedMass:2014osg}. It is important to note that the analysis of all the cases listed above is performed taking the result corresponding to $\lambda=\lambda_{\star}$ as discussed in subsection~\ref{subsec:ETMC_lattice_calculation}, the only exception being when we change the $A_{tr}$ parameter. In these two cases we take the results corresponding to values of $A[g^{\lambda}]/A[0]$ smaller than $A_{tr}$.}
\label{fig:decay_rate}
\end{figure} 

In fig.~\ref{fig:decay_rate} we show our results for the total decay rate, with the different points corresponding to different input parameters used in the analysis, as described in the figure's caption. The plot shows clearly that all results are compatible with each other. In order to take into account all the results showed in the figure, we use eq.~(28) of ref.~\cite{EuropeanTwistedMass:2014osg} to get an estimate of the central value and its standard deviation, corresponding to the filled red dots in the plot, and we quote that value as our final result for the total decay rate. This procedure is repeated for all other observables considered in this work.

The final ETMC results for all the physical observables, divided into four different channels, are shown together with the OPE results in figures~\ref{fig:diffqSM}, \ref{fig:diffqETMC}, \ref{fig:mom1ETMC}, and~\ref{fig:mom2cETMC}.

\subsection{Extrapolation to  $\sigma= 0$}
\label{subsec:sigma_to_zero_extrapolation}

\begin{figure}[tbp]
\includegraphics[width=0.5\textwidth]{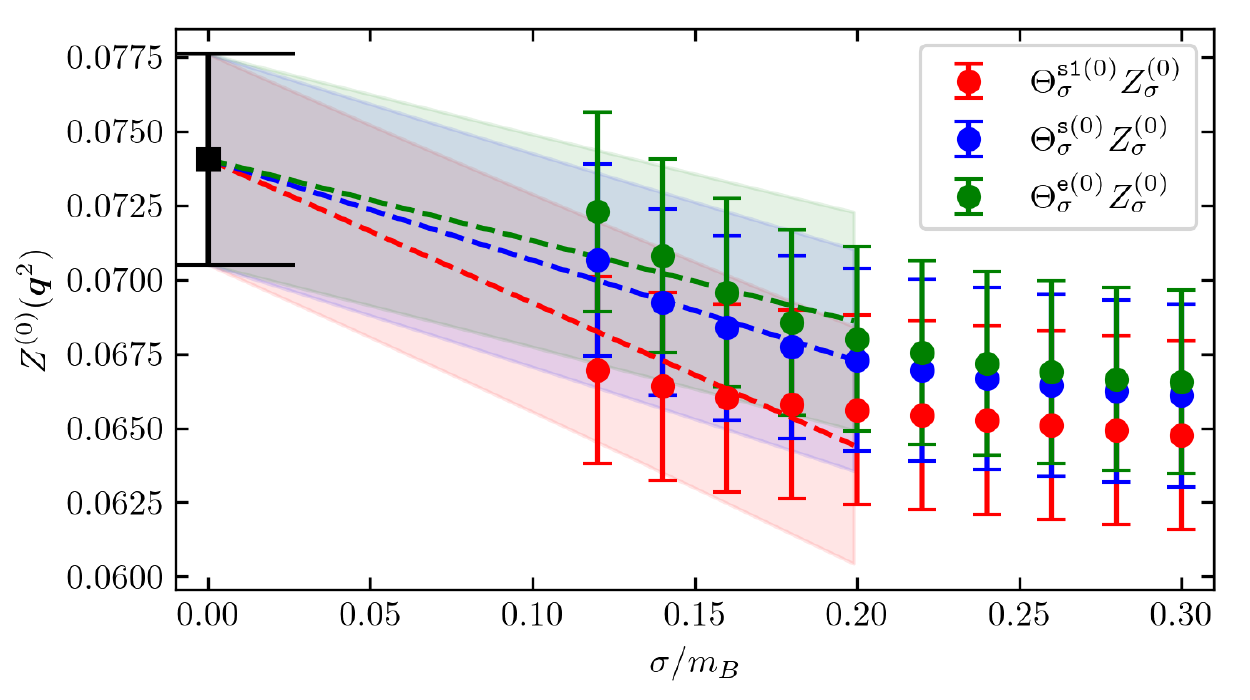}
\includegraphics[width=0.5\textwidth]{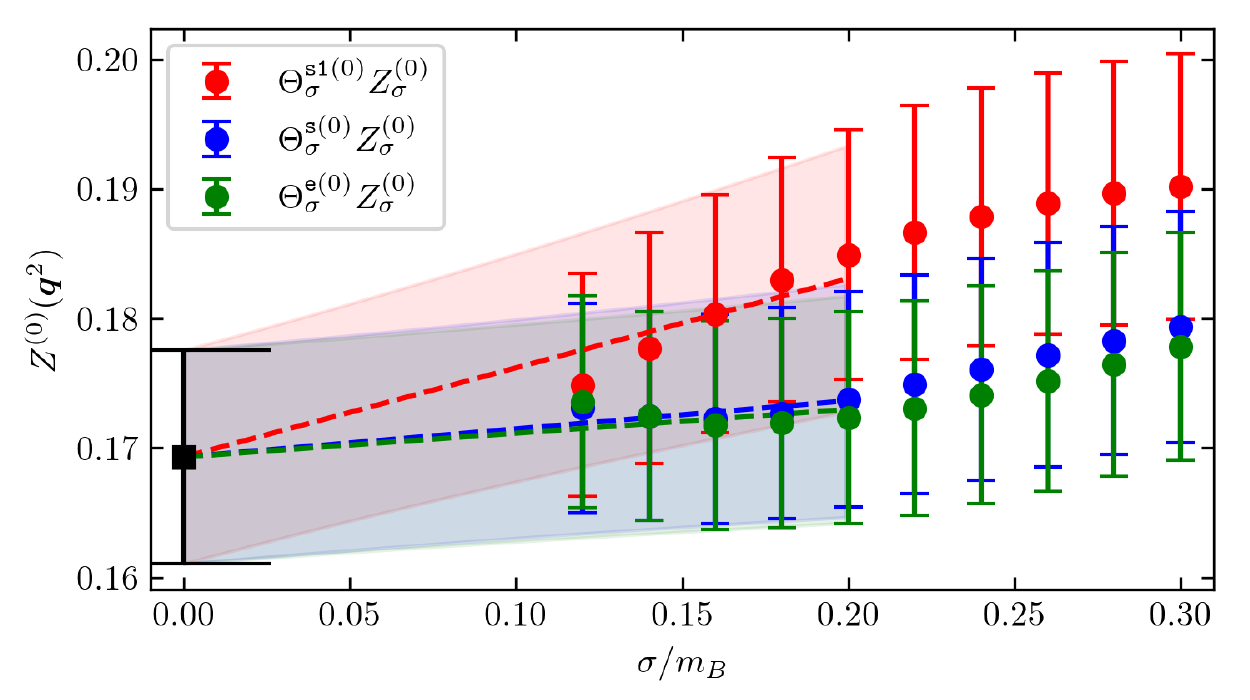}\\
\includegraphics[width=0.5\textwidth]{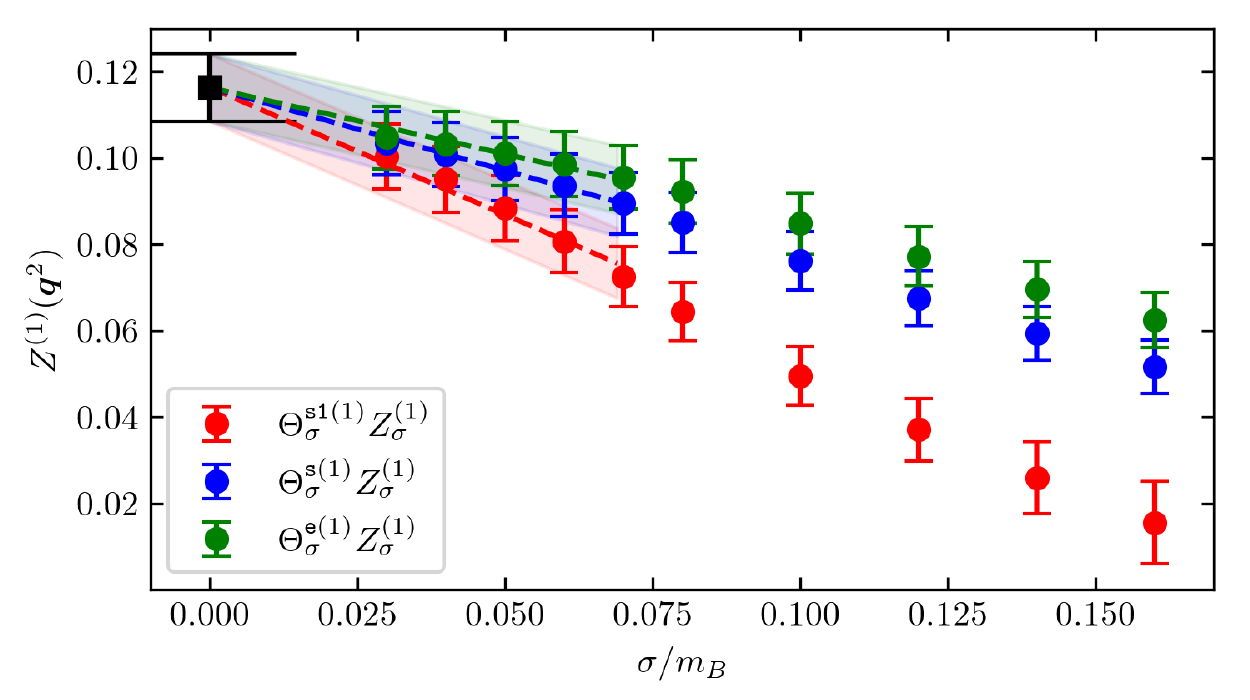}
\includegraphics[width=0.5\textwidth]{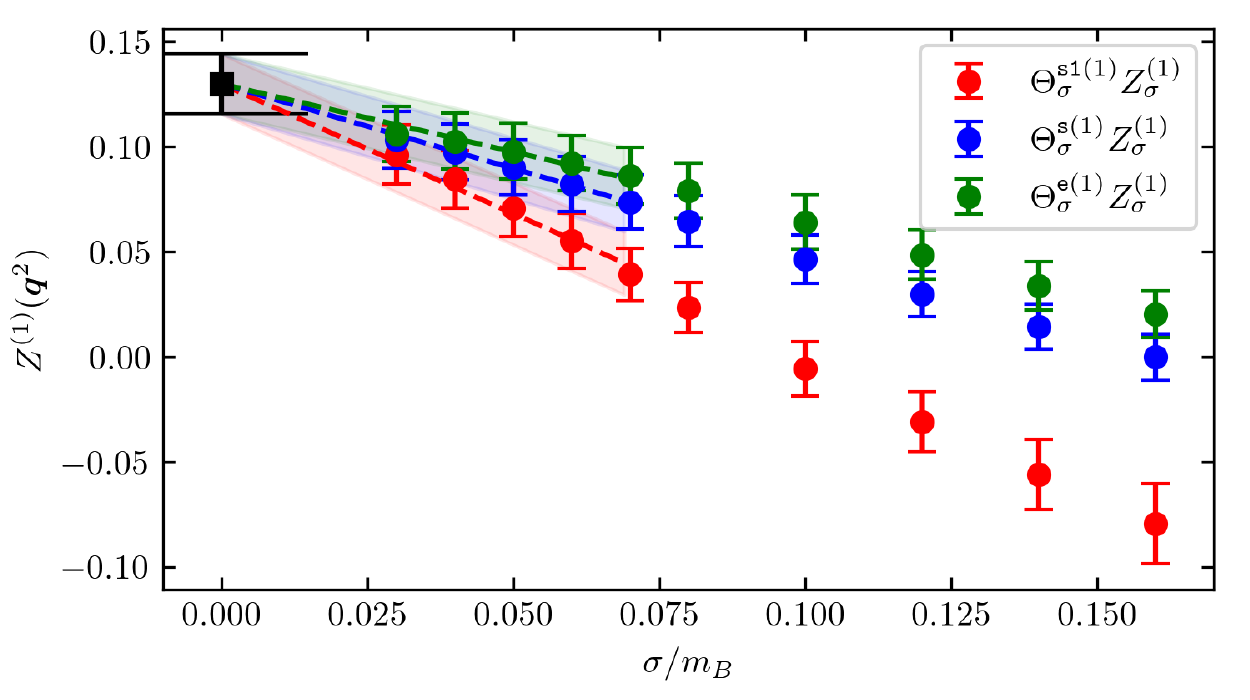}\\
\includegraphics[width=0.5\textwidth]{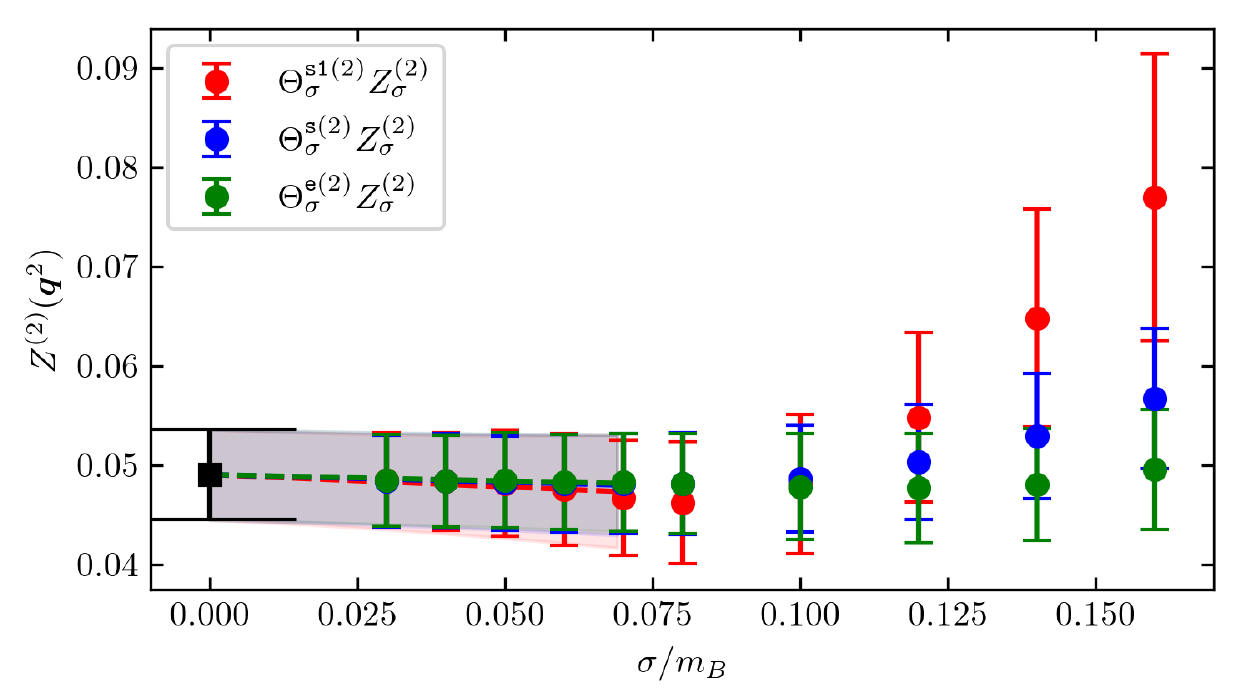}
\includegraphics[width=0.5\textwidth]{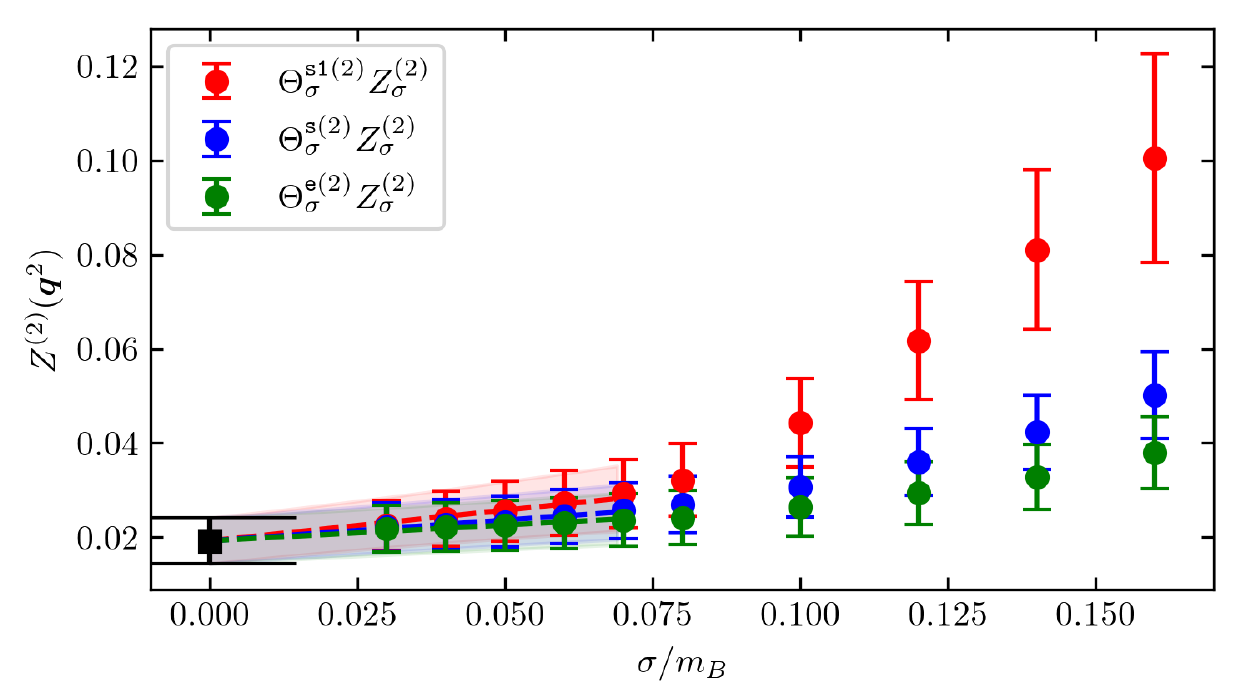}
\caption{Combined $\sigma\to 0$ extrapolations of three contributions $Z^{(l)}(\bm{q}^2)$ to the differential decay rate, see eq.~(\ref{eq:Xint2}). The plots on the left correspond to $\vert \bm{q} \vert\simeq 0.5$~GeV while those on the right to $\vert \bm{q} \vert\simeq 0.7$~GeV. The reconstruction of the kernels $\Theta^{(0)}_\sigma(\omega_{\mathrm{max}}-\omega)$ is more difficult from the numerical point of view w.r.t. the case of the kernels $\Theta^{(l)}_\sigma(\omega_{\mathrm{max}}-\omega)$ with $l=1,2$. In all cases we have obtained results at 10 different values of $\sigma$ that, in the case of $\Theta^{(0)}_\sigma(\omega_{\mathrm{max}}-\omega)$ span the region $\sigma\in [0.12 m_{B_s},0.3 m_{B_s}]$ while in the other case we have $\sigma\in [0.03 m_{B_s},0.16 m_{B_s}]$. In all cases we include the five smallest values of $\sigma$ into a combined linear extrapolation to quote our results at $\sigma=0$.}
\label{fig:sigmato0}
\end{figure} 

The ETMC data are produced at several values of the smearing parameter $\sigma$ and, for each of the target kernels $\Theta^{(l)}(x)$ with three different smeared versions of the $\theta$-function in eq.~(\ref{eq:thetasss}). These are used in a combined $\sigma\to 0$ extrapolation for each contribution to the differential decay rate and to the leptonic and hadronic moments. 

Before presenting the results of the $\sigma\to 0$ extrapolation an important remark is needed. As discussed in section~\ref{sec:formulation_of_the_method_and_application_to_observables} the limits of zero smearing radius and of infinite volume do \emph{not} commute. Because of the quantized energy spectrum on a finite volume, the $\sigma\to 0$ extrapolation must be performed only after the infinite-volume limit. Under the reasonable assumption that smeared QCD spectral densities are affected by exponentially suppressed finite-volume effects, and given the exploratory nature of the present work, we shall assume below that finite-volume effects are negligible with respect to our statistical uncertainties. This assumption can only be verified with simulations on larger volumes, a task that we leave for future work on the subject. Taking  this issue into account, the $\sigma\to 0$ extrapolation discussed below has to be considered as a feasibility study that, as we work at unphysical meson masses and fixed cutoff, we consider interesting and promising. 

In fig.~\ref{fig:sigmato0} we show the $\sigma\to 0$ extrapolations of the three contributions $Z^{(l)}_\sigma(\bm{q}^2)$ to the differential decay rate for $\vert \bm{q} \vert\simeq 0.5$~GeV (plots on the left) and $\vert \bm{q} \vert\simeq 0.7$~GeV (plots on the right). The reconstruction of the kernels $\Theta^{(0)}_\sigma(\omega_{\mathrm{max}}-\omega)$ is more challenging from the numerical point of view with respect to the case of the kernels $\Theta^{(l)}_\sigma(\omega_{\mathrm{max}}-\omega)$ with $l=1,2$. In all cases studied in this work we have obtained results at 10 different values of $\sigma$ that, for the kernel $\Theta^{(0)}_\sigma(\omega_{\mathrm{max}}-\omega)$ span the region $\sigma\in [0.12 m_{B_s},0.3 m_{B_s}]$ while for the other kernels we have $\sigma\in [0.03 m_{B_s},0.16 m_{B_s}]$. 
For all the values of $\bm{q}^2$ we have included the five smallest $\sigma$ values into a combined linear extrapolation to obtain the central values and statistical errors that we quote for our results at $\sigma=0$. As evident from the plots in fig.~\ref{fig:sigmato0} there is a reassuring convergence of the results corresponding to the different kernels for small values of $\sigma$. The five points included in the fit are always in the linear regime and the $\chi^2/\mbox{d.o.f.}$ for all the combined $\sigma\to 0$ linear extrapolations performed in this work never exceed 1.

\begin{figure}[tbp]
\includegraphics[width=0.8\textwidth]{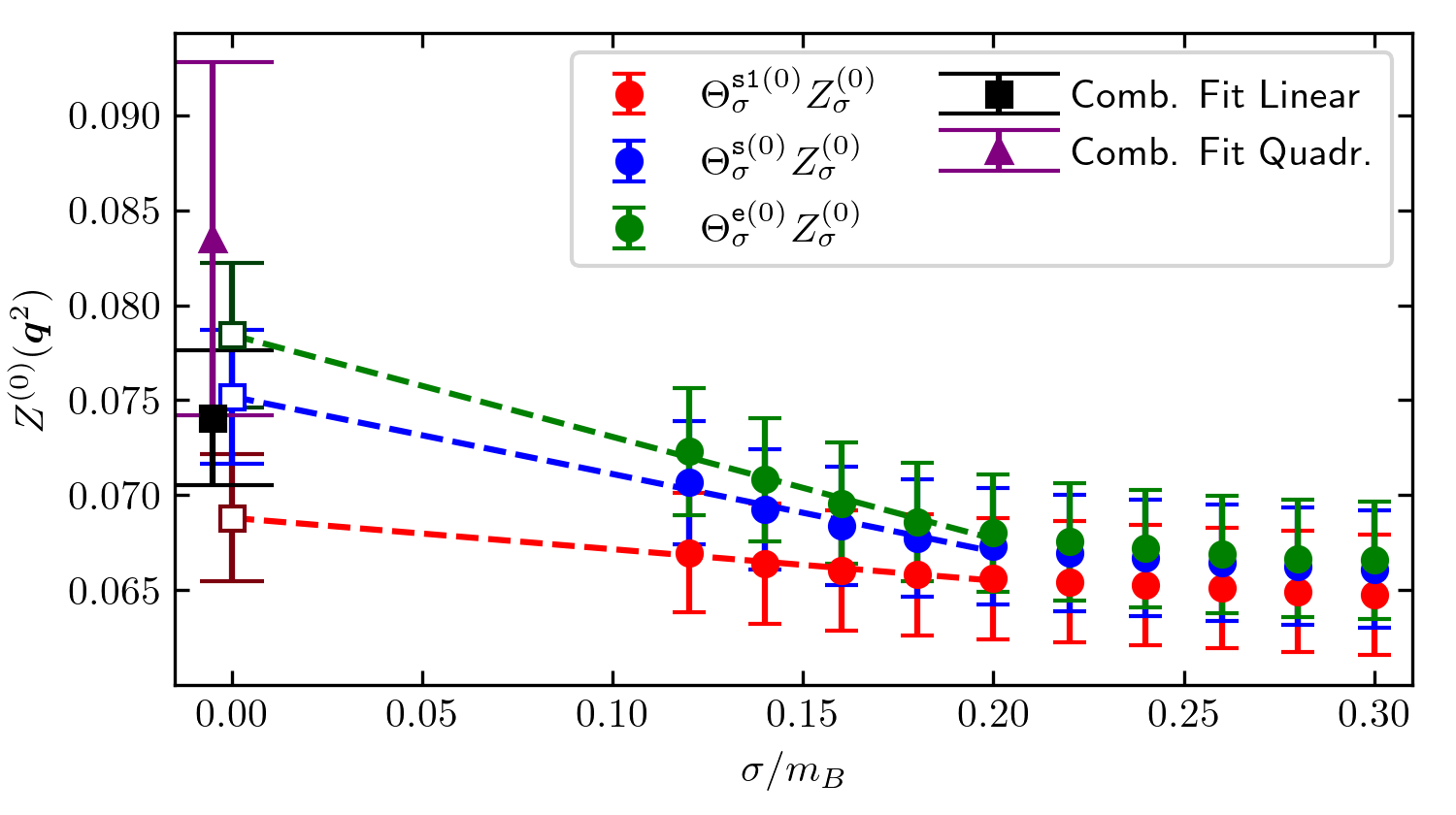}
\caption{Systematics associated with the $\sigma\to 0$ extrapolation of $Z^{(0)}(\bm{q}^2)$ at $\vert \bm{q} \vert\simeq 0.5$~GeV, the same set of data shown in the top--left panel of fig.~\ref{fig:sigmato0}. The unconstrained linear extrapolations of the different sets of data, corresponding to the three different smearing kernels, are shown together with the results of the combined linear extrapolation of the five points at the smaller values of $\sigma$ (black point) and of the combined quadratic extrapolation including all ten values of $\sigma$ (violet point). The black and violet points have been slightly displaced on the horizontal axis to help the eye.}
\label{fig:sigmato0sys}
\end{figure} 

The systematics associated with the $\sigma\to 0$ extrapolations has been quantified (see also the caption of fig.~\ref{fig:decay_rate}) by performing unconstrained linear extrapolations of the five points at the smaller values of $\sigma$ and combined quadratic extrapolations of all points, i.e. with ten values of $\sigma$. This procedure is illustrated in fig.~\ref{fig:sigmato0sys} where we show, for the same set of data appearing in the top--left panel of fig.~\ref{fig:sigmato0}, the unconstrained linear extrapolations and the result of the combined quadratic extrapolation (violet point). As can be seen in this plot, the results of the three different unconstrained extrapolations are compatible within the quoted errors and also compatible with our central value result (black point). Following the procedure explained in the caption of fig.~\ref{fig:decay_rate}, i.e.\  estimating the systematics associated with the extrapolation by adding in quadrature the statistical error of the black point and the difference between the central values of the black and violet points, largely takes into account the spread of the results coming from the different extrapolations, including the unconstrained ones. The same procedure has been repeated for all the sets of data analyzed in this work and similar plots can be shown in all cases.

\begin{figure}[tbp]
\includegraphics[width=0.5\textwidth]{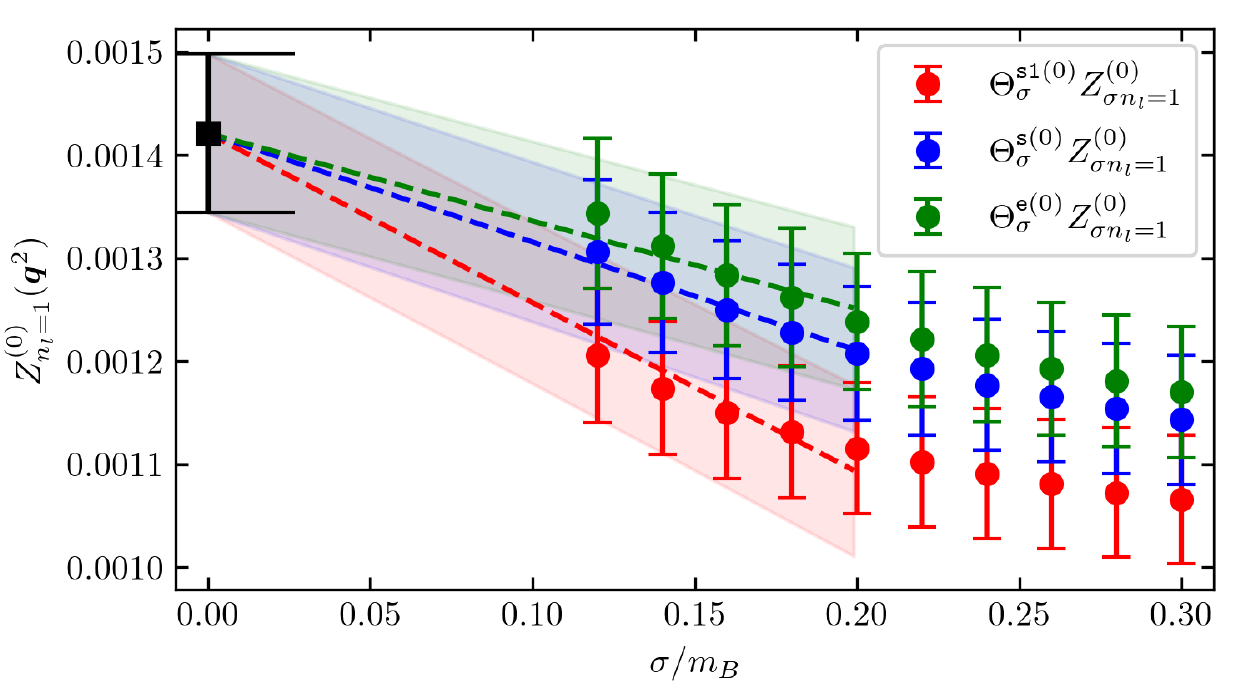}
\includegraphics[width=0.5\textwidth]{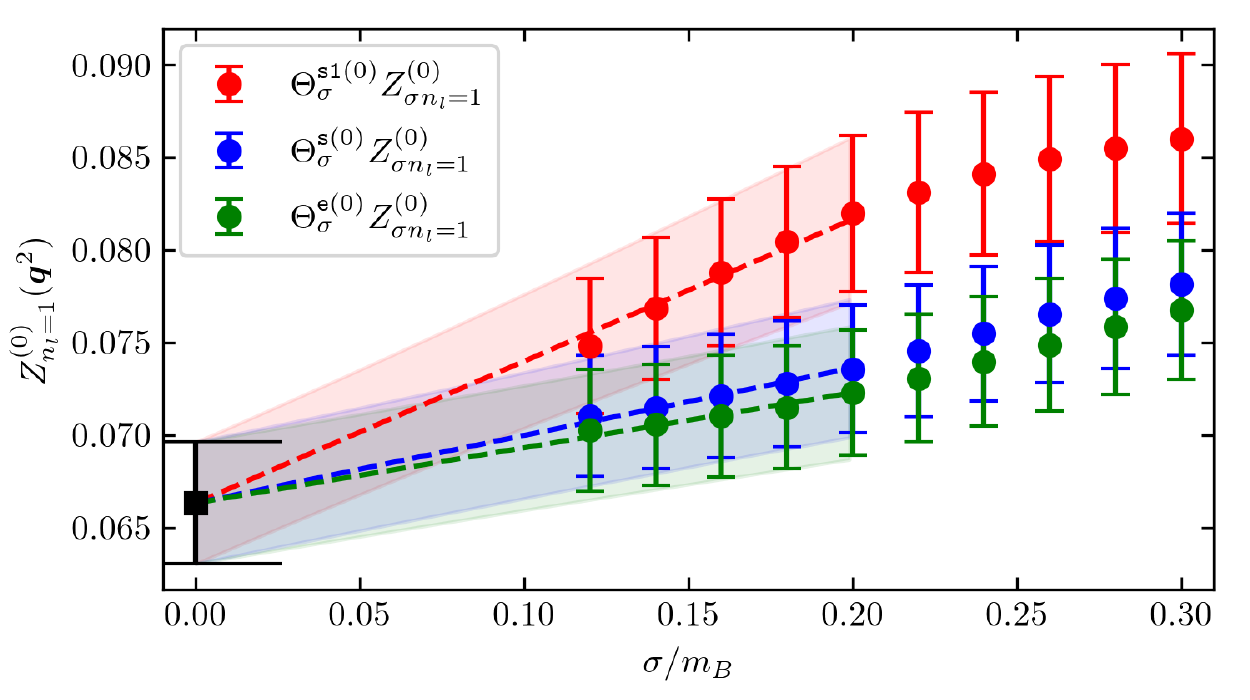}\\
\includegraphics[width=0.5\textwidth]{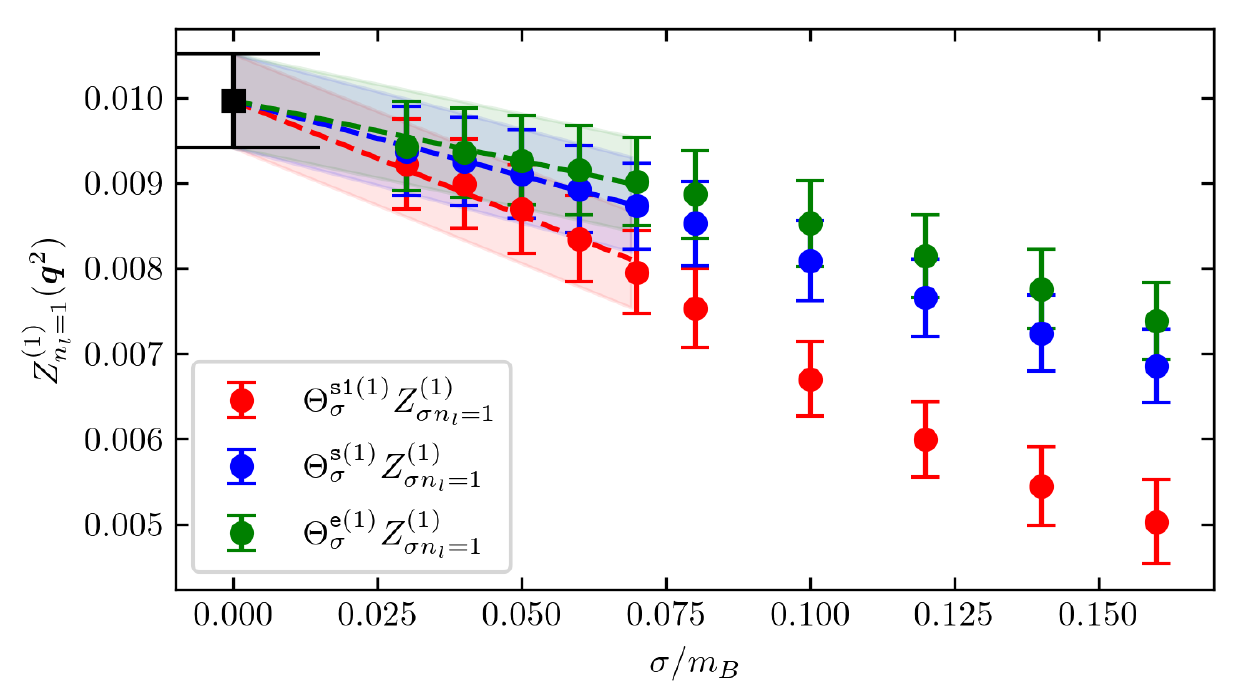}
\includegraphics[width=0.5\textwidth]{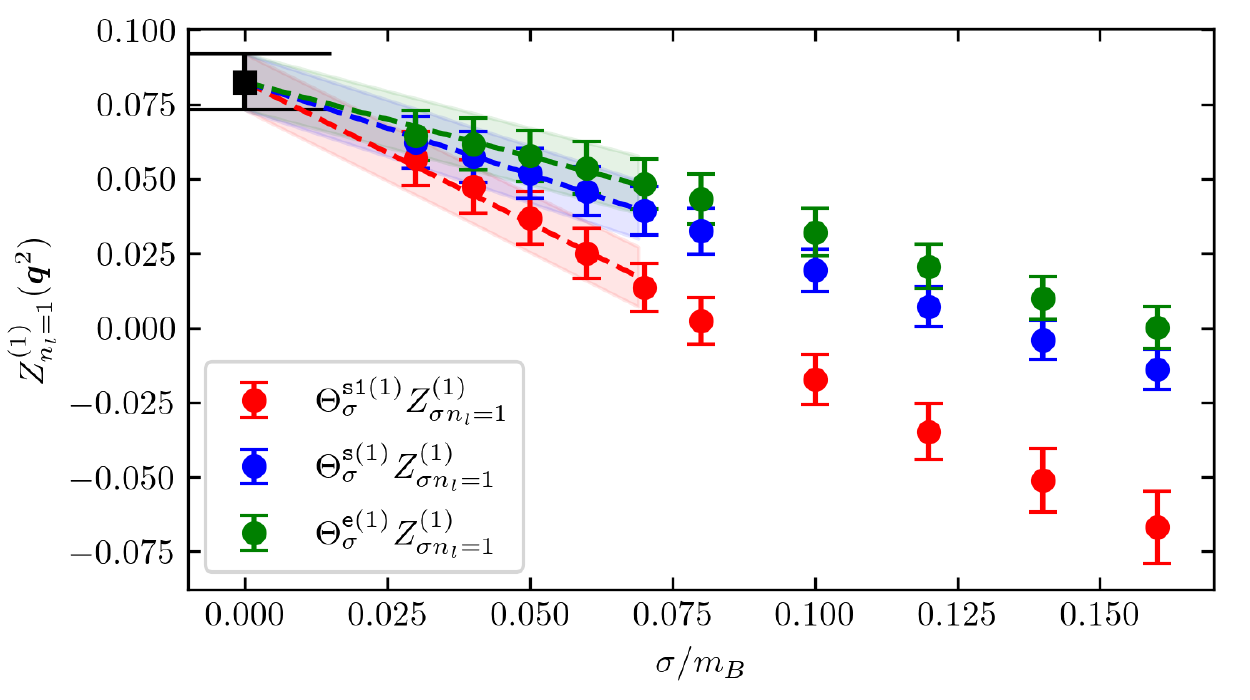}\\
\includegraphics[width=0.5\textwidth]{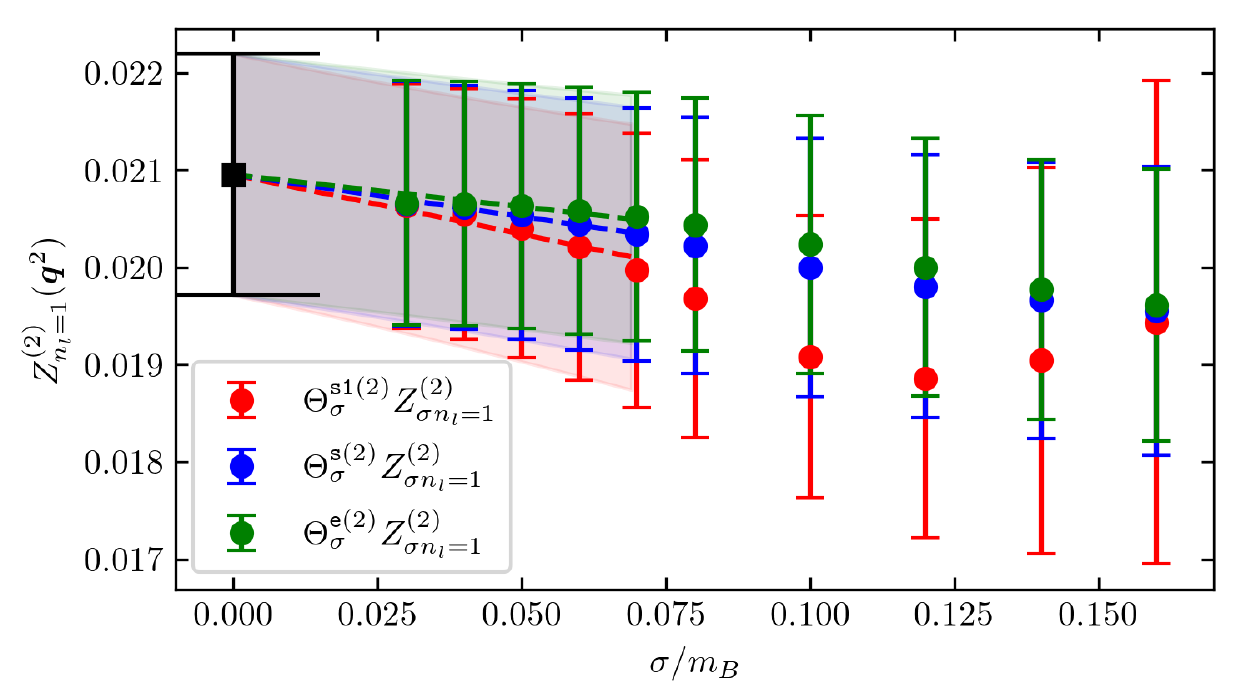}
\includegraphics[width=0.5\textwidth]{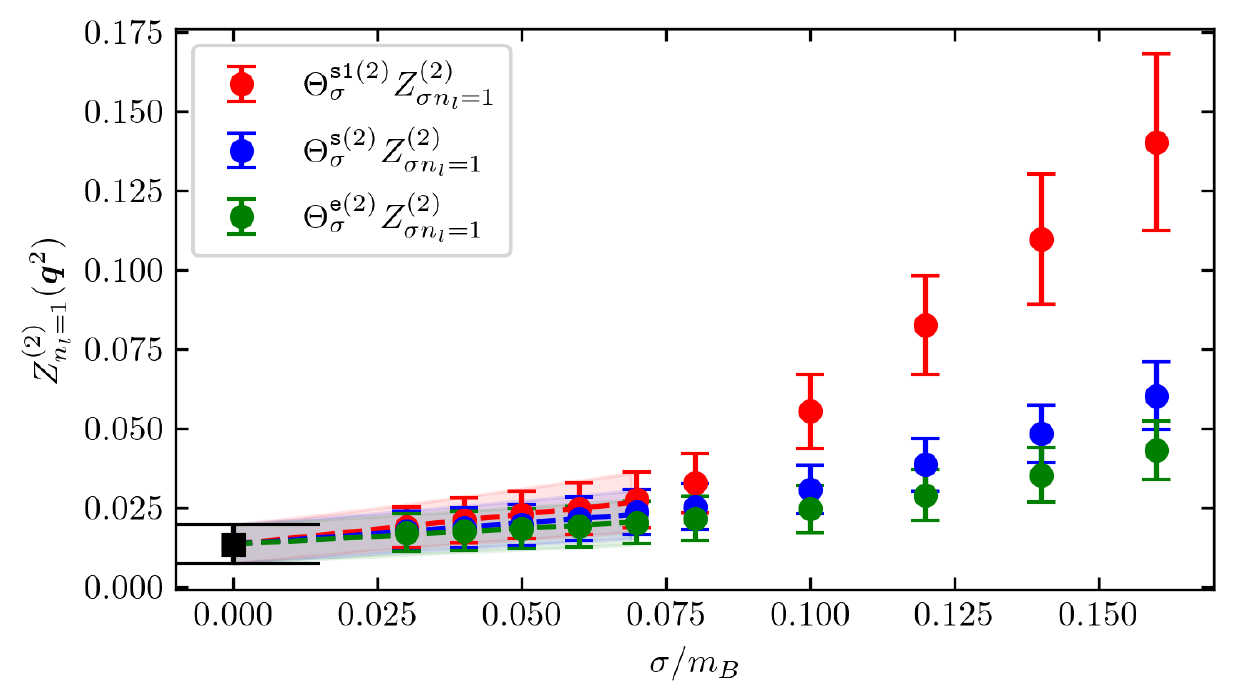}\\
\includegraphics[width=0.5\textwidth]{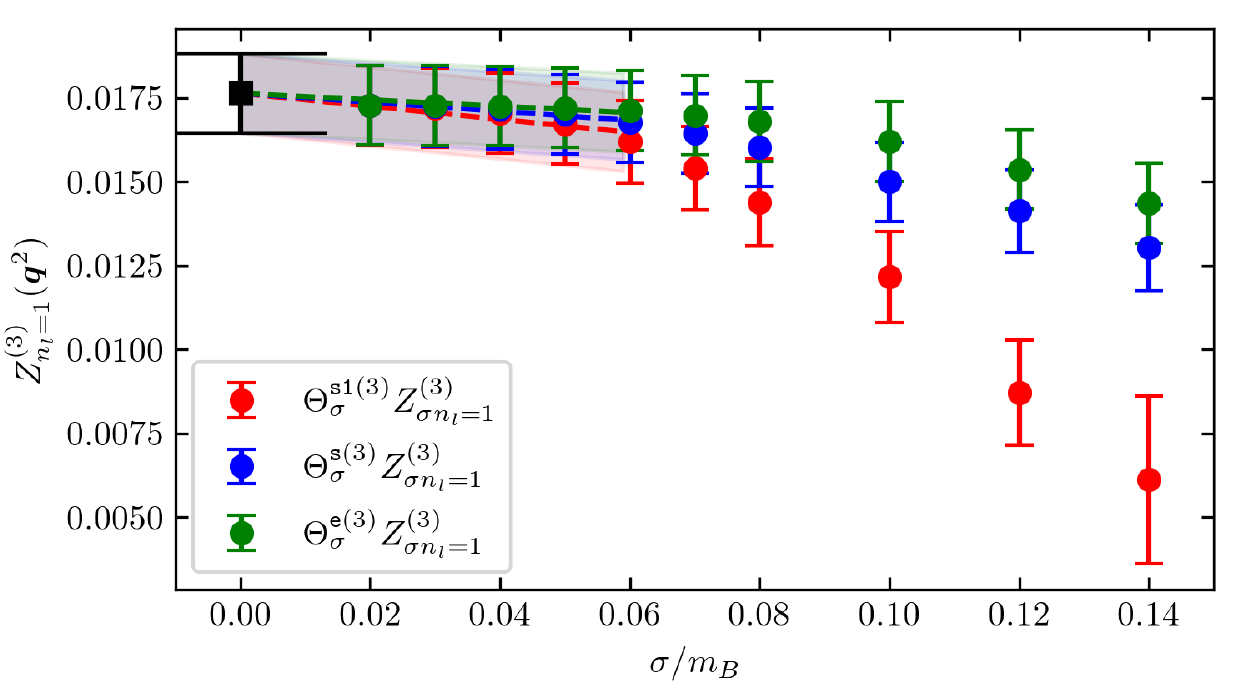}
\includegraphics[width=0.5\textwidth]{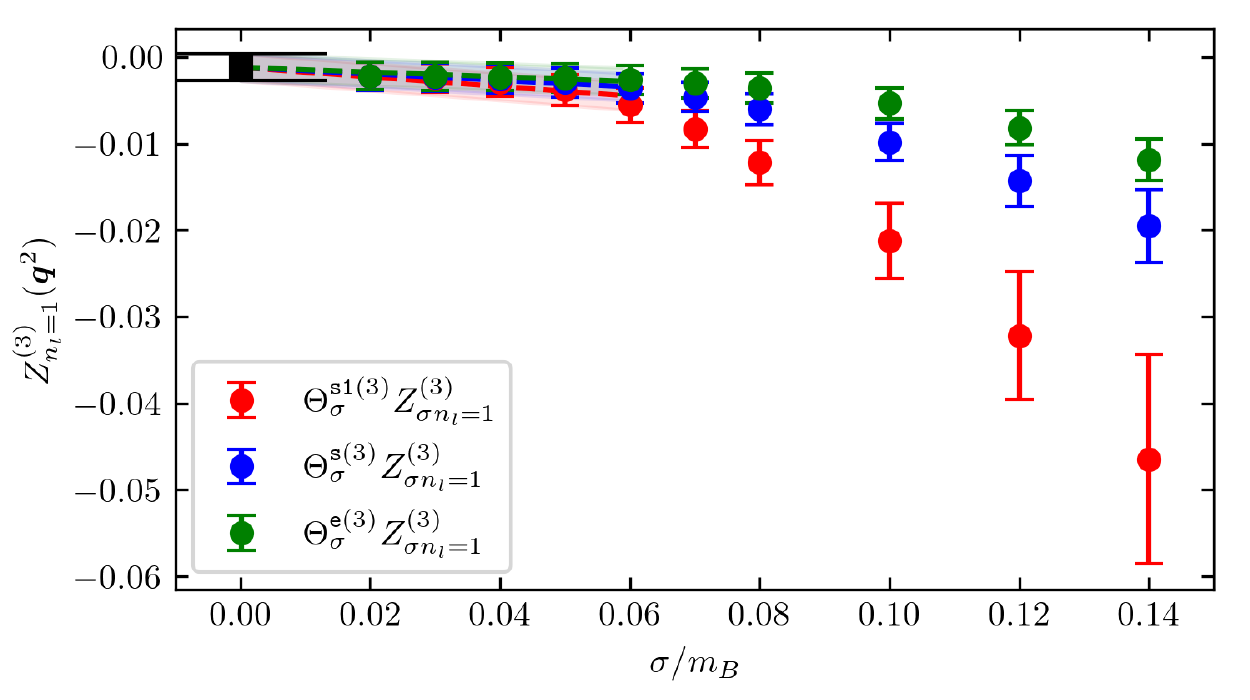}
\caption{Combined $\sigma\to 0$ extrapolations of the four contributions $Z^{(l)}_{n_\ell=1}(\bm{q}^2)$ to the first leptonic moment, see eq.~(\ref{eq:X1int2}). The plots on the left correspond to $\vert \bm{q} \vert\simeq 0.26$~GeV while those on the right to $\vert \bm{q} \vert\simeq 0.78$~GeV.}
\label{fig:E1sigmato0}
\end{figure} 

In fig.~\ref{fig:E1sigmato0} we show the $\sigma\to 0$ extrapolations of the four different terms that enter the calculation of the leptonic moment $L_1(\bm{q}^2)$.



\section{Operator-product expansion and comparison with lattice results}
\label{sec:OPE_and_comparison_with_lattice}
\newcommand{\mupi}{\mu_\pi^2}
\newcommand{\mug}{\mu_G^2}
\newcommand{\rd}{\rho_D^3}
\newcommand{\rls}{\rho_{LS}^3}
\def \be{\begin{equation}}
\def \ee{\end{equation}}


As inclusive semileptonic $B$ decays are described by an OPE, observables which are sufficiently inclusive admit a double expansion in $\alpha_s$ and in inverse powers of $m_b$~\cite{Chay:1990da,Bigi:1992su,Bigi:1993fe,Blok:1993va,Manohar:1993qn}, or more precisely of the energy release, which is of the order of $m_b-m_c$.  Schematically, for an observable $M$ we have  
\begin{flalign}
M=& M^{(0)} + M^{(1)} a_s  +M^{(2)} a_s^2  + \Big(M_\pi^{(0)}+M_\pi^{(1)} a_s\Big) \frac{\mupi}{m_b^2}\nonumber\\
&+ \Big(M_G^{(0)}+ M_G^{(1)} a_s\Big) \frac{\mug}{m_b^2}+
 M_D^{(0)} \frac{\rd}{m_b^3}+ M_{LS}^{(0)} \frac{\rls}{m_b^3} + \dots 
 \label{genericOPE}
 \end{flalign}
 where $a_s=\alpha_s(\mu)/\pi$ is the QCD coupling evaluated at a  scale $\mu\sim m_b$
 and the ellipsis represents higher-order terms in $a_s$ and in $1/m_b$.  The parameters $\mupi$, $\mug$, $\rd$, $\rls$ are expectation values of  dimension-5 and dimension-6 local operators in the physical $B$ meson. For instance, 
\be
  \mupi (\mu_k)  = \frac1{2M_B}\langle B |  \bar b_v\,\vec \pi^2 \,b_v|B
  \rangle_{\mu_k},  \qquad
  \mug(\mu_k) = \frac1{2M_B}\langle B| \bar b_v\frac{i}2 \sigma_{\mu\nu}
  G^{\mu\nu} b_v|B\rangle_{\mu_k}
\ee
where $\vec \pi= -i \vec D$, while $D^\mu$ is the covariant derivative, 
$b_v(x)=e^{-i m_b v\cdot x}b(x)$ is the $b$ field deprived of its high-frequency modes, and
$G^{\mu\nu}$ is the gluon-field tensor. In the so-called kinetic scheme~\cite{Bigi:1996si,Czarnecki:1997sz,Fael:2020njb}, the Wilsonian cutoff $\mu_k\sim $ 1~GeV is introduced to factorise long- and short-distance contributions.
Indeed, the OPE disentangles the physics associated with {\it soft} scales of order
$\Lambda_{\rm QCD}$ (described by the above parameters) from that associated with {\it hard} scales $\sim m_b$, which determine the Wilson coefficients $M_i$ that admit an  expansion in $\alpha_s$. Quite importantly, the power corrections start at $O(\Lambda^2_{\rm QCD}/m_b^2)$ and are therefore comparatively suppressed. The kinetic scheme provides a short-distance, renormalon-free definition of $m_b$ and of the
OPE parameters by introducing the cutoff $\mu_k$ to factor out the infrared contributions from the perturbative calculation.

The smearing provided by the phase-space integration, discussed in section~\ref{sec:formulation_of_the_method_and_application_to_observables},   is in general sufficient to guarantee the convergence of the OPE for the quantities introduced in eqs.~(\ref{eq:omega_integ}) and (\ref{eq:MX2})--(\ref{eq:Eldiff}), which can then be expressed in the form~(\ref{genericOPE}). The OPE calculation proceeds therefore as in refs.~\cite{Gambino:2004qm,Blok:1993va,Manohar:1993qn}.   There are however two specific points related to the kinematics chosen in the lattice calculation that need to be mentioned.  First,  while the hard scale that governs the OPE is generally $m_b-m_c$, there are regions of the phase space, e.g. at small recoil $|\bm q|\sim 0$, where it is rather $m_c$, possibly implying  a slower convergence of the expansion. Second, near the maximum value of $\bm q^2$ the smearing interval in $\omega$ closes up and one cannot expect the OPE to provide reliable results.

\subsection{Details of the OPE calculation and related uncertainties}
\label{subsec:details_of_OPE_and_uncertainties}
From a technical point of view, the OPE provides  a double expansion like the one in eq.~(\ref{genericOPE}) for the hadronic tensor $W^{\mu\nu}$  defined in eq.~(\ref{eq:Wmunu})  that can be used to compute the total rate, the moments, and any sufficiently inclusive quantity. The coefficients of the expansion involve the Dirac delta $\delta(r^2-m_c^2)$ and its derivatives, which upon integration over the {\it quark} (partonic) phase space lead to results valid for sufficiently inclusive observables. It is customary to use the decomposition of $W^{\mu\nu}$ into Lorentz-invariant form factors as in eq.~(\ref{eq:Wdecomp})
and to identify the four-velocities of the $B$ meson and of the $b$ quark, $v=p/m_{B}= p_b/m_b$. In this section we will use eq.~(\ref{eq:Wdecomp}) replacing 
$m_B$ with the $b$ quark mass $m_b$ and employing 
a hat for quantities that are normalised to $m_b$. 
In the case of massless leptons considered in this work, the form factors $W_{4,5}$ do not contribute to the decay amplitude.

The lowest order of the expansion for the relevant $W_i$ and the $1/m_b^2$ corrections
can be found in refs.~\cite{Blok:1993va, Manohar:1993qn}, while analytic expressions for the  $O(\alpha_s)$  terms are given in refs.~\cite{Aquila:2005hq,Alberti:2012dn}.  The $O(1/m_b^3)$ corrections have been first computed in ref.~\cite{Gremm:1996df}. Higher power corrections have been investigated in ref.~\cite{Mannel:2010wj}, but involve a large number of new and poorly known parameters. They appear to be sufficiently suppressed at the physical $m_b$~\cite{Gambino:2016jkc}; we will not consider them  but they represent an important source of theoretical uncertainty in our low $m_b$ setup. 
The $O(\alpha_s/m_b^2)$ corrections to the $W_i$ are also known~\cite{Alberti:2012dn,Alberti:2013kxa}, while for the total rate we also have $O(\alpha_s/m_b^3)$ corrections~\cite{Mannel:2015jka, Mannel:2019qel}. 
Numerical results for the  $O(\alpha_s^2 \beta_0)$ contributions are also available~\cite{Aquila:2005hq}, while the complete  $O(\alpha_s^2)$ are available only for the total rate and for a few moments~\cite{Pak:2008cp,Pak:2008qt,Melnikov:2008qs,Biswas:2009rb}. 
Finally, the $O(\alpha_s^3)$ correction to the total rate has been recently computed in ref.~\cite{Fael:2020tow}.

While these corrections have generally been computed in the $V-A$ case realised in the SM,
the decomposition in $VV, AA$ and $AV=VA$ components is potentially useful in our case, and has been made manifest for the $O(1/m_b^{2,3})$ and $O(\alpha_s)$ corrections, see refs.~\cite{Blok:1993va,Alberti:2015qmj,Colangelo:2020vhu}. 
In the calculation of the $\bm{q}^2$ spectrum and of the differential moments we will therefore
consider only power corrections up to and including $O(1/m_b^3)$ and the $O(\alpha_s)$ perturbative corrections.
However in the calculation of the total width and of the total moments we will restrict to the SM case and will employ all the known corrections.

Following section~\ref{sec:formulation_of_the_method_and_application_to_observables}, we take the three-momentum $\bm{q} $ to point along the $k$ direction and the $i$ and $j$ directions to be perpendicular to that. The components of the hadronic tensor along these directions are given by
	\begin{flalign*}
		W^{00} &= -W_1 +W_2 + \hat q_0^2 W_4 +2 \hat q_0 W_5 \;, \\
		W^{ii}\, &= W^{jj}= W_1 \;, \\
		W^{kk} &= W_1 + \hat q_k^2 W_4 \;, \\
		W^{0i} &= W^{i0} =  W^{ik} = W^{ki} =  W^{jk} = W^{kj} = 0 \;,\\
		W^{0k} &= W^{k0} = \hat q_0 \hat q_k W_4 +\hat q_k W_5 \;, \\
		W^{ij} &= -W^{ji} = -i\epsilon^{ij0k}\hat q_k W_3 \;.
	\end{flalign*}
	
In the OPE the decay occurs at the quark level: $p_b=p'+ p_\ell+p_\nu$, where $p_b$ and $p'$ 
are the momenta of the initial $b$ quark and of a final hadronic state made of a $c$ quark and $n\ge 0$  perturbative gluons.
At the leading order in $\alpha_s$ and in $1/m_b$, this is a free-quark decay into an on-shell $c$ quark, which implies that
the $W_i$ are proportional to  $\delta(p'^2-m_c^2)= \delta(\hat u)/m_b^2$, where $\hat u = (p'^2-m_c^2)/m_b^2$.
We can rewrite this $\delta$ function   
in terms of the energy of the final $c$ quark,
	\be
		\delta(\hat u) 
		= \frac{1}{2\sqrt{\bm{\hat q}^2+\rho}}\left[
		\delta\left(\hat \chi - \sqrt{\bm{\hat q}^2+\rho} \right)+\delta\left(\hat \chi + \sqrt{\bm{\hat q}^2+\rho} \right) \right]\;,\label{delta}
	\ee
where $\rho=m_c^2/m_b^2$ and $\hat \chi$ is the parton-level
energy of the final hadronic state in units of $m_b$, which is related to the total hadronic energy 
$\omega$ by 
$\omega= m_b \hat\chi + \Lambda$, 
with $\Lambda=M_B-m_b$.
Similarly, the invariant hadronic mass $M_X^2$ is related to the partonic variables by
\[M_X^2=(p_B-q)^2 = (p_b+ \Lambda v-q)^2=  m_b^2 \hat u +2 m_b \Lambda \hat \chi + \Lambda^2 + m_c^2\, .\]
Only the first term of eq.~(\ref{delta}) contributes  to the physical process of interest and can be readily integrated over $\hat \chi$. At $O(1/m_b^{2,3})$ one has to deal with $\delta'(\hat u)$, $\delta''(\hat u)$ and $\delta'''(\hat u)$ that upon integration subject to kinematic constraints lead to new singularities.
A typical case is provided by the interplay between the $\delta'$ and the requirement that $q^2\ge0$: 
\begin{eqnarray}
\int f(\hat u)\,  \theta(q^2)\, \delta'(\hat u)\, d \hat u&=& \int f(\hat u) \, \theta(1+\rho+\hat u -2 \sqrt{\rho+ \bm{\hat q}^2+\hat u}) \, \delta'(\hat u)\, d \hat u\nonumber\\
&=& - f'(0) \theta(1+\rho -2 \sqrt{\rho+ \bm{\hat q}^2}) + f(0) \frac{1+\rho}2 \delta(\bm{\hat q}^2-\bm{\hat q}^2_{\mathrm{max}})\;.
\end{eqnarray} 
The singularity at the partonic endpoint  of the $\bm{q}^2$ spectrum,  $\bm{\hat q}^2_{\mathrm{max}}=(1-\rho)^2/4$, appears because one reaches the maximum energy exactly on the mass-shell of the charm quark.

We apply exactly the same setup to compare with both JLQCD and ETMC data, 
adjusting only the heavy-quark masses to the two cases.  The unphysically light 
$b$ quark mass and the OPE parameters are expressed in the kinetic scheme with $\mu_k$ = 1~GeV, while the $c$ quark mass is expressed in the $\overline{\rm MS}$ scheme at 2~GeV.
In the case of the JLQCD data we employ $m_b(\mathrm{1~GeV})$= 2.70(4)~GeV,  obtained from matching the observed 
$m_{B_s}$ with the results of \cite{Gambino:2017vkx,Gambino:2019vuo}, and $\overline{m_c}^{(4)}(\mathrm{2~GeV})$ = 1.10(2)~GeV.
In the case of the ETMC data we employ $\overline{m_c}^{(4)}(\mathrm{2~GeV})$ = 1.186(41)~GeV and 
$\overline{m_b}^{(4)}(\mathrm{2~GeV})$ = 2.372(82)~GeV (with 100\% correlated uncertainties),
and   translate the latter
  into the kinetic scheme using the three-loop conversion formula \cite{Fael:2020njb} 
implemented in version 3.1 of {\it RunDec} \cite{Herren:2017osy} obtaining $m_b(\mathrm{1~GeV})$ = 2.39(8)~GeV. The strong coupling employed in the conversion and elsewhere is $\alpha_s^{(4)}(\mathrm{2~GeV})$ = 0.301.

For the OPE parameters that appear in eq.~(\ref{genericOPE}) we start from the results of the most recent fit to the semileptonic moments \cite{Bordone:2021oof}, which refer  to the physical $B$ meson, with a much heavier $b$ quark and without a strange 
spectator. The difference induced in these parameters by the strange spectator at the physical point has been 
investigated in \cite{Gambino:2019vuo,Bigi:2011gf,Bordone:2022qez}, where it was found that  spectroscopic and lattice data  approximately suggest a $20\%$ upward shift in $\mu_\pi^2$ and $\mu_G^2$, while heavy-quark sum rules hint at a similar or even stronger $\SU(3)$ flavour-symmetry breaking in $\rho_D^3$.
The dependence on the mass of the heavy quark, on the other hand, can be analysed 
by observing that $\mu_\pi^2$ and $\mu_G^2$ satisfy a heavy-quark expansion
\be
\mu_\pi^2= \mu_\pi^2|_\infty -\frac{\rho^3_{\pi\pi}+\frac12 \rho^3_{\pi G}}{m_b} +\dots,\qquad
\mu_G^2= \mu_G^2|_\infty +\frac{\rho^3_{S}+\rho^3_{A}+\frac12 \rho^3_{\pi G}}{m_b} +\dots
\label{nonloc}
\ee
where $\rho^3_{\pi\pi}$, $\rho^3_{\pi G}$, $\rho^3_{S}$, $\rho^3_{A}$ are
expectation values of non-local operators, of which little is known, see
ref.~\cite{Gambino:2017vkx}. If they were of the same order of magnitude of $\rho_D^3$ and $\rho^3_{LS}$, i.e. about 0.1--0.2~GeV$^3$, they could shift $\mu_\pi^2$ and $\mu_G^2$ by 0.02--0.1~GeV in going from the physical value of $m_b$ to $m_b\sim$ 2.5~GeV, which amounts to a $5$--$25\%$ shift.
We show the  inputs of our calculation in table~\ref{tab:OPEinputs}. 
While the heavy-quark masses are slightly different between the two setups,
we adopt the same expectation values in both cases. Their central values take into account the
shift related to the strange spectator, while the uncertainties follow  from the uncertainty of the fit of  ref.~\cite{Bordone:2021oof}, the $\SU(3)$ symmetry breaking, and the lower $b$ mass.

\begin{table}[tbp]
  \begin{center} \begin{tabular}{cc}\hline
  $m_b^{kin}$ ({\small JLQCD}) &  $2.70\pm 0.04$  \\
 $\overline m_c(\mathrm{2~GeV})$ ({\small JLQCD})  & $1.10  \pm 0.02$\\
 $m_b^{kin}$ ({\small ETMC}) &  $2.39\pm 0.08$  \\
 $\overline m_c(\mathrm{2~GeV})$ ({\small ETMC})  & $1.19  \pm 0.04$\\
  $\mupi $ & $0.57 \pm 0.15$\\
  $\rd$ & $0.22\pm 0.06$\\
  $\mug(m_b)$ & $0.37\pm 0.10$\\
  $\rls$  & $-0.13\pm  0.10$\\
  $\alpha_s^{(4)}(\mathrm{2~GeV})$ &  $0.301\pm 0.006$\\
  \hline
    \end{tabular} \end{center}
       \caption{ \label{tab:OPEinputs}
    Inputs for our OPE calculation.
All parameters are in GeV at the appropriate power and all, except $m_c$, in the kinetic scheme at $\mu=1$~GeV. The heavy-quark masses for the ETMC setup are 100\% correlated. As a remnant of the semileptonic fit, we include a 50\% correlation between $\mupi$ and $\rd$.}
\end{table}

Beside the parametric uncertainty of the inputs, our results are subject to an uncertainty due  
the truncation of the expansion in eq.~(\ref{genericOPE}) and to possible violations of quark-hadron duality.
We estimate the former 
by varying the OPE parameters, the heavy-quark masses, and $\alpha_s$ in an uncorrelated way and adding the relative uncertainties in quadrature. In particular, we 
shift  $m_{b,c}$ by $6$~MeV, $\mu^2_{\pi,G}$ by $15\%$, and $\rho^3_{D, LS}$ by $25\%$.
These corrections should mimic the effect of higher-power corrections. Since in the case of the $\bm{ q}^2$ spectrum and differential  moments we restrict ourselves to $O(\alpha_s)$ corrections, we include the relative uncertainty in the same way, shifting $\alpha_s$ by $0.15$, which corresponds to a $50\%$ uncertainty. In the case of the total width and total moments, higher-order perturbative corrections are known and the perturbative uncertainty can be reduced, as discussed below.

\subsection{Comparison with lattice results}
\subsubsection{$q^2$ spectrum and differential moments}

\begin{figure}[tbp]
  \centering
  \includegraphics[width=12cm]{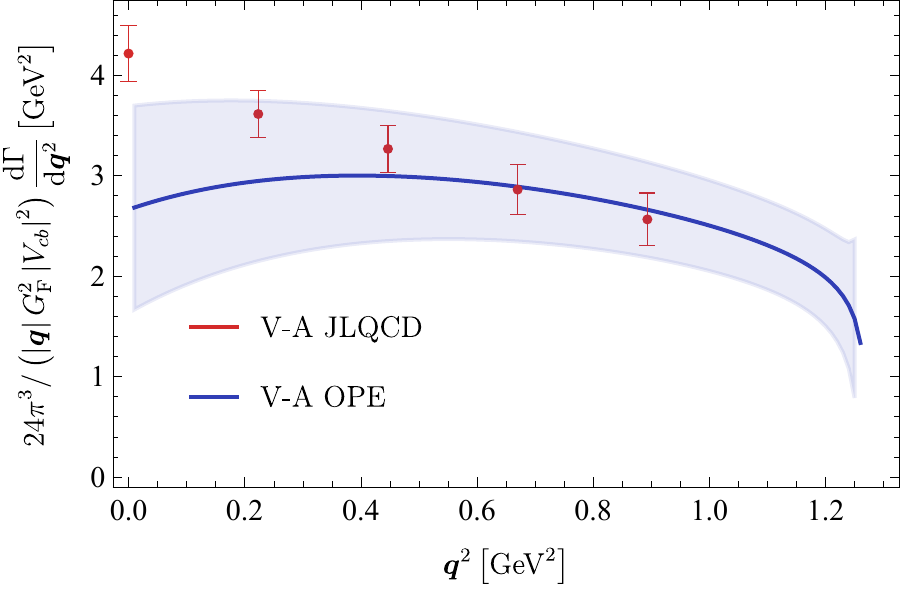}
  \includegraphics[width=12cm]{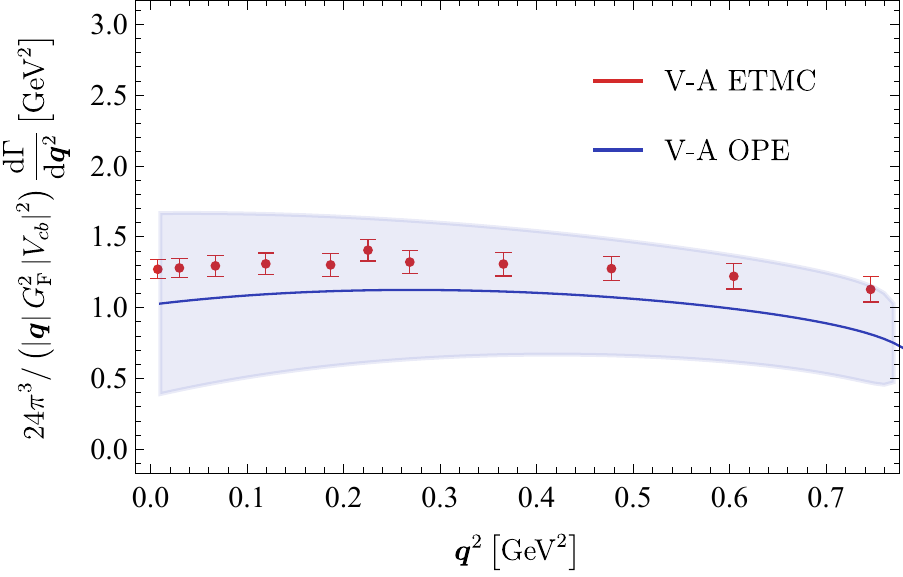}
  \caption{Differential $\bm{q}^2$ spectrum, divided by $|\bm{q}|$, in the SM.
    Comparison of OPE with JLQCD (top panel) and ETMC (bottom panel) data are shown.
  }
  \label{fig:diffqSM}
\end{figure}

We start our comparison of lattice and OPE results with the $\bm{q}^2$ spectrum and the differential  moments introduced in eq.~(\ref{eq:MX2diff}) and in eq.~(\ref{eq:Eldiff}). Only the
$O(\alpha_s)$ perturbative corrections are included in this case. 
Figure~\ref{fig:diffqSM} shows the $\bm{q}^2$ spectrum in the SM, namely with a  $V-A$ current.
Despite the large uncertainty of the OPE prediction, about $30\%$ in the JLQCD case and $50\%$ in the ETMC case, the overall agreement is good. The OPE uncertainty is dominated by the 
power corrections. We also stress that close to the partonic endpoint, corresponding to 1.27~GeV$^2$ and 0.82~GeV$^2$ in the two cases,  we do not expect the OPE calculation to be reliable, as discussed above. 
The corresponding hadronic endpoints are 1.35~GeV$^2$ and 0.75~GeV$^2$, respectively.

\begin{figure}[tbp]
  \centering
  \includegraphics[width=12cm]{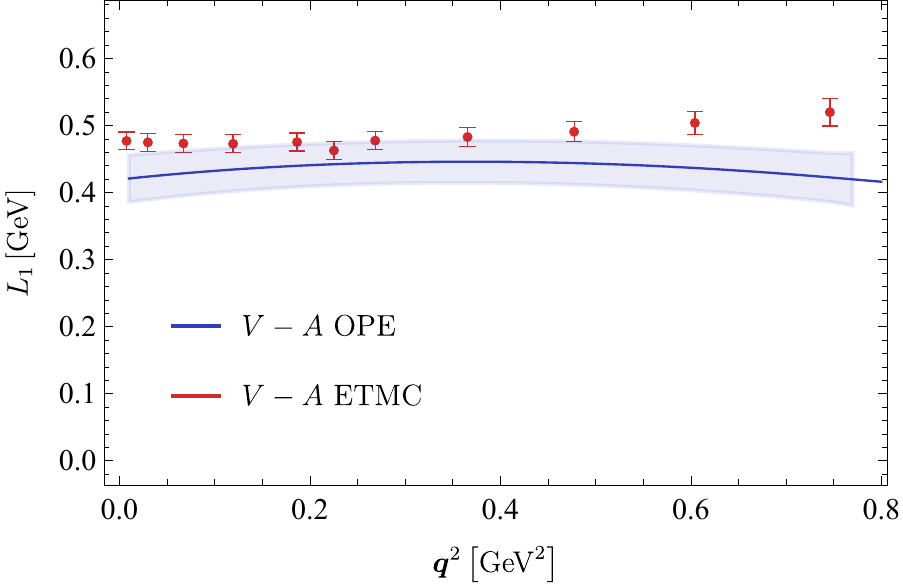}
   \caption{Differential lepton energy mean value, $L_1(\bm{q}^2)$,  in the SM.
    The comparison of OPE with  ETMC  data is shown.
  }
  \label{fig:L1SM}
\end{figure}
The uncertainties affecting both calculations can be greatly reduced by considering the differential moments. In particular, the OPE uncertainty  becomes  smaller because of the cancellations between power corrections to the numerator and to the denominator. To expose the cancellations we expand the ratios in powers of $\alpha_s$ and $1/m_b$.
In figure~\ref{fig:L1SM} we show the first differential lepton energy moment, $L_1(\bm{q}^2)$,  in the SM, comparing the OPE with ETMC data.
 As expected, the relative uncertainty of both the OPE calculation and of the lattice data
is much smaller than in the bottom panel of figure~\ref{fig:diffqSM} and we observe good agreement at low and moderate $\bm{q}^2$. 

\begin{figure}[tbp]
  \centering
\includegraphics[width=7.5cm]{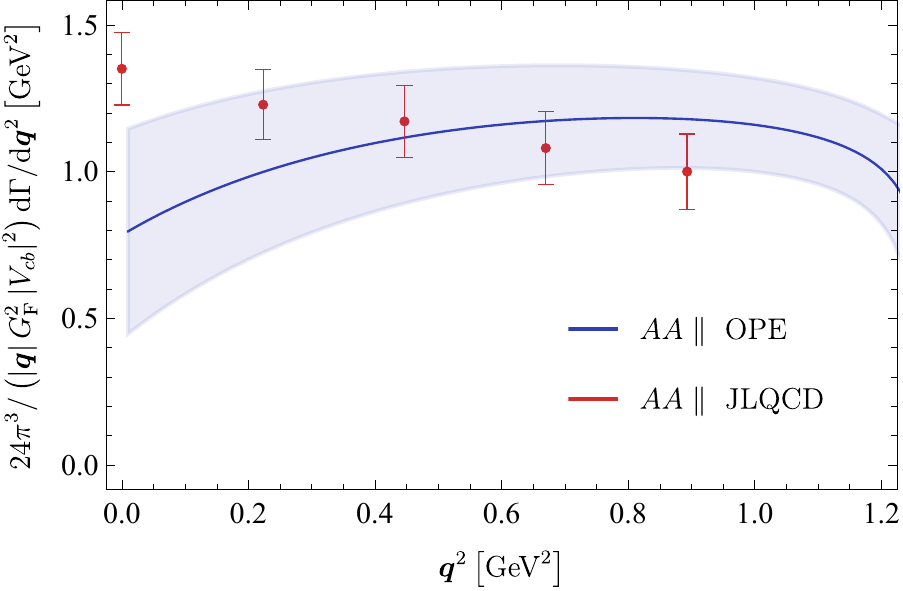}
  \includegraphics[width=7.5cm]{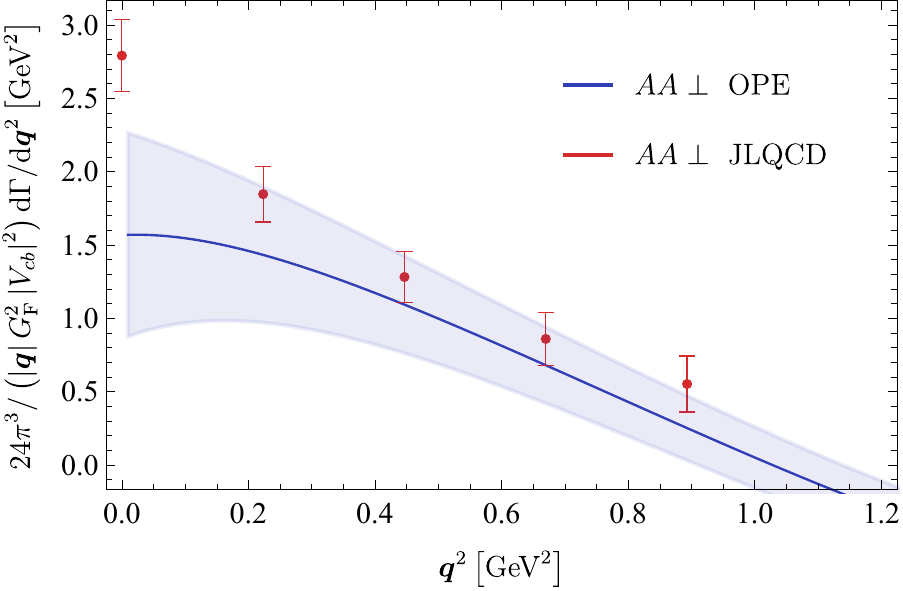}\\
  \includegraphics[width=7.5cm]{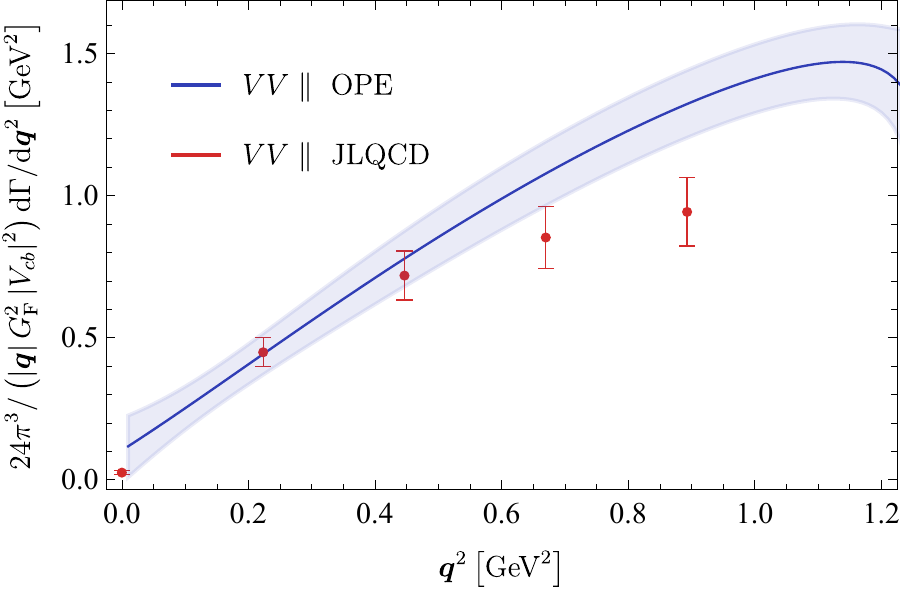}
  \includegraphics[width=7.5cm]{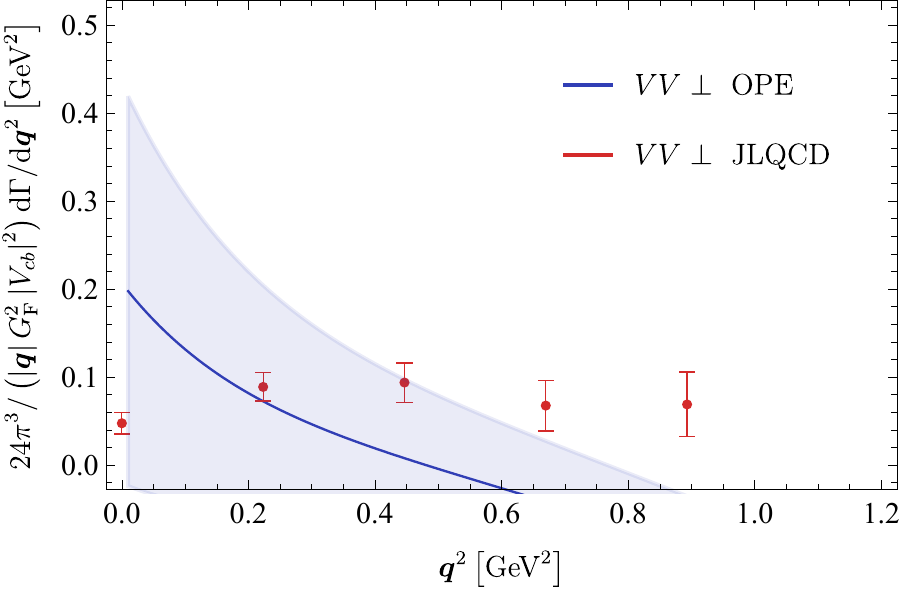}
  \caption{Differential $\bm{q}^2$ spectrum, divided by $|\bm{q}|$, in the various channels. The plots show the comparison  between OPE and JLQCD data.  }
  \label{fig:diffqJLQCD}
\end{figure}

\begin{figure}[tbp]
  \centering
\includegraphics[width=7.5cm]{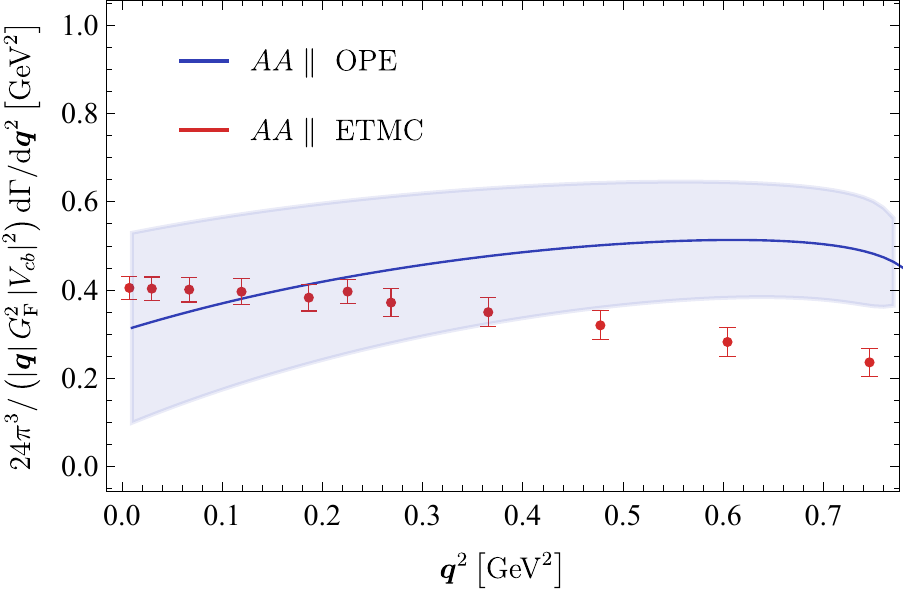}
  \includegraphics[width=7.5cm]{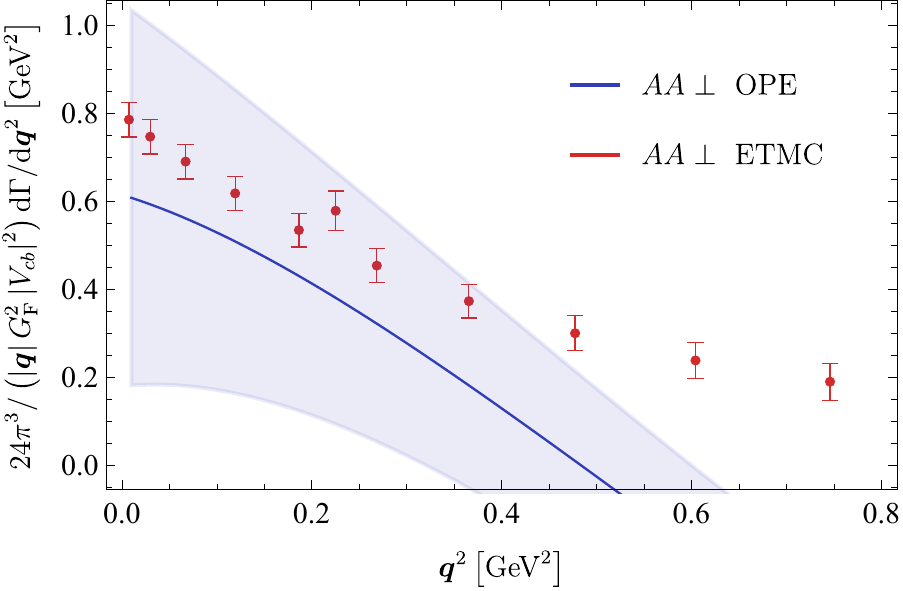}\\
  \includegraphics[width=7.5cm]{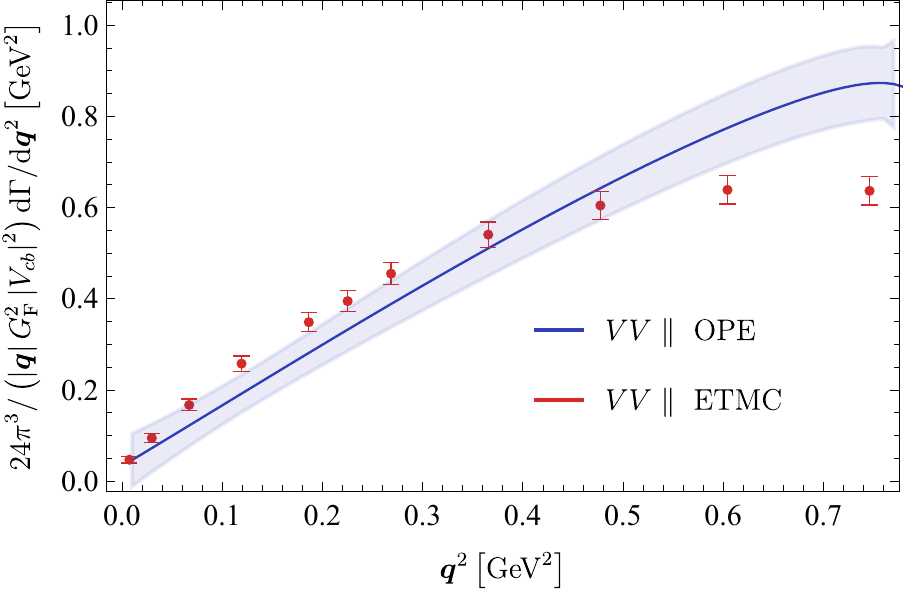}
  \includegraphics[width=7.5cm]{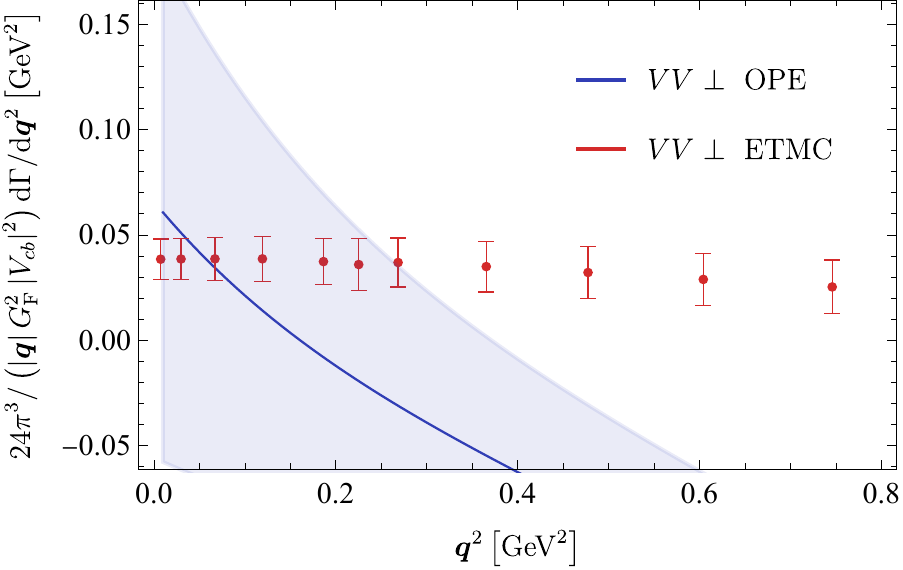}
  \caption{Differential $\bm{q}^2$ spectrum, divided by $|\bm{q}|$, in the various channels. The plots show the comparison  between OPE and ETMC data.  }
  \label{fig:diffqETMC}
\end{figure}

Figs.~\ref{fig:diffqJLQCD} and \ref{fig:diffqETMC} show the $\bm{q}^2$ spectrum in the individual  channels. Comparing them with figure~\ref{fig:diffqSM} we see that in the individual channels the agreement between OPE and lattice results is poorer than in their sum, especially at large $\bm{q}^2$. This is to be expected and (unless discretisation and/or finite-volume effects turn out to have a sizeable impact on the lattice results) is likely to be a manifestation of duality violations. Notice that 
the OPE central predictions for the $AA_\perp$ and $VV_\perp$ channels turn negative at large and moderate $\bm{q}^2$, respectively, and that for $\bm{q}^2>$ 0.6~GeV$^2$ 
the spectrum is always negative  within errors. This unphysical feature suggests that our error estimates 
are not adequate at large $\bm{q}^2$. The contribution to the $VV_\perp$ channel, moreover, is 
particularly small and very sensitive to large power corrections.

\begin{figure}[tbp]
  \centering
\includegraphics[width=7.5cm]{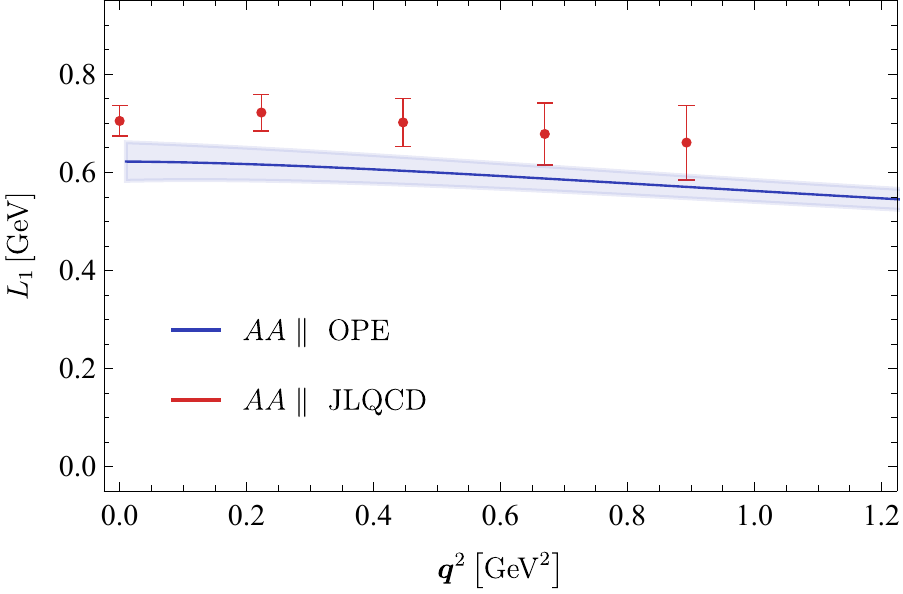}
\includegraphics[width=7.5cm]{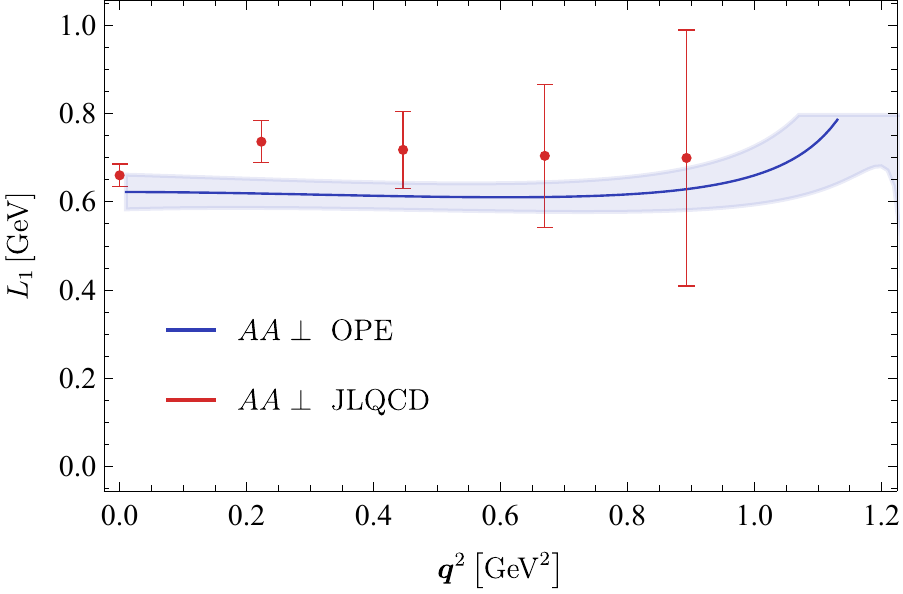}\\
\includegraphics[width=7.5cm]{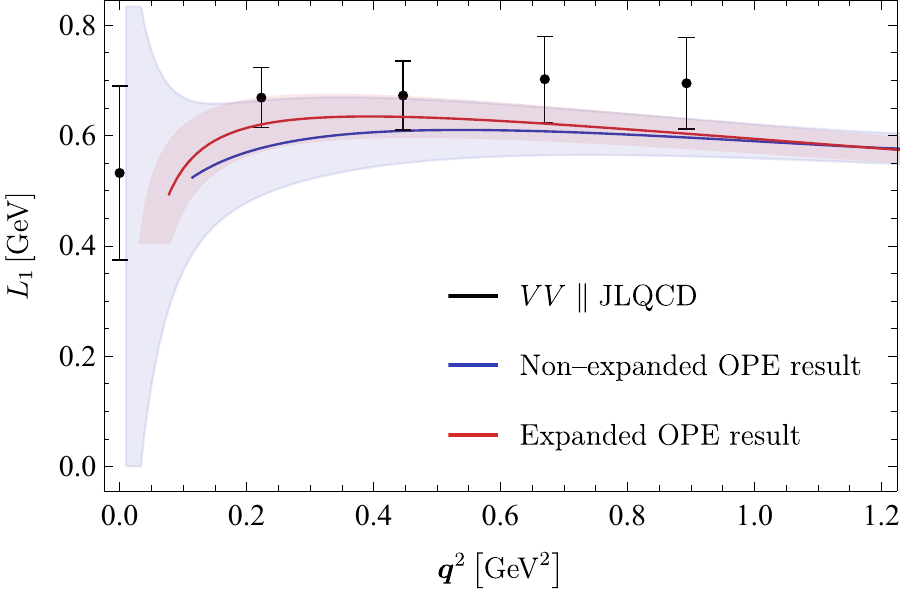}
\includegraphics[width=7.5cm]{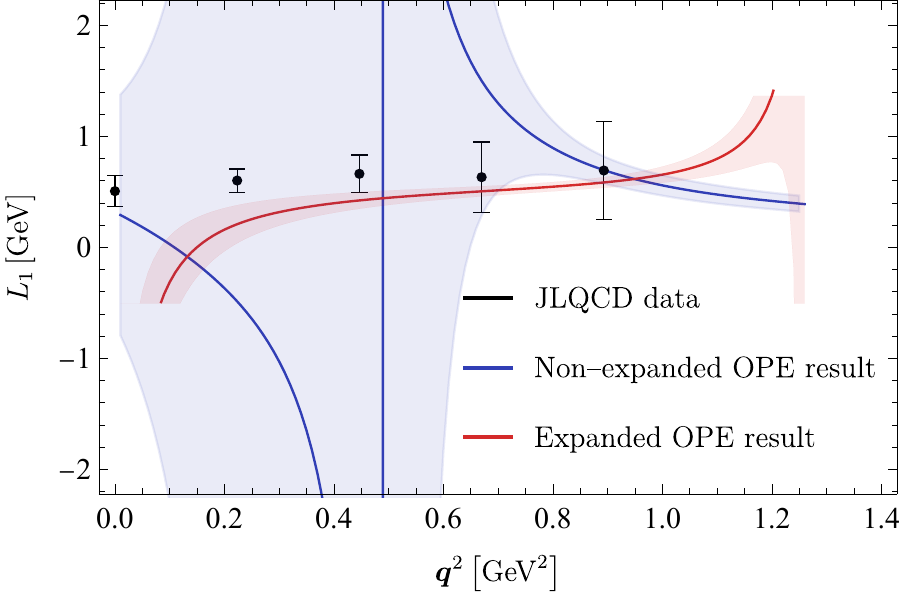}
  \caption{Differential first leptonic moment in the various channels. The plot shows the comparison  between OPE and JLQCD data.  }
  \label{fig:mom1JLQCD}
\end{figure}
\begin{figure}[tbp]
  \centering
\includegraphics[width=7.5cm]{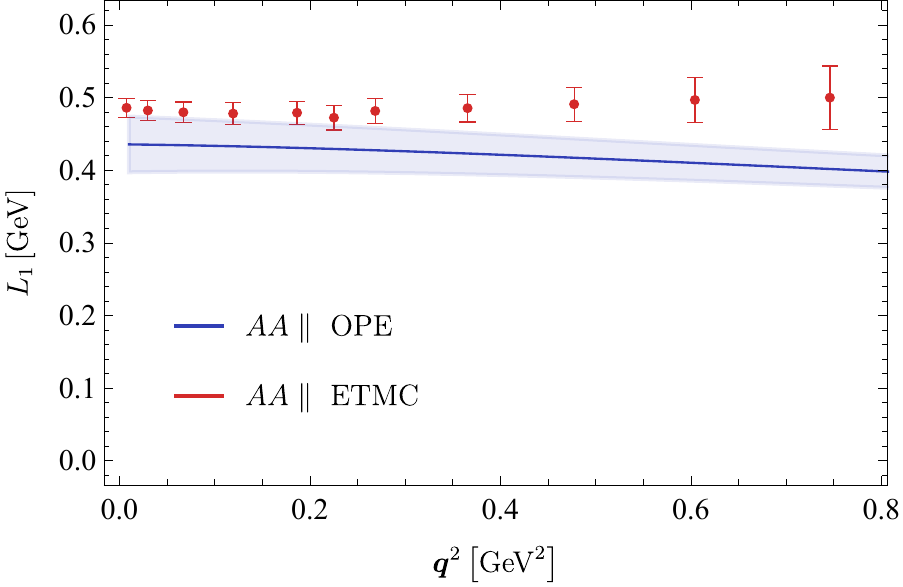}
  \includegraphics[width=7.5cm]{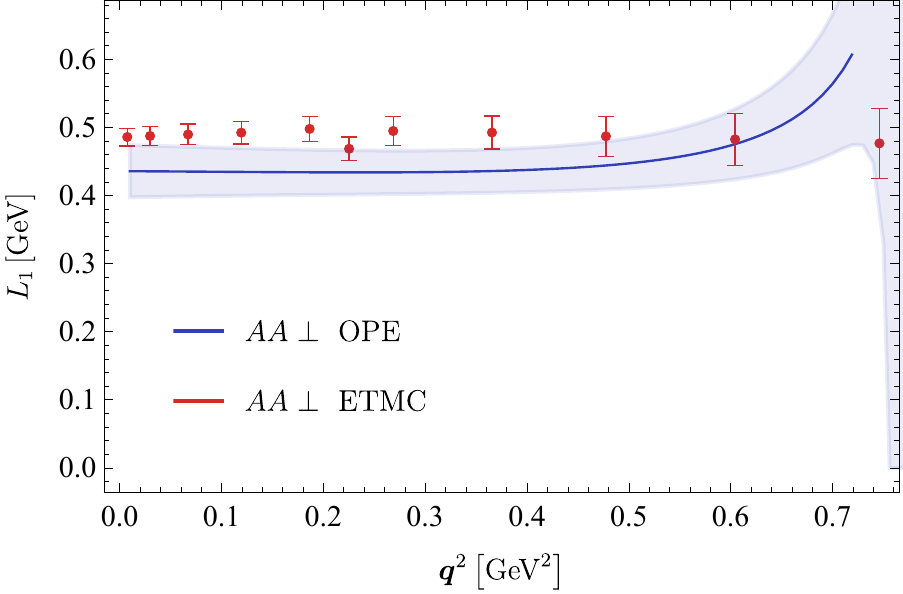}\\
  \includegraphics[width=7.5cm]{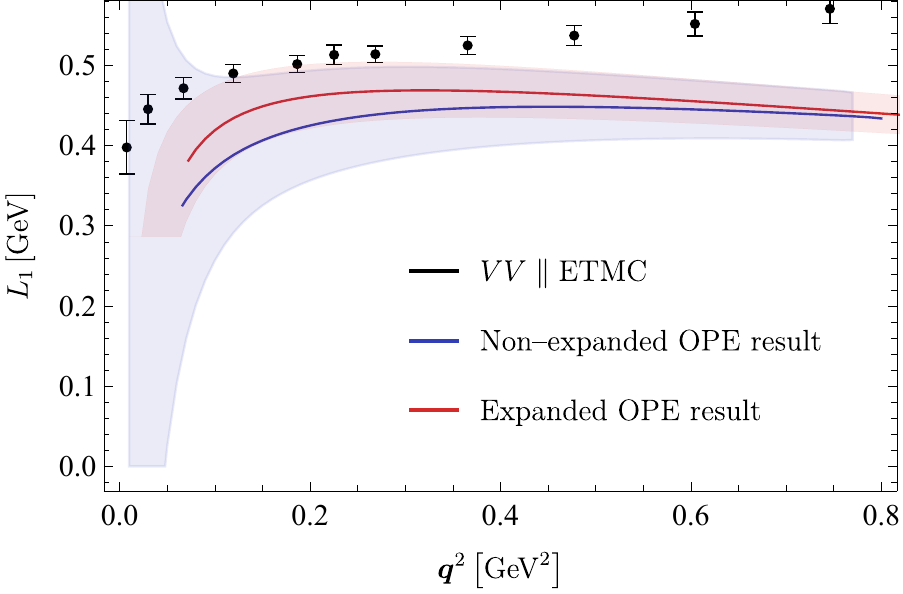}
  \includegraphics[width=7.5cm]{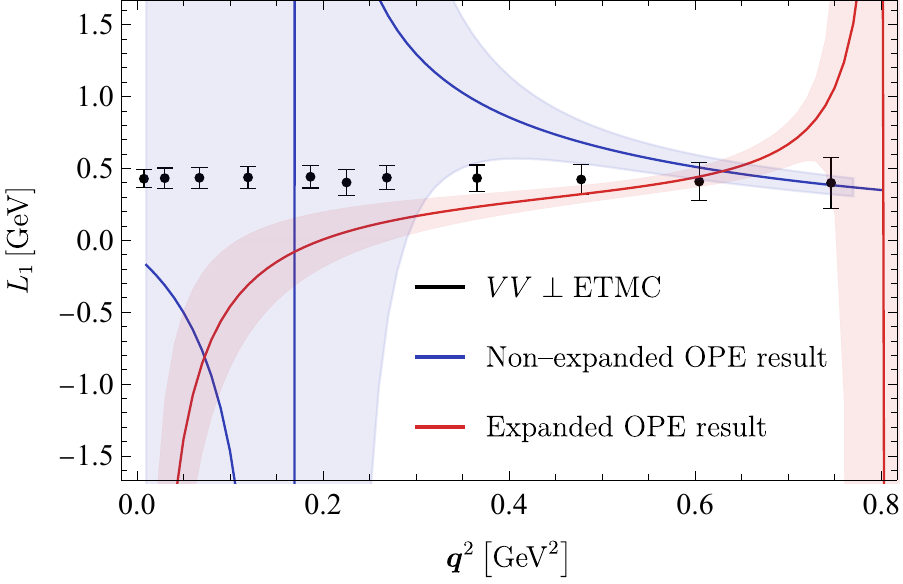}
  \caption{Differential moment $L_1(\bm{q}^2)$  in the various channels. The plots show the comparison  between OPE and ETMC data.  }
  \label{fig:mom1ETMC}
\end{figure}

Figs.~\ref{fig:mom1JLQCD} and \ref{fig:mom1ETMC} show  $L_1(\bm{q}^2)$  in the individual  channels, for the  JLQCD and ETMC cases.  
In general, we observe 
good agreement with the lattice data, especially at low $\bm{q}^2$. However, the expansion 
in powers of $\alpha_s $ and $1/m_b$ of the denominator  is not justified when the lowest-order contribution to the denominator becomes particularly small or has a zero, like in the $VV_\parallel$ and $VV_\perp$ channels. In these cases we also 
show the unexpanded version of the ratio, whose uncertainty is much larger, but we stress that 
away from the singularities the expanded form is preferable, and this appears to be confirmed by better agreement with the lattice data. 

\begin{figure}[tbp]
  \centering
\includegraphics[width=7.5cm]{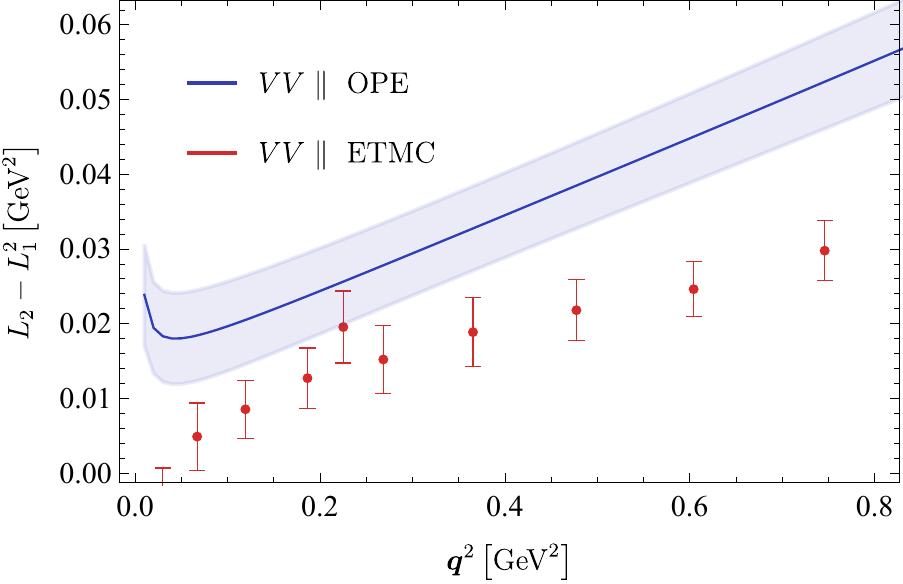}
  \includegraphics[width=7.5cm]{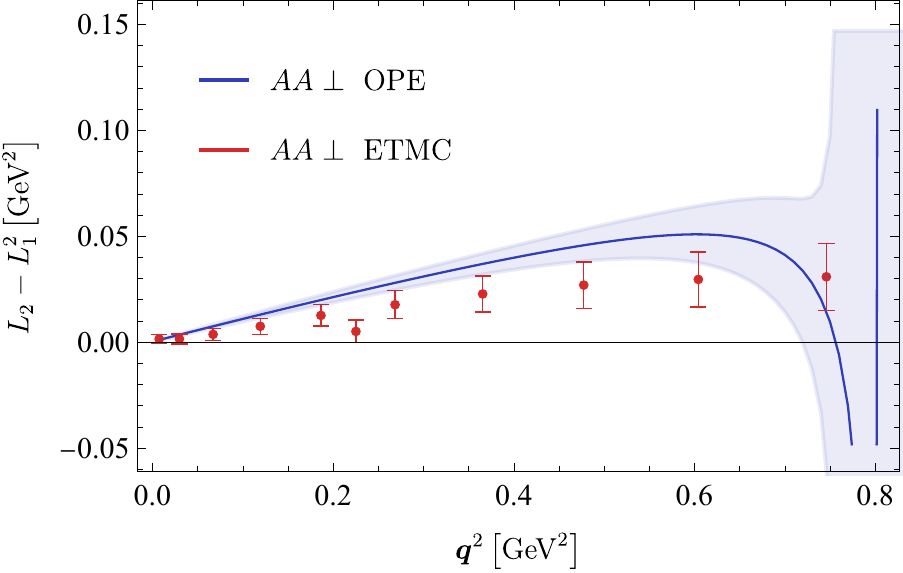}\\
  \includegraphics[width=7.5cm]{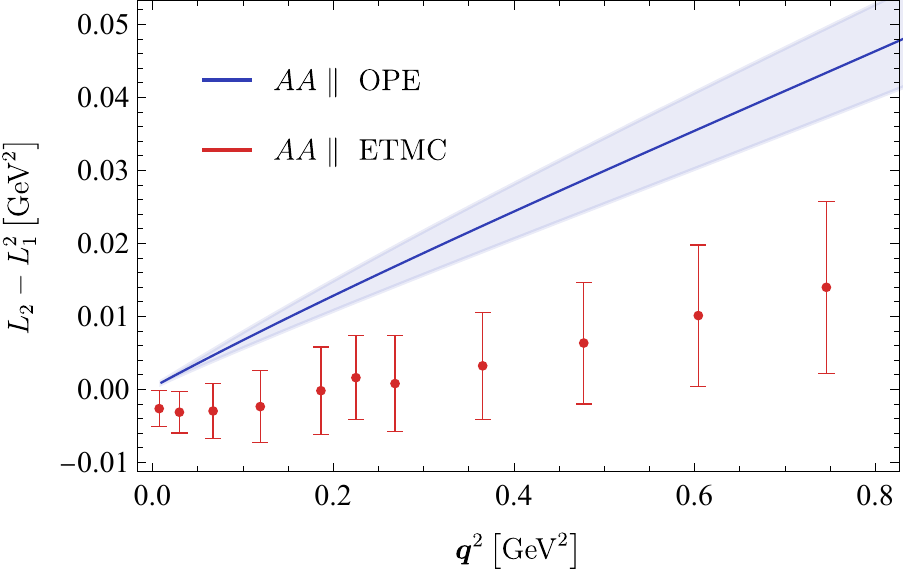}
   \caption{Differential moment $L_{2c}=L_2-L_1^2$  in the various channels. The plots show the comparison  between OPE and ETMC data.  }
  \label{fig:mom2cETMC}
\end{figure}

Figure~\ref{fig:mom2cETMC} shows the second central moment computed at different values of $\bm{q}^2$ in the ETMC case. We do not display the $VV_\perp$ channel, for which the OPE
result would have a very large uncertainty.  In the case of $L_{2c}(\bm{q}^2)$ the OPE does not reproduce the lattice 
results within uncertainties, except for very small $\bm{q}^2$. It is certainly possible that our method to estimate the OPE uncertainty fails 
here as a result of multiple cancellations between large contributions to $L_2$ and $L_1^2$ which are not necessarily replicated by higher-order contributions. On the other hand, 
it has not yet been possible to estimate discretisation and finite volume effects on
our lattice results, and the additional systematics could affect this particular quantity in a relevant way. For this quantity we do not display the comparison with the JLQCD data, which agree with the OPE but have  very large uncertainties.
 
\begin{figure}[tbp]
  \centering
  \includegraphics[width=7.5cm]{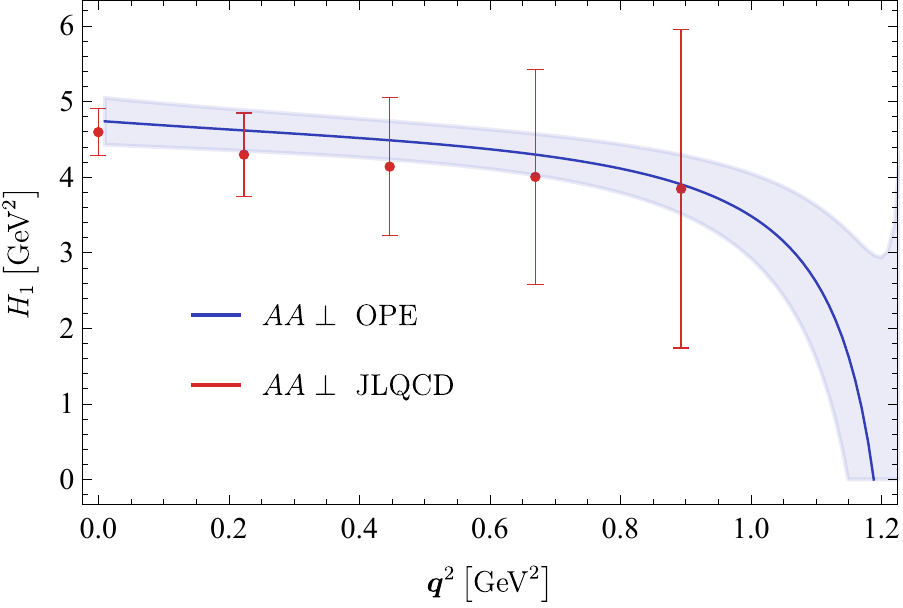}
  \includegraphics[width=7.5cm]{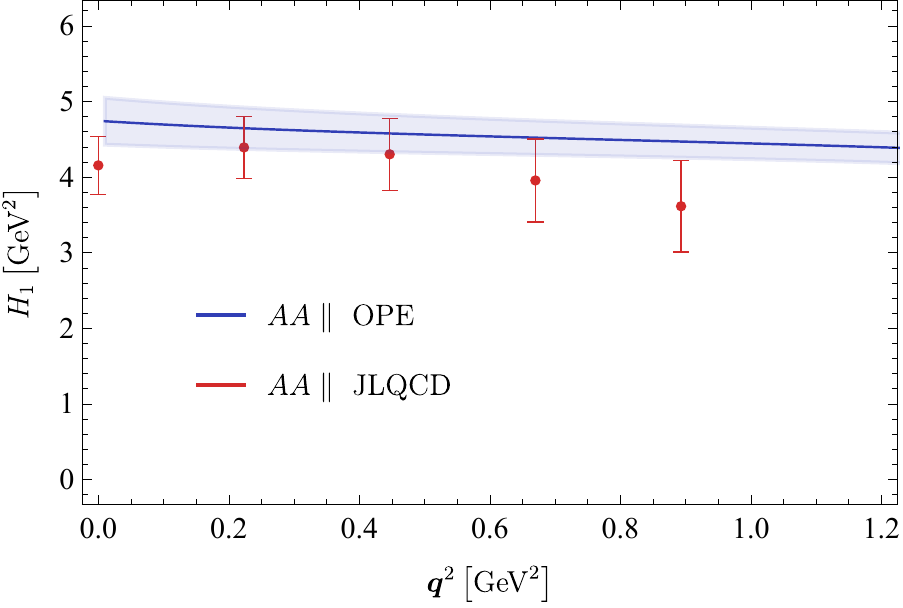}
\\
\includegraphics[width=7.5cm]{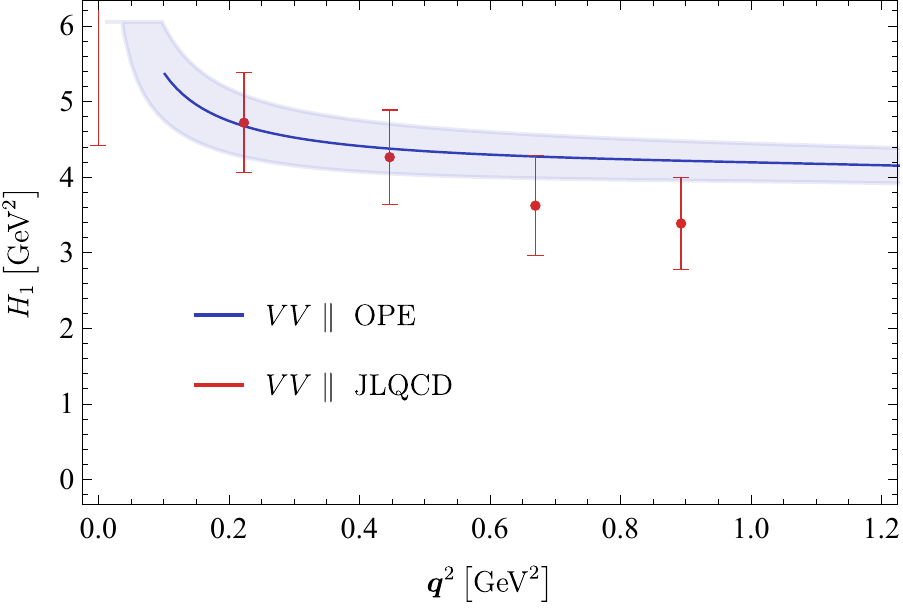}
   \caption{Differential moment $H_1(\bm{q}^2)=\langle M_X^2\rangle(\bm{q}^2)$  in the various channels. The plots show the comparison  between OPE and JLQCD data.  }
  \label{fig:H1JLQCD}
\end{figure}

\begin{figure}[tbp]
  \centering
  \includegraphics[width=7.cm]{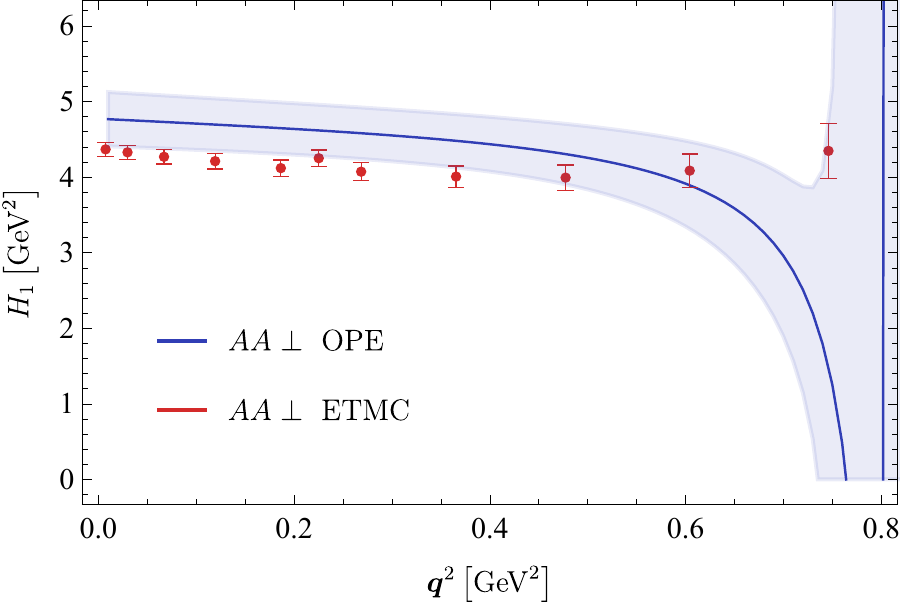}
  \includegraphics[width=7.cm]{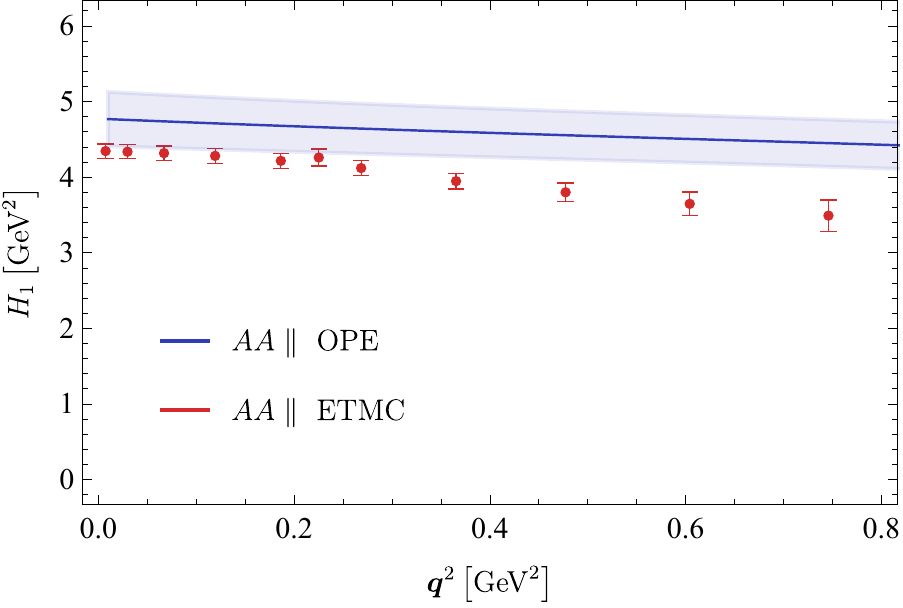}
\\
\includegraphics[width=7.cm]{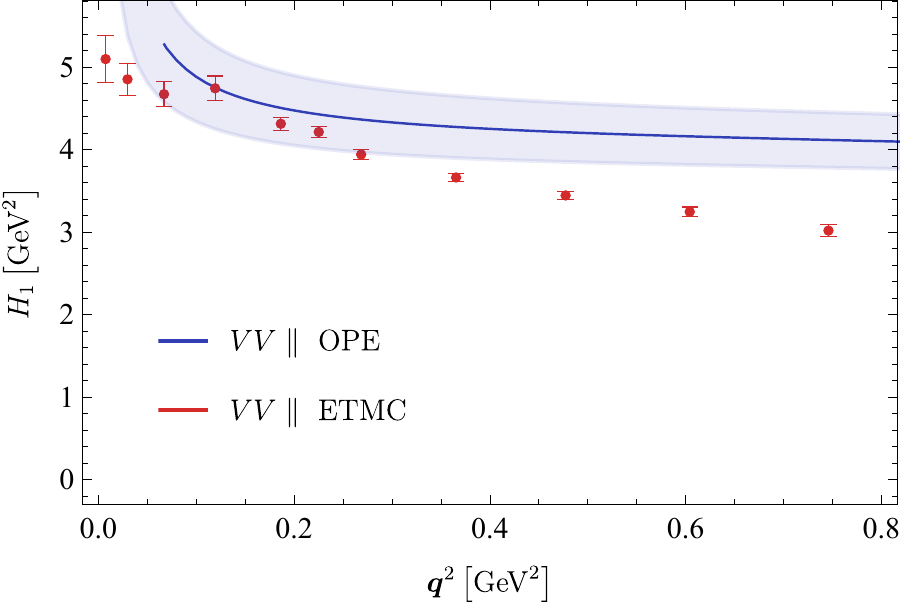}
   \caption{Differential moment $H_1(\bm{q}^2)=\langle M_X^2\rangle(\bm{q}^2)$  in the various channels. The plots show the comparison  between OPE and ETMC data.  }
  \label{fig:H1ETMC}
\end{figure}

\begin{figure}[tbp]
  \centering
  \includegraphics[width=7.cm]{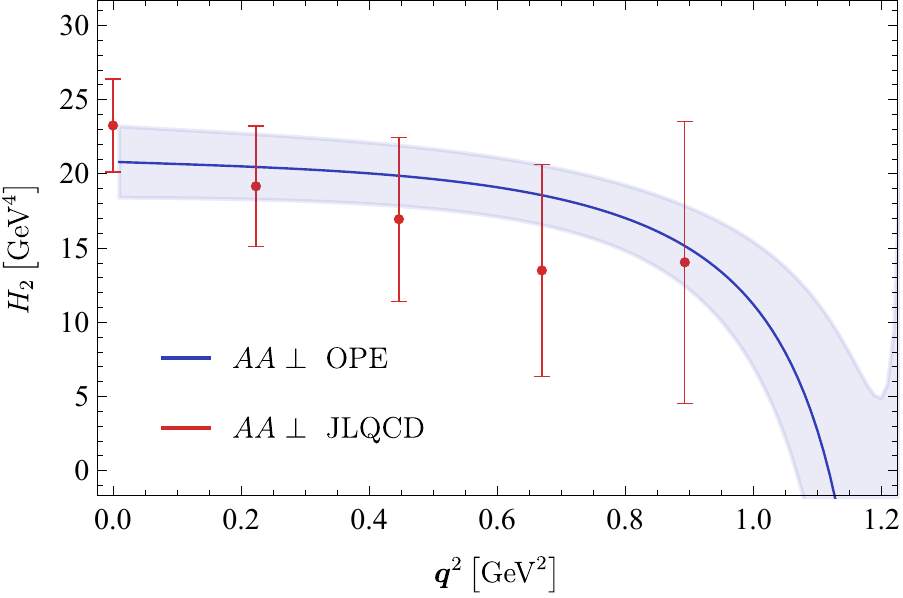}
  \includegraphics[width=7.cm]{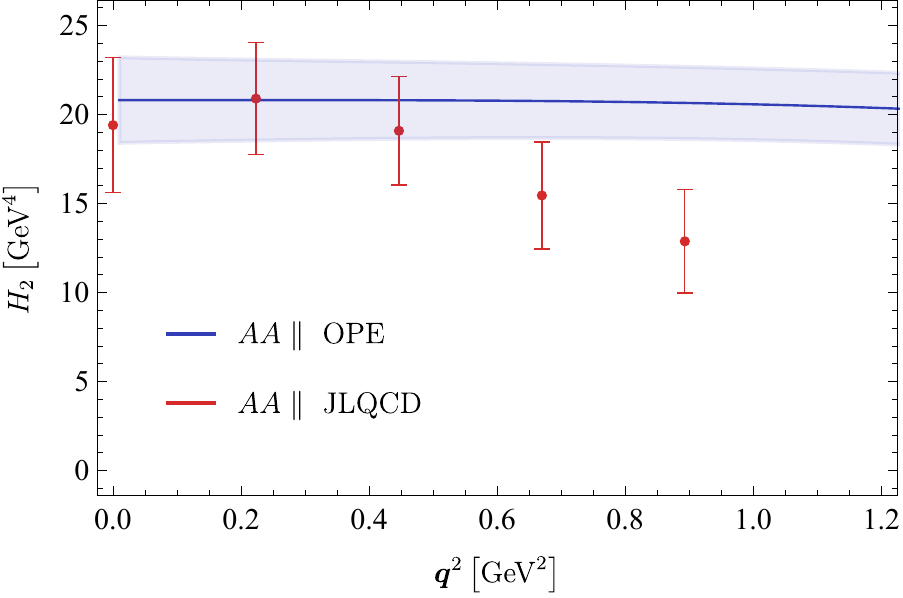}
\\
\includegraphics[width=7.cm]{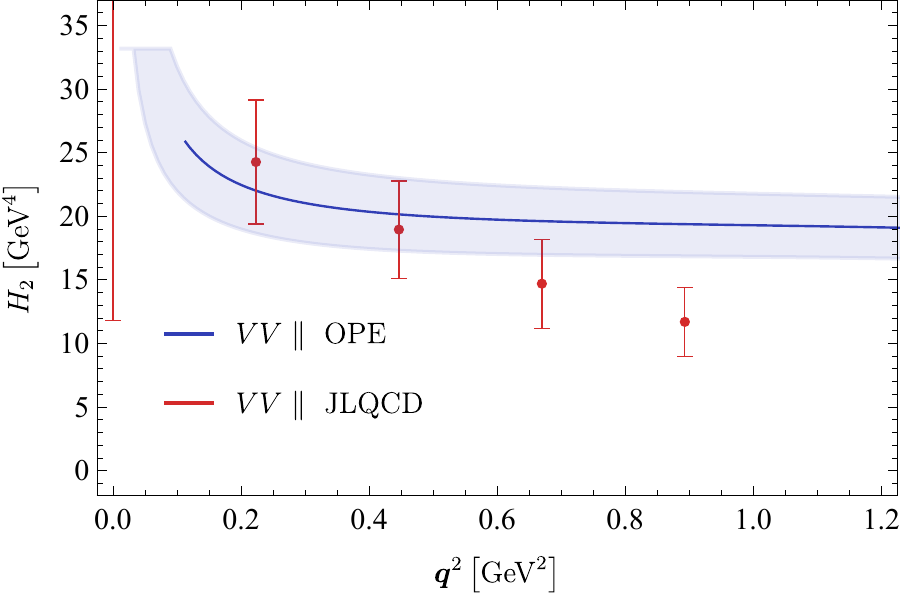}
   \caption{Differential moment $H_2(\bm{q}^2)=\langle M_X^4\rangle(\bm{q}^2)$  in the various channels. The plots show the comparison  between OPE and JLQCD data.  }
  \label{fig:H2JLQCD}
\end{figure}

We also looked at the moments of the hadronic invariant mass. In figure~\ref{fig:H1JLQCD} we show the mean hadronic mass $\langle M_X^2\rangle$ as a function of $\bm{q}^2$ computed from JLQCD configurations in comparison with the OPE predictions. 
Again, we do not display the $VV_\perp$ channel because of the large OPE uncertainty. We observe excellent agreement except at large $\bm{q}^2$, but the 
lattice uncertainty is larger here than in the case of the leptonic moments. Analogous plots for the ETMC calculation are shown in  fig.~\ref{fig:H1ETMC}.
In figure~\ref{fig:H2JLQCD} we also show  $\langle M_X^4\rangle$ as a function of $\bm{q}^2$ with JLQCD data.

\subsubsection{Total width and moments}
We perform a comparison between  OPE predictions and lattice results also in the case of  the total semileptonic width and of the global moments introduced in eq.~(\ref{eq:MX2}) and in eq.~(\ref{eq:El}).
In this case the OPE results are going to be slightly more accurate as we can take advantage of 
existing two- and even three-loop calculations \cite{Fael:2020tow}. 
We can also test the relevance of the singularity at $\bm{q}^2_{\mathrm{max}}$.
The lattice results for the $\bm{q}^2$ spectrum can be interpolated by polynomials or piecewise polynomials,
leading to the results shown in table~\ref{tab:totalJLQCD} and in table~\ref{tab:totalETMC}.
 As the $\bm{q}^2$-spectrum is peaked near $\bm{q}^2_{\mathrm{max}}$, see figure~\ref{fig:decay_rate},
the total width is particularly sensitive to that region. In the JLQCD case the limited 
number of $\bm{q}^2$ points makes the extrapolation to the highest $\bm{q}^2$ values more uncertain, with clear implications on the estimate of the total width. On the other hand, it
is difficult to estimate such uncertainty, hence table~\ref{tab:totalJLQCD}
shows only the statistic uncertainty.
	
In the OPE the total width receives large and concurring power and perturbative corrections, which reflect in a $\sim$ 20--40\% uncertainty. This is at variance with what happens in the case of the physical $b$ quark, for which a recent estimate of the total uncertainty is about $2\%$~\cite{Bordone:2021oof}. Indeed, the convergence of the OPE expansion 
deteriorates rapidly as $m_b$ decreases approaching $m_c$, even from 2.7 to 2.4 GeV.
To illustrate this point we show the various contributions to the semileptonic width in the ETMC case:
\begin{equation}
\frac{\Gamma}{|V_{cb}^2|} = \Big[ 3.03 - 0.32_{pert} 
- 0.65_{\mug}-0.09_{\mupi}-0.66_{\rd} -0.10_{\rls}+\dots
\Big] \times 10^{-13}\, \mathrm{GeV}\;,
\end{equation}
where the perturbative contribution includes $O(\alpha_s^3)$ and the non-perturbative contributions include the $O(\alpha_s)$ corrections to the Wilson coefficients. We estimate the perturbative uncertainty by varying the scale of $\alpha_s$ between 1.5 and 3.0 GeV.
Notice that more than half of the uncertainty on the width reported in tables~\ref{tab:totalJLQCD} and \ref{tab:totalETMC} is due to the large uncertainty 
on the heavy quark masses, in both  the JLQCD and ETMC cases.

\begin{table}[tbp]
  \begin{center} \begin{tabular}{c|ll}\hline
 &JLQCD & OPE
   \\ \hline
 $\Gamma/|V_{cb}^2|  \times 10^{13}$ (GeV)  &  $4.46(21)$ &  5.7(9)  \\
 $\langle E_\ell \rangle$ (GeV) & 0.650(40)&  0.626(36) \\
 $\langle M_X^2\rangle$ (GeV$^2$)&  3.75(31)& 4.22(30)  \\
   \hline
    \end{tabular} \end{center}
       \caption{ \label{tab:totalJLQCD}  Total width and moments in the JLQCD case.
  }
\end{table}
\begin{table}[tbp]
  \begin{center} \begin{tabular}{c|ll}\hline
  & ETMC & OPE 
  \\ \hline
 $\Gamma/|V_{cb}^2|  \times 10^{13}$ (GeV)   & 0.987(60) & 1.20(46) \\
 $\langle E_\ell \rangle$ (GeV) &   0.491(15)& 0.441(43) \\
  $\langle E_\ell^2 \rangle$ (GeV$^2$) &   0.263(16)& 0.207(49) \\
 $\langle E_\ell^2 \rangle-\langle E_\ell \rangle^2$(GeV$^2$)  & 0.022(16) & 0.020(8)\\
 $\langle M_X^2\rangle$ (GeV$^2$)&  3.77(9)&  4.32(56)\\
   \hline
    \end{tabular} \end{center}
       \caption{ \label{tab:totalETMC}  Total width and moments in the ETMC case.
  }
\end{table}

For what concerns the leptonic moments, only the $O(\alpha_s^2)$ corrections
have been computed, either numerically for physical values of the heavy-quark masses~\cite{Melnikov:2008qs}, or analytically in an expansion  up to $O(r^7)$ in powers of $r=m_c/m_b$~\cite{Pak:2008cp}. Unfortunately, this expansion converges slowly  and does not provide
reliable results for $r\sim 0.5$, which is the value relevant in the ETMC case. We therefore show results computed to $O(\alpha_s)$ 
and include the $O(\alpha_s \mu_{\pi,G}^2/m_b^2)$
corrections discussed in ref.~\cite{Alberti:2013kxa} as well. The first moment
in the ETMC case is given by
\begin{equation}
\langle E_\ell \rangle =  \Big[   0.533 - 0.051_{\mug}+0.021_{\mupi}
- 0.051_{\rd}-0.003_{\rls}-0.008_{\alpha_s}+\dots
\Big] \,\mathrm{GeV},
\end{equation}
where both  power and  perturbative corrections are smaller than in the total width.
Similarly, the second central moment $L_{2c}=\langle E_\ell^2 \rangle -\langle E_\ell \rangle^2$ is given by
 \begin{equation}
L_{2c}=  \Big[   0.0297 - 0.0057_{\mug}+0.0103_{\mupi}
- 0.0167_{\rd}+0.0006_{\rls}+0.0021_{\alpha_s}+\dots
\Big] \,\mathrm{GeV}.
\end{equation}
As shown in tables~\ref{tab:totalJLQCD} and \ref{tab:totalETMC},  there is reasonable agreement between OPE and both JLQCD and ETMC data in all cases.
As a general comment, we stress that the large contributions of $\rd$ are related to a kinematically enhanced Wilson coefficient and do not necessarily imply 
similarly large higher-power corrections.

Finally, the OPE prediction for the first hadronic mass moment in the JLQCD case is
\begin{equation}
\langle
M_X^2\rangle=
 \Big[ 3.84- 0.36_{\mupi} + 0.23_{\mug} + 0.41_{\rd}   + 0.05_{\rls}+0.04 _{\alpha_s}  +\dots \Big]\,
\mathrm{GeV}^2,
\end{equation}
where we do not include  the $O(\alpha_s/m_b^2)$ corrections and consequently enlarge the uncertainty slightly. The OPE prediction for the first hadronic moment is in reasonable agreement with both the JLQCD and ETMC values, see table~\ref{tab:totalJLQCD} and table~\ref{tab:totalETMC}.

\subsection{Determination of the OPE parameters}
\label{subsec:determination_of_OPE_parameters}
As different physical quantities have a different dependence on the OPE parameters, it is 
possible to constrain their values using lattice data. The analytic expressions for the 
power corrections to the differential
$\bm{q}^2$ distribution and for the moments, which encode this dependence, are rather lengthy and are provided in an ancillary Mathematica file. 

To illustrate this point, let us consider a few examples using simpler
approximate formulas, and focussing on the differential leptonic  moments at moderately low $\bm{q}^2$,
where the OPE is more reliable. We choose  
 a  $\bm{q}^2$ value for which we have lattice data, $\bm{q}^2_*$ = 0.1865~GeV$^2$.
 In the ETMC setup, the OPE prediction for $L_1(\bm{q}^2_*)$ can be approximated by
\begin{eqnarray}
L_1^{VV\parallel}(\bm{q}^2_*)\!&\simeq&\! 0.5597+  \frac12 \delta_b - 0.47 \delta_c + 0.056 \mug - 0.19 \mupi - 
0.094 \rd - 0.057 \rls\;, \nonumber \\
L_1^{AA\parallel}(\bm{q}^2_*)\!&\simeq&\! 0.5455+  \frac12 \delta_b - 0.47 \delta_c - 0.141 \mug - 0.074 \mupi - 0.069 \rd + 0.043 \rls \;, \nonumber \\
L_1^{AA\perp}(\bm{q}^2_*)\!&\simeq& \!0.5448+  \frac12 \delta_b - 0.47 \delta_c -0.175 \mug - 0.033 \mupi - 
0.101 \rd + 0.039 \rls \;, \nonumber
\end{eqnarray}
where $\delta_{b}=m_b-2.39$, $\delta_c=m_c-1.19$, and all quantities are expressed in GeV to the appropriate power. Notice that the lowest order expression for the differential leptonic moments is universal, namely does not depend on the channel. 
We do not consider the $VV_\perp$ channel because, as discussed above, the expanded form does not provide a good approximation.
The analogous expressions for the second central moments are
\begin{eqnarray}
 L_{2c}^{VV\parallel}(\bm{q}^2_*)&\simeq& 0.005   + 0.010 \mupi +
0.052 \rd - 0.015 \rls\;, \nonumber \\
 L_{2c}^{AA\parallel}(\bm{q}^2_*)&\simeq&
0.009 + 0.010 \mupi - 0.058 \rd + 0.011 \rls\;, \nonumber \\
 L_{2c}^{AA\perp}(\bm{q}^2_*)&\simeq&
0.019 +0.010 \mupi - 
0.026 \rd -0.002 \rls\;. \nonumber
\end{eqnarray}
Each of these moments has a different dependence on the non-perturbative parameters
and they can be used in a fit to the lattice ETMC results to obtain constraints on those parameters. In fact,
using only these six inputs with their theoretical uncertainty does not lead to any 
improvement on the constraints given in table~\ref{tab:OPEinputs}.
Considering 
additional  $\bm{q}^2$ points enhances the sensitivity to the non-perturbative parameters,
but one has to estimate the correlation among the theoretical uncertainties at adjacent 
$\bm{q}^2$ points. One can also include in the fit the data for the $\bm{q}^2$ distribution in the different channels, 
as well as additional moments like the hadronic mass 
moments. A global fit to lattice data is however beyond the scope of this paper, especially because our estimate of the lattice systematic uncertainty is incomplete. We stress that the limiting factor here is not the statistical uncertainty of the present ETMC calculation, but the theoretical uncertainty we attach to 
the OPE predictions. In this respect the unphysical case we have considered, with the partonic energy 
release (of the order of $m_b-m_c$) about a factor 2 (JLQCD) or 3 (ETMC) smaller than in reality, is 
strongly penalising. At the physical point the OPE enjoys a much better convergence and the 
prospects for constraining the non-perturbative 
parameters are better than it appears from this exploratory study.

\begin{figure}[tbp]
  \centering
\includegraphics[width=7.5cm]{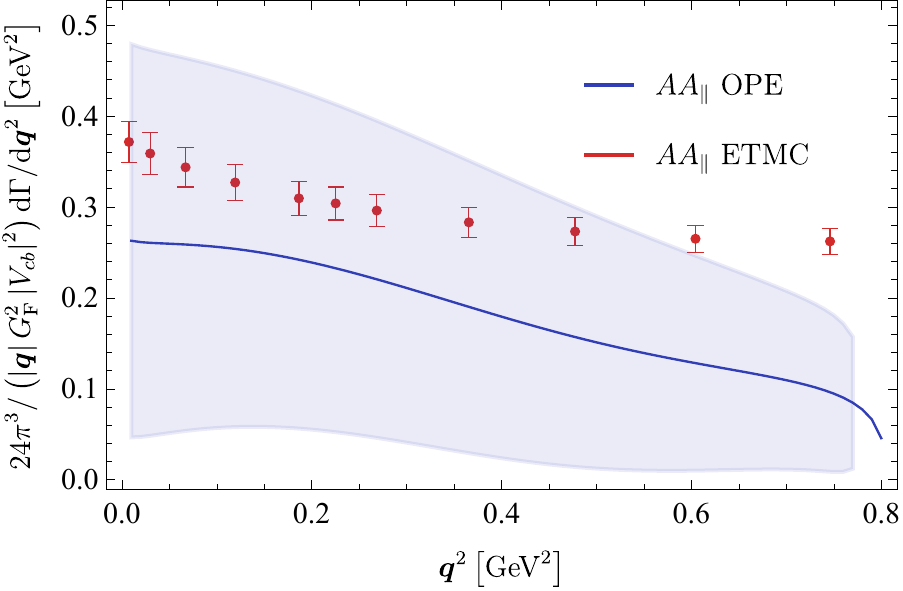}
  \includegraphics[width=7.5cm]{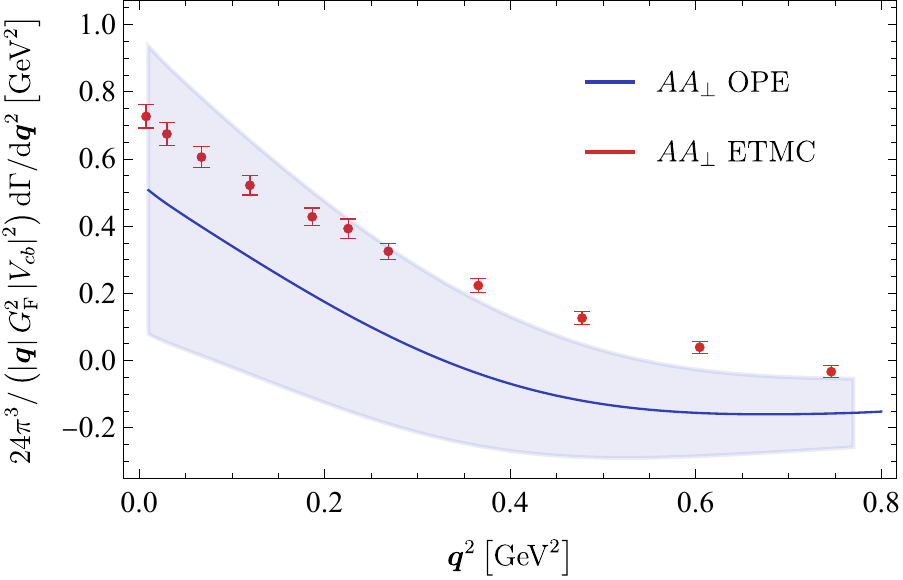}\\
  \includegraphics[width=7.5cm]{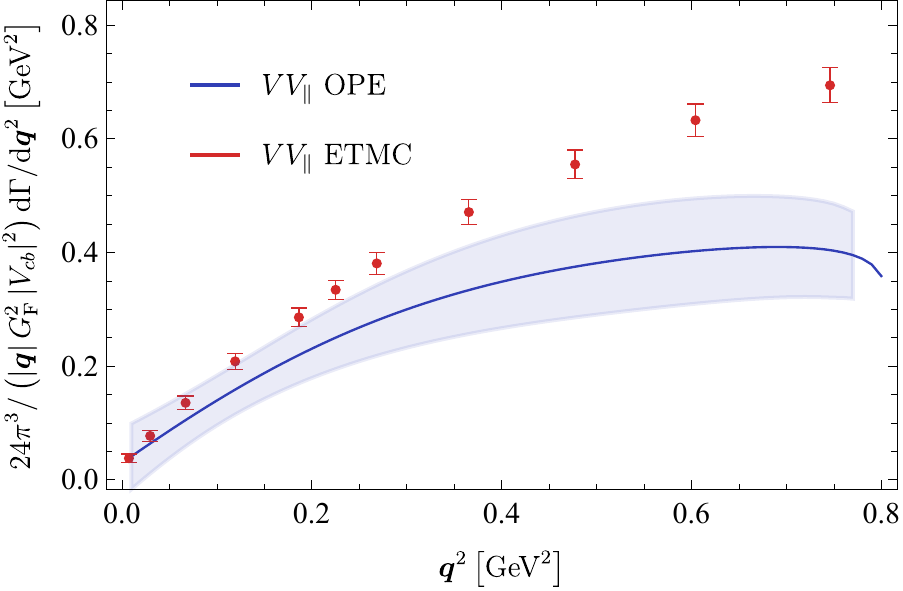}
   \includegraphics[width=7.5cm]{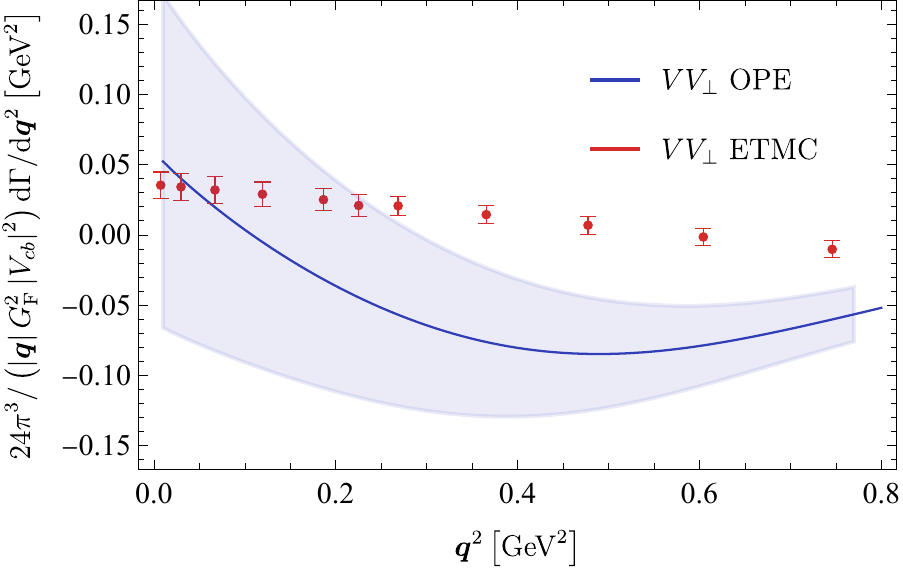}
   \caption{ Differential $\bm{q}^2$ spectrum computed with a sigmoid approximation to the 
   kernel with $\sigma=0.12 m_{B}$. The plots show the comparison  between OPE and ETMC data.  }
  \label{fig:smoothkernel}
\end{figure}

\subsection{Computations with a smooth kernel}
\label{subsec:smooth_kernel}
In sections~\ref{sec:formulation_of_the_method_and_application_to_observables} and \ref{sec:numerical_implementation_in_lattice_QCD} we have seen that the reconstruction of the discontinuous kernel is one of the 
main problems in the calculation of physical quantities. As far as the comparison with the OPE 
is concerned, however, the kernel does not need to be discontinuous. Indeed, one can 
compute inclusive (unphysical) quantities in the OPE employing a smooth kernel ($\sigma\neq 0$) and
compare them directly with the analogous quantities computed on the lattice. In this way it is possible to check that  the level of agreement between the two calculations is not affected by the $\sigma\to 0$ limit, and to extract information on the non-perturbative parameters of the OPE, as well as on the heavy quark masses, from slightly more precise lattice data.

\begin{figure}[tbp]
  \centering
\includegraphics[width=7.5cm]{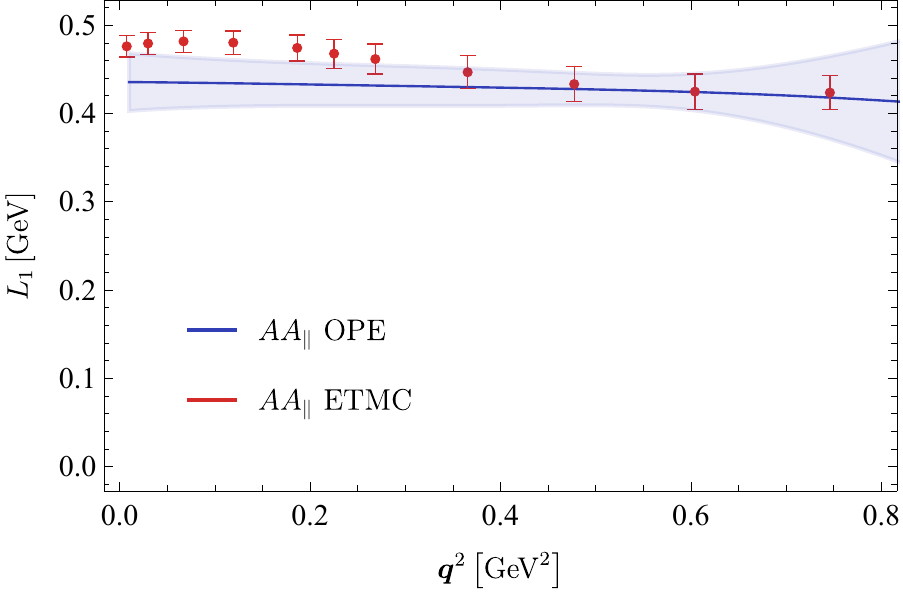}
  \includegraphics[width=7.5cm]{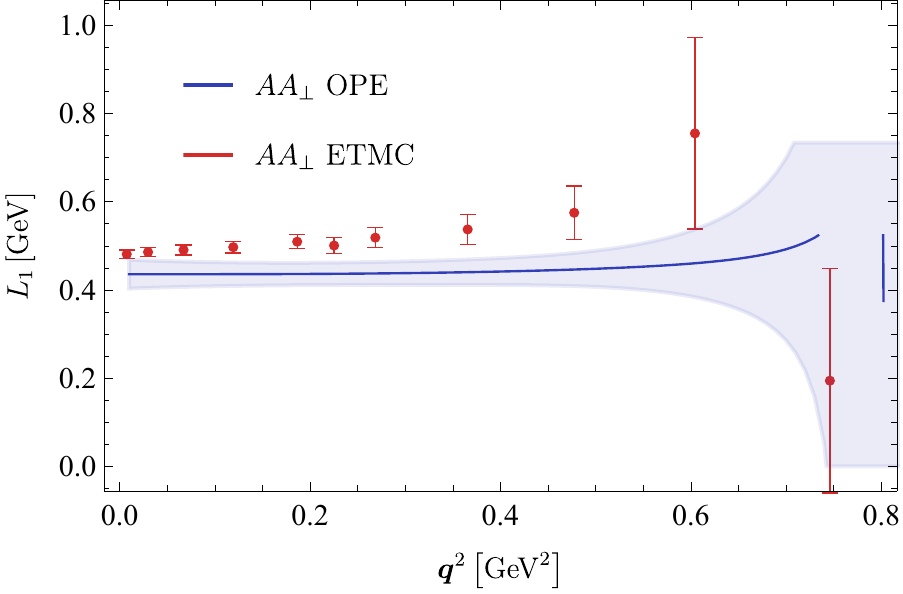}\\
  \includegraphics[width=7.5cm]{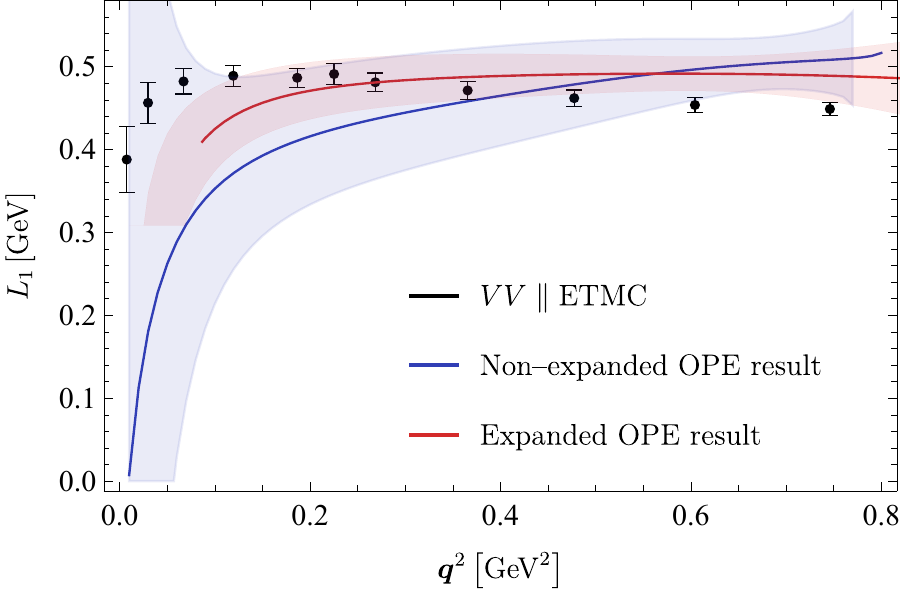}
   \caption{ Differential first lepton moments computed with a sigmoid approximation to the 
   kernel with $\sigma=0.12 m_B$ . The plots show the comparison  between OPE and ETMC data.  }
  \label{fig:smoothL1}
\end{figure}

\begin{figure}[tbp]
  \centering
\includegraphics[width=7.5cm]{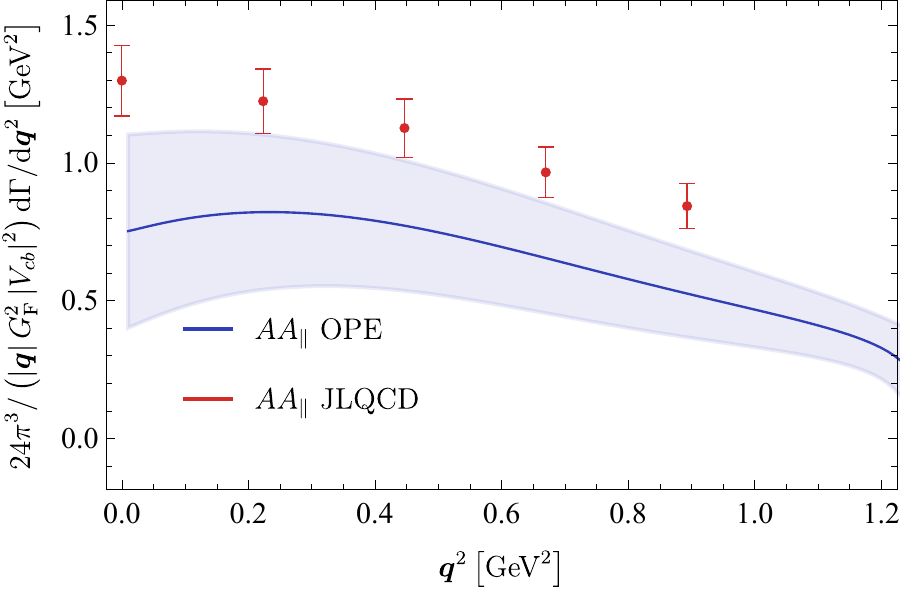}
  \includegraphics[width=7.5cm]{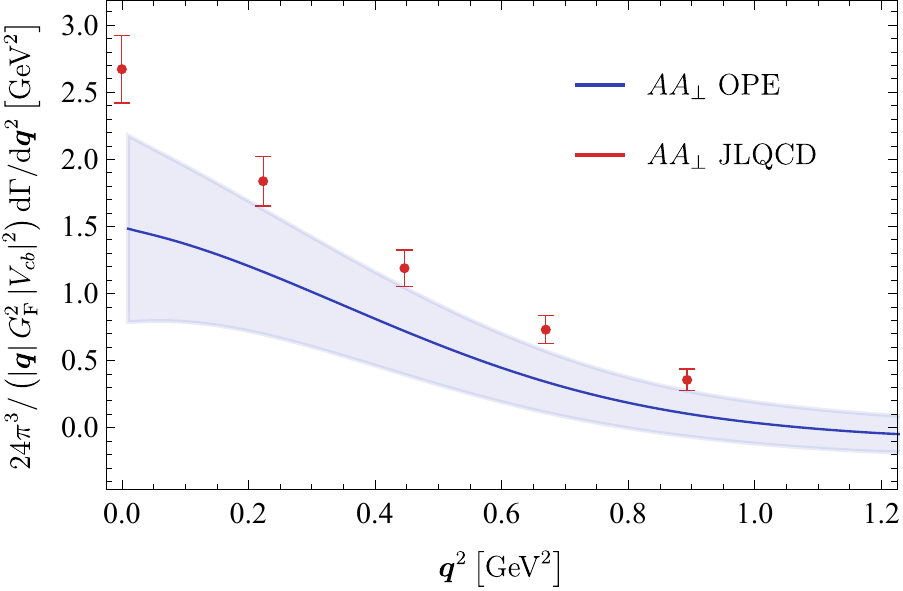}\\
  \includegraphics[width=7.5cm]{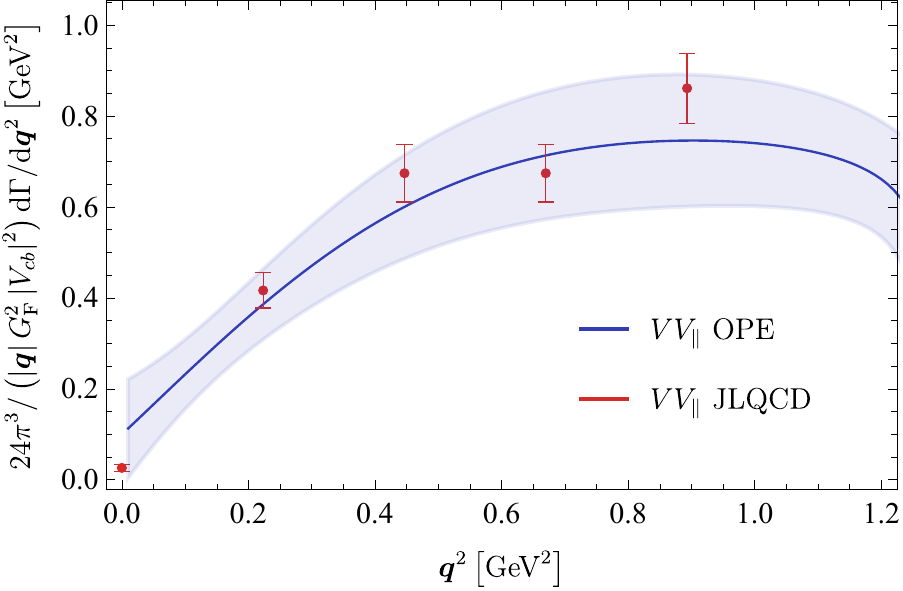}
   \includegraphics[width=7.5cm]{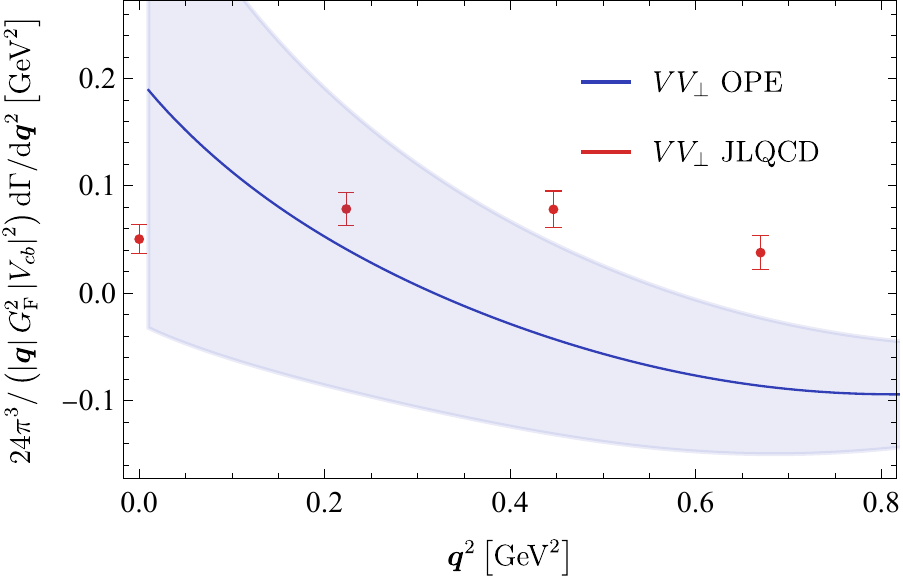}
   \caption{ Differential 
 $\bm{q}^2$ spectrum computed with a sigmoid approximation to the 
   kernel with $\sigma=0.1/a$.   
    The plots show the comparison  between OPE and JLQCD data.  }
  \label{fig:smoothXbarJLQCD}
\end{figure}

In figure~\ref{fig:smoothkernel} we show the $\bm{q}^2$ spectrum in the different channels computed on the lattice using the sigmoid approximation $\theta_\sigma^{\tt s}$ of eq.~(\ref{eq:thetasss}) for $\theta(\omega_{\mathrm{max}}-\omega)$  with $\sigma =0.12 m_{B}$.
 In the OPE calculation, where the partonic kinematics holds, 
we replace $\theta(1-\hat \eta - \sqrt{\bm{q}^2})$ by the sigmoid
$\theta_\sigma^{\tt s}(1-\hat \eta - \sqrt{\bm{q}^2})$
using  $\sigma =0.12 m_{B}$.
At low $\bm{q}^2$
the agreement between OPE and ETMC data is similar to that in figure~\ref{fig:diffqETMC},
while at large $\bm{q}^2$ there is marginal improvement, as expected because the smearing
occurs over a larger $\omega$ range.
In figure~\ref{fig:smoothL1} we show the first differential leptonic moment $L_1(\bm{q}^2)$ in the different channels, excluding $VV_\perp$ because of the large uncertainties in the OPE calculation. With respect to figure~\ref{fig:mom1ETMC} we observe a marked improvement of the agreement between OPE and ETMC data at large $\bm{q}^2$ in the $AA_\parallel$ and $VV_\parallel$ channels, while in the $AA_\perp$ channel the agreement is slightly worse.
Finally, in figure~\ref{fig:smoothXbarJLQCD} we show the $\bm{q}^2$ spectrum in the different channels computed from the JLQCD configurations using the sigmoid approximation $\theta_\sigma^{\tt s}$ with $\sigma =0.1/a$. Here the overall agreement between lattice calculations and OPE is similar to figure~\ref{fig:diffqJLQCD}, but now the $\bm{q}^2$ dependence of the lattice data is closer to the OPE result, obtained using $\theta_\sigma^{\tt s}(1-\hat \eta - \sqrt{\bm{q}^2})$
with $\sigma =0.1$.


\section{Discussion and future prospects}
\label{sec:discussion_and_future_prospects}

In this article we have presented the first comprehensive investigation of inclusive semileptonic 
$B$-meson decays on the lattice. Using the method of ref.~\cite{Gambino:2020crt} we have 
computed various inclusive observables with gauge-field ensembles generated by the JLQCD and ETM collaborations for unphysically light values of the $b$ quark mass (about $2.7$~GeV and $2.4$~GeV, respectively) and $m_c$ close to its physical value. In this exploratory study we have not performed the continuum and infinite-volume limits.

An important feature of the method we have adopted is that it requires the approximation of the energy-integral kernel. The kinematics of the inclusive semileptonic decay involves a discontinuity at the boundary of the phase space, for which a reasonable approximation with the Euclidean correlator obtained on the lattice is impractical. The problem can be dealt with using a sequence of smooth kernels, parametrized by a smearing width $\sigma$, which converge to the physical phase space in the limit $\sigma\to 0$. As emphasized in section~\ref{sec:formulation_of_the_method_and_application_to_observables}, the $\sigma\to 0$ limit does not commute with the infinite-volume limit that has to be taken first. Under the assumption that finite volume effects are negligible with respect to the statistical errors associated with our lattice results, we have studied the $\sigma\to 0$ extrapolation in detail and found that it does not induce a significant uncertainty.

We have compared the JLQCD results with the contributions of the charmed ground states, estimated from a JLQCD calculation of the $B_s \to D_s^{(*)}$ form factors for the same values of the heavy-quark masses (details are given in the appendix~\ref{app:ground_state}). Due to the proximity between the charm and bottom masses and to the limited phase space available in the decay, the inclusive results are nearly saturated by the ground-state contributions. Although the correlator themselves show the presence of excited states, their contribution to the inclusive rate is relatively small. The ETMC results obtained at even lower $b$ quark mass are also expected to be largely dominated by the ground states.

While the JLQCD and ETMC results cannot be compared directly as they are obtained at different $b$ quark masses, they can be both compared with the expectations from the OPE, assuming that discretisation and finite-volume effects are negligible. When the OPE can be considered reliable, the agreement with both JLQCD and ETMC results is generally good, while we observe possible indications of quark-hadron duality violation at large $\bm{q}^2$.
The variance of the lepton energy distribution also shows a clear and unexpected deviation, which could be due to underestimated uncertainties in our OPE calculation or to non-negligible lattice systematics. To the best of our knowledge, this is the first time that the onset of quark-hadron duality is studied on the lattice.
For $m_b\sim$ 2.4~GeV, the OPE converges much more slowly than at the physical point, but the normalised moments allow us to perform a relatively clean comparison with the lattice data.
 
We have found that the calculation of the total width and of other global quantities like the moments of the lepton energy or of the hadronic invariant-mass distribution depends crucially on the number of $\bm{q}^2$ points that can be computed on the lattice. In the ETMC calculation the flexibility 
due to the use of twisted boundary conditions has allowed us to reach an accuracy of 6\% on the total width and 3\% on the first leptonic moment. These uncertainties do not yet include several lattice systematics that need to be considered, but are dominated by statistical uncertainties and could be improved with a dedicated effort. This is an aspect which will become important for future phenomenological applications, which should also focus on reaching the physical $b$ mass. 

Finally, we have shown that one can constrain the non-perturbative parameters in the OPE  analysis from our results. We have not attempted a fit to the lattice data in the unphysical setup we have considered, as this is penalised by large uncertainties from higher-dimensional operators. With larger $b$-quark masses these uncertainties will be reduced and the data obtained at different values of $m_b$ will provide an additional handle on the non-local matrix elements that appear in eq.~(\ref{nonloc}).

There are certainly many issues to be improved or investigated in order to 
get results of direct phenomenological relevance. First, we have not yet studied the 
continuum and infinite-volume limits. Although we have presented a rather detailed discussion of the systematics associated with the reconstruction of the smearing kernels, including the required extrapolation at vanishing smearing radius, this last step is only permitted after having checked the onset of the infinite-volume limit. The continuum and infinite-volume limits can only be taken by performing calculations at different values of the lattice spacing and on different physical volumes, a task that is beyond the exploratory nature of this study and that we postpone to future work on this subject.

Second, the calculation has to be performed at the physical $b$ and light quark masses. Simulations with physical pion masses are nowadays possible and, for instance, a collection of $N_f=2+1+1$ ensembles with physical light, strange and charm quark masses has been produced by the ETM collaboration at different values of the lattice spacing and with different physical volumes. Although it is not possible to simulate directly a physical $b$ quark on these ensembles (because of potentially dangerous cutoff effects), the problem can nevertheless be approached by using well-established techniques such as the ETMC ratio method~\cite{ETM:2009sed}, based on ratios of the observable of interest computed at nearby heavy-quark masses. The ratio method has been already applied to determine the mass of the $b$ quark, the leptonic decay constants, the bag parameters of $B_{(s)}$ mesons and the matrix elements of dimension-four and dimension-five operators appearing in the Heavy Quark Expansion of pseudoscalar and vector meson masses~\cite{ETM:2011zey,ETM:2013jap,ETM:2016nbo,Gambino:2017vkx,Gambino:2019vuo}. Its main advantages can be summarised as follows:  $i)$
 $B$-physics computations can be carried out using the same relativistic action setup with which the lighter-quark computations are performed; 
 $ii)$  an extra simulation at the static point limit is not necessary, while the exact information about it is automatically incorporated in the construction of the ratios of the observable; $iii)$ the use of ratios greatly helps in reducing the discretisation effects.  However, there is an important subtlety. In order to apply the ratio method (or any other method based on extrapolations in the $b$-quark mass) in the case of the inclusive decay rates  one has to cope with the fact that at unphysical (lighter) values of the $b$-quark mass the phase-space available to the decay shrinks. This implies that some of the hadronic channels that are open at the physical value of $m_b$ are \emph{totally} excluded from the phase-space integral at $m_h<m_b$. The important point to be noticed here is that this happens when the integration limits are imposed sharply, i.e. by using the exact Heaviside functions that implement the phase-space constraint. The problem is totally analogous to the ordered double-limit required in order to deal with the finite-volume distortion of the hadronic spectral density. Indeed, we envisage applying the ratio method to $\Gamma\equiv\Gamma(m_b)$ \emph{before} taking the $\sigma\to 0$ extrapolation: while $\Gamma(m_h)$ is (at least in principle) a distribution in $m_h$, $\Gamma_\sigma(m_h)$ is certainly a smooth function that can safely be extrapolated at the physical value of $m_b$.  
Moreover, we already have simulations with $m_h \sim 0.8 m_b$, and it can be reasonably argued that for such large masses the missing (mostly continuum) states scale with $m_h$.

Although we have compared the lattice results with the OPE, a more direct and effective validation of our method would come from a comparison with experimental data, such as those for the branching ratio and for the electron energy spectrum in inclusive semileptonic decays of the $D$ or $D_s$ mesons~\cite{CLEO:2009uah,BESIII:2021duu}. Here the challenge is to get accurate results at physical light-quark masses, while the charm quark can be simulated directly on present lattices.  
Beside validating the method without extrapolations in the heavy-quark mass, a calculation of charm decays might shed light on the following two open and phenomenologically relevant questions. $i)$
 To what extent is the OPE applicable to charm decays? 
$ii)$  What is the role played by weak annihilation (WA) contributions?
The first question refers to the onset of quark-hadron duality, and a detailed study of charm decays in connection with their OPE description may yield an insight on this conceptual issue. Answering the second question may help us quantifying the role played by WA contributions in charmless semileptonic $B$ decays, hence improving the  inclusive determination of $|V_{ub}|$. If one could reproduce the lepton energy spectrum of the $D_s$ inclusive semileptonic decays that is  measured experimentally, a more ambitious future application would be a direct calculation of $B\to X_u \ell\nu$.

Finally, one may wonder whether the foreseeable precision will be sufficient for a precision determination of $|V_{cb}|$ and for interesting phenomenology. Indeed, present  experimental errors for  $B\to X_c \ell \nu$ are $1.4\%$ on the branching ratio and a few per mille
on the first few moments of the lepton energy distribution. The lattice precision is unlikely to get close to that, at least initially.
On the other hand, on a relatively short time-scale lattice calculations of inclusive semileptonic decays might be able to enhance the predictive power of the OPE by accessing other quantities that are inaccurate or beyond the reach of current experiments and are highly sensitive to the non-perturbative parameters, allowing us to validate and improve the results of the semileptonic fits on which the OPE predictions are based.

\acknowledgments
The numerical calculations of the JLQCD collaboration were performed
on SX-Aurora TSUBASA at the High Energy Accelerator Research
Organization (KEK) under its Particle, Nuclear and Astrophysics
Simulation Program, as well as
on the Oakforest-PACS supercomputer operated by the Joint Center for Advanced
High Performance Computing (JCAHPC).
We thank the members of the JLQCD collaboration for sharing the computational framework and lattice data, and Takashi~Kaneko in particular for providing the numerical data for the exclusive decay form factors.
The numerical simulations of the ETM collaboration were run on machines of the Consorzio Interuniversitario per il Calcolo Automatico dell'Italia Nord Orientale (CINECA) under the specific initiative INFN-LQCD123.
The work of S.H. is supported in part by JSPS KAKENHI Grant Number 18H03710 
and by the Post-K and Fugaku supercomputer project through the Joint
Institute for Computational Fundamental Science (JICFuS).
The work of P.G., S.M., F.S., S.S. is supported by the Italian Ministry of Research (MIUR) under grant PRIN 20172LNEEZ. This project has received funding from the Swiss National Science Foundation (SNF) under contract 200020\_204428. We warmly thank Agostino~Patella for his participation at a very early stage of this work.

\appendix

\section{Contributions from the ground states}
\label{app:ground_state}
Among the complete set of states inserted in eq.~(\ref{eq:Wmunu}), we
consider the contribution of the lowest-lying states, which are the
$S$-wave states, i.e. $D$ and $D^*$ mesons.
(Here and in the following, we omit the subscript $s$ for brevity.)
The corresponding matrix elements can be parametrized by the form
factors as
\begin{eqnarray}
  \label{eq:DVB}
  \langle D(v')|V^\mu|B(v)\rangle
  & = &
        h_+(w) (v+v')^\mu + h_-(w) (v-v')^\mu\;,
  \\
  \label{eq:DstarVB}
  \langle D^*(v',\epsilon)|V^\mu|B(v)\rangle
  & = &
        -h_V(w) \epsilon^{\mu\nu\rho\sigma} v_\nu v'_\rho
        \epsilon_\sigma^*\;,
  \\
  \label{eq:DstarAB}
  \langle D^*(v',\epsilon)|A^\mu|B(v)\rangle
  & = &
       - i h_{A_1}(w) (1+w) \epsilon^{*\mu}
        \nonumber\\
  & &  +i\left[
        h_{A_2}(w) v^\mu + h_{A_3}(w) v'^\mu \right]
        \epsilon^*\cdot v\;,
\end{eqnarray}
where $\epsilon^*$ denotes the polarization vector of the vector $D^*$
meson.
We use the HQET definition of the meson states, so that the kinematics
is parametrized by the velocities $v$ and $v'$ (with $p=m_Bv$ and $p'=m_{D^{(*)}}v'$)
and $w=v\cdot v'$. In the rest frame of the $B$ meson $\vec v'=-\vec q/m_{D^{(*)}}$.

\begin{figure}[tbp]
  \centering
  \includegraphics[width=11cm]{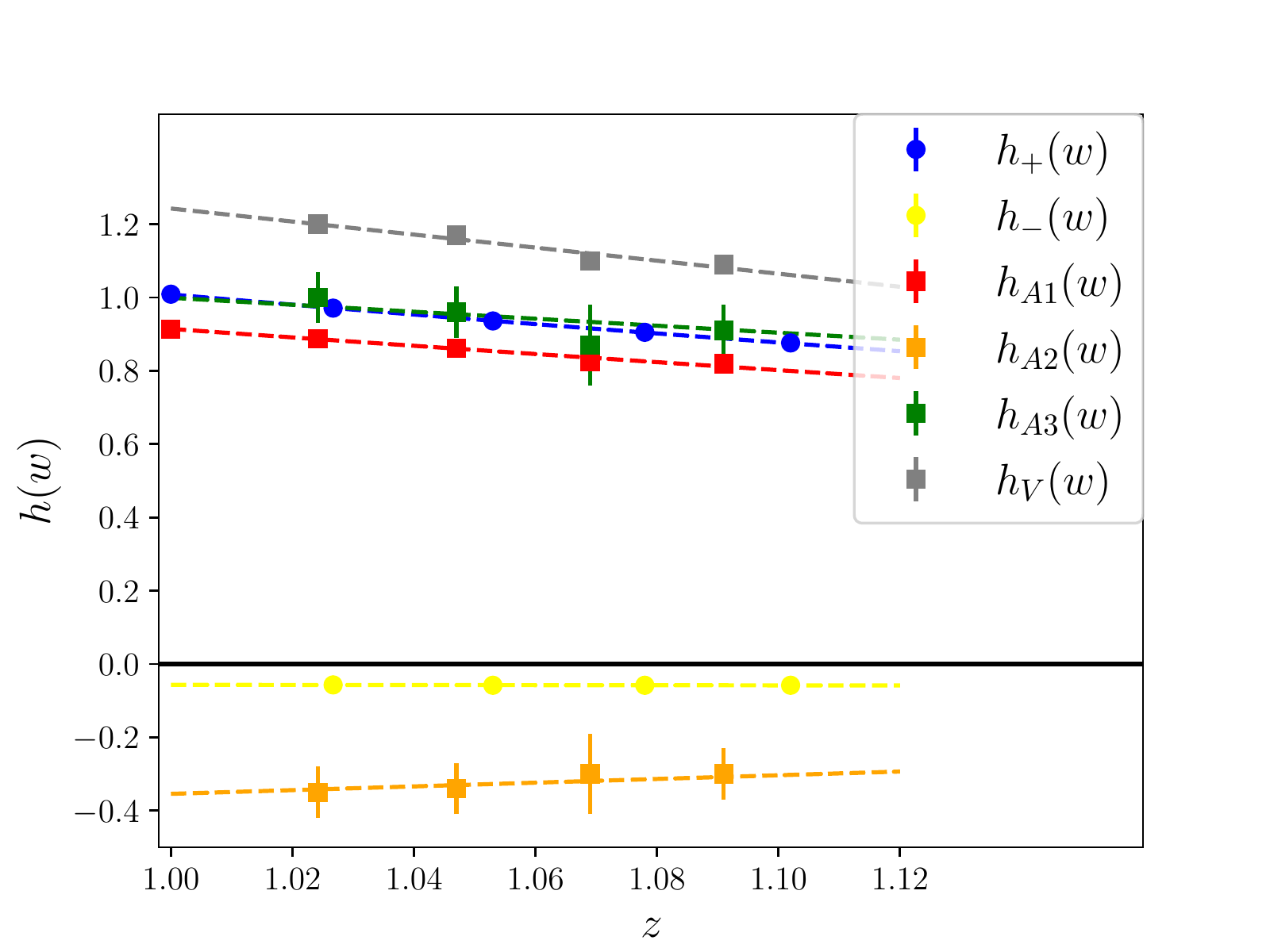}
  \caption{
    Form factors computed from three-point functions.
  }
  \label{fig:FFs}
\end{figure}

From a separate calculation of the $B\to D^{(*)}$ from factors on the
lattice with the same setup as we use for the inclusive decays,
we numerically obtain the form factors of the form
\begin{equation}
  h_X(w) = c_X^{(0)} + c_X^{(1)}(w-1) + c_X^{(2)}(w-1)^2
\end{equation}
after fitting the lattice data.
The fit is shown in fig.~\ref{fig:FFs}, and the numerical
coefficients $c_X^{(0)}$, $c_X^{(1)}$, $c_X^{(2)}$ 
are listed in table~\ref{tab:c_X}.

\begin{table}[tbp]
  \centering
  \begin{tabular}{c|ccc}
    $X$ & $c_X^{(0)}$ & $c_X^{(1)}$ & $c_X^{(2)}$\\
    \hline\hline
    $+$ & 1.0082(26) & $-$1.40(12) & 1.0(1.2) \\
    $-$ & $-$0.057(11) & $-$0.01(17) & \\
    \hline
    $A_1$ & 0.9143(34) & $-$1.17(15) & 0.4(1.6) \\
    $A_2$ & $-$0.354(75) & 0.5(1.2) & \\
    $A_3$ & 0.999(75) & $-$1.0(1.2) & \\
    $V$ & 1.243(13) & $-$1.78(20) & \\
    \hline
  \end{tabular}
  \caption{
    Numerical coefficients $c_X^{(i)}$ to parametrize the form
    factors of $B\to D$ ($X=+$ and $-$) and $B\to D^*$ ($A_1$, $A_2$,
    $A_3$ and $V$) decays.
    A polynomial expansion of the form
    $h_X(w) = c_X^{(0)} + c_X^{(1)}(w-1) + c_X^{(2)}(w-1)^2$
    is introduced.
  }
  \label{tab:c_X}
\end{table}

Now, we insert the parametrizations given in eqs.~(\ref{eq:DVB}), (\ref{eq:DstarVB}), and (\ref{eq:DstarAB})
into eq.~(\ref{eq:Wmunu}) and perform the
$\omega$ integral, which merely picks the ground state through
$\delta(p_0-q_0-E_{D^{(*)}})=\delta(\omega-E_{D^{(*)}})$.
For $\bar{X}\equiv\sum_{l=0}^2 X^{(l)}$, we obtain
\begin{equation}
  \bar{X}^{VV\parallel} =
  \frac{\bm{q}^2}{4m_DE_D} \left( (m_B+m_D)h_+-(m_B-m_D)h_- \right)
\end{equation}
for the $D$ meson contribution, which corresponds to the partial decay
rate
\begin{eqnarray}
  \Gamma^{B\to D}
  & = & \frac{G_F^2|V_{cb}|^2}{8\pi^3} \int\!d\bm{q}^2
        \frac{|\bm{q}|}{3} \cdot \frac{\bm{q}^2(m_B+m_D)^2}{4m_DE_D}
        \left[h_+-\frac{m_B-m_D}{m_B+m_D}h_-\right]^2
        \nonumber\\
  & = & \frac{G_F^2|V_{cb}|^2m_B^5}{48\pi^3}
        \int\! dw\, (w^2-1)^{3/2} r^3(1+r)^2
        \left[h_+-\frac{1-r}{1+r}h_-\right]^2,
\end{eqnarray}
where $w=v\cdot v'=\sqrt{1+\bm{q}^2/m_D^2}=E_D/m_D$.
The last line is a well-known formula for the $B\to D\ell\nu$ decay rate. 

The vector meson $D^*$ contributes in three channels: $AA_\parallel$,
$AA_\perp$, $VV_\perp$.
The contributions are
\begin{eqnarray}
  \bar{X}^{AA\parallel}
  & = & \frac{1}{4m_{D^*}E_{D^*}}
        \biggl[ (m_B-E_{D^*})E_{D^*} h_{A1}(1+w)
        \nonumber\\
  & &   + \bm{q}^2\left(h_{A1}(1+w)-h_{A2}-\frac{m_B}{m_{D^*}}h_{A3}\right)
        \biggl]^2\;,
  \\
  \bar{X}^{AA\perp}
  & = & \left[ (m_B-m_{D^*})^2-2m_B(E_{D^*}-m_{D^*}) \right]
        \frac{(1+w)^2}{2w} h_{A1}^2\;,
  \\
  \bar{X}^{VV\perp}
  & = & \left[ (m_B-m_{D^*})^2-2m_B(E_{D^*}-m_{D^*}) \right]
        \frac{\bm{q}^2}{2m_{D^*}E_{D^*}} h_V^2\;.
\end{eqnarray}
Adding them together, we obtain
\begin{eqnarray}
\label{eq:B_to_Dstar_intermediate}
  \Gamma^{B\to D^*}
  & = & \frac{G_F^2|V_{cb}|^2}{8\pi^3}
        \int\!d\bm{q}^2\frac{|\bm{q}|}{3}
        \left\{
        (q_0^2-\bm{q}^2)
        \left[
        \frac{(1+w)^2}{2w}h_{A1}^2+\frac{\bm{q}^2}{2m_{D^*}E_{D^*}}h_V^2
        \right] \right.
        \nonumber\\
  & & \left.
      + \frac{1}{4m_{D^*}E_{D^*}}
      \biggl[
      (m_B-E_{D^*})E_{D^*} h_{A1}(1+w)
      \right.
      \nonumber\\
  & & \left.
      +\bm{q}^2\left(h_{A1}(1+w)-h_{A2}-\frac{m_B}{m_{D^*}}h_{A3}\right)
      \biggl]^2
      \right\}\;,
\end{eqnarray}
where $r=m_{D^*}/m_B$ and $E_{D^*}=m_{D^*}w=m_Brw$,
while $\bm{q}^2=m_{D^*}^2(w^2-1)=m_B^2r^2(w^2-1)$, and
$q_0^2-\bm{q}^2=(m_B-m_{D^*})^2-2m_B(E_{D^*}-m_{D^*})=m_B[(1-r)^2-2r(w-2)]$.
Eq.~(\ref{eq:B_to_Dstar_intermediate}) can then be rewritten as
\begin{eqnarray}
  \Gamma^{B\to D^*}
  & = & \frac{G_F^2|V_{cb}|^2m_B^5}{48\pi^3}
        \int\!dw\, (w^2-1)^{1/2}r^3(1-r)^2(w+1)^2 |h_{A1}|^2
        \nonumber\\
  & & \times\left\{
      2\frac{r^2-2rw+1}{(1-r)^2}\left[1+\frac{w-1}{w+1}R_1^2\right]
      +\left[1+\frac{w-1}{1-r}(1-R_2)\right]^2\right\}\;,
\end{eqnarray}
with
$R_1\equiv h_V/h_{A1}$ and
$R_2\equiv (h_{A3}+rh_{A2})/h_{A1}$,
which confirms a well-known formula.

From this analysis, the contributions of the $S$-wave ground states,
$D$ and $D^*$, to the integrands
$\bar{X}^{VV\parallel}$, $\bar{X}^{VV\perp}$,
$\bar{X}^{AA\perp}$, and $\bar{X}^{AA\parallel}$
can be identified.

The contribution of the $VA$ and $AV$ insertions vanishes for the
total decay rate as well as for the hadronic mass moments, but it is
non-zero for the lepton energy moments.
In the SM the contribution of the $AV$ interference from the ground state $B\to D^*$ to
the first leptonic moment can be written as
\begin{equation}
  \bar{X}^{AV} = -\left[ (m_B-E_{D^*})^2-\bm{q}^2 \right]
  \frac{\bm{q}^2}{4E_{D^*}} (1+w) h_V h_{A1}\;.
\end{equation}

\bibliography{inclusive}

\end{document}